\documentclass[12pt,preprint]{aastex}
\usepackage{cite}

\shorttitle{The WIRED Survey} 
\shortauthors{Debes et al.}
\begin{document}
\title{The WIRED Survey II:\ Infrared Excesses in the SDSS DR7 White Dwarf Catalog}
\author{John H. Debes\altaffilmark{1,2,3}, D. W. Hoard\altaffilmark{4}, Stefanie Wachter\altaffilmark{5}, David T. Leisawitz\altaffilmark{1}, \& Martin Cohen\altaffilmark{6}}

\altaffiltext{1}{Goddard Space Flight Center, Greenbelt, MD 20771}
\altaffiltext{2}{NASA Postdoctoral Program Fellow}
\altaffiltext{3}{Space Telescope Science Institute, Baltimore, MD 21218}
\altaffiltext{4}{Spitzer Science Center, California Institute of Technology, Pasadena, CA 91125}
\altaffiltext{5}{Infrared Processing and Analysis Center, California Institute of Technology, Pasadena, CA 91125} 
\altaffiltext{6}{Monterey Institute for Research in Astronomy, Marina, CA 93933}

\begin{abstract}
With the launch of the {\em Wide-field Infrared Survey Explorer} ({\em WISE\/}), a new era of detecting planetary debris and brown dwarfs around white dwarfs (WDs) has begun with the {\em WISE} InfraRed Excesses around Degenerates (WIRED) Survey.  The WIRED Survey is sensitive to substellar objects and dusty debris around WDs out to distances exceeding 100~pc, well beyond the completeness level of local WDs.  In this paper, we present a cross-correlation of the preliminary Sloan Digital Sky Survey (SDSS) Data Release 7 (DR7) WD Catalog between the {\em WISE}, Two-Micron All Sky Survey (2MASS), UKIRT Infrared Deep Sky Survey (UKIDSS), and SDSS DR7 photometric catalogs.  From $\sim18,000$ input targets, there are {\em WISE} detections comprising 344 ``naked'' WDs (detection of the WD photosphere only), 1020 candidate WD+M dwarf binaries, 42 candidate WD+brown dwarf (BD) systems, 52 candidate WD+dust disk systems, and 69 targets with indeterminate infrared excess.  We classified all of the detected targets through spectral energy distribution model fitting of the merged optical, near-IR, and {\em WISE} photometry.  Some of these detections could be the result of contaminating sources within the large ($\approx6\arcsec$) {\em WISE} point spread function; we make a preliminary estimate for the rates of contamination for our WD+BD and WD+disk candidates, and provide notes for each target-of-interest.  Each candidate presented here should be confirmed with higher angular resolution infrared imaging or infrared spectroscopy.  We also present an overview of the observational characteristics of the detected WDs in the {\em WISE} photometric bands, including the relative frequencies of candidate WD+M, WD+BD, and WD+disk systems.

\end{abstract}


\section{Introduction}

The {\em Wide-field Infrared Survey Explorer} ({\em WISE\/}) is a NASA medium class Explorer mission that was launched on 14 Dec 2009 \citep{wright10}.  {\em WISE} mapped the entire sky simultaneously in four infrared (IR) bands centered at 3.4, 4.6, 12, and 22~$\mu$m (denoted {\em W1}, {\em W2}, {\em W3}, and {\em W4}, respectively).  {\em WISE} has several main goals, namely to take a census of cool stars and brown dwarfs (BDs) close to the Sun, to probe the dustiest galaxies in the Universe, and to catalog the Near Earth Object population \citep{wright10}.  The {\em WISE} mission will also provide crucial information about a diverse range of phenomena in the IR sky at a sensitivity 100 times better than that of the {\em Infrared Astronomical Satellite} ({\em IRAS\/}), which was launched almost 30 years ago and performed the first all sky IR survey \citep{iras84}.  

The {\em WISE} InfraRed Excesses around Degenerates (WIRED) survey is designed to detect IR excesses around white dwarfs (WDs) using photometry from the {\em WISE} catalog.  Dust, low mass companions, and cyclotron radiation from accreting magnetic WDs all emit at mid-IR wavelengths, providing a rich variety of sources to be discovered.  There are over 2100 spectroscopically identified WDs in the  McCook \& Sion catalog \citep{mccook99,hoard07} and $\sim$18,000 identified in the preliminary Sloan Digital Sky Survey (SDSS) Data Release 7 (DR7) WD catalog \citep{kleinman11}.  Because of the all sky coverage of {\em WISE}, WIRED will provide a more systematic search for IR excesses around WDs than those performed with targeted {\em Spitzer Space Telescope} \citep{werner04} observations.

Brown dwarfs discovered in orbit around WDs are particularly important for providing ``ground truth'' to spectral models of BDs and exoplanets, since WDs have well-defined cooling ages that are relatively simple to determine (e.g., \citealt{bergeron95} and references therein). Thus, a precise determination can be made of the age of a substellar companion.  Unlike observations of young substellar objects, the flux from an older BD is less sensitive to the initial conditions of formation or to errors in the age, providing a firm benchmark for a mass estimate \citep{dayjones11}.  These systems will help observers understand the rash of newly discovered massive exoplanets that are being directly imaged or detected through the secondary eclipses of transiting exoplanet systems.  

The discovery and characterization of additional WD+BD binaries offers a valuable opportunity to learn about the common envelope phase of stellar evolution by observing the state of the system that is left behind (e.g., \citealt{farihihoard10}).  The orbital distribution of BD companions to WDs might be affected by post-main sequence evolution, which can be compared to the orbital distribution of BDs around main sequence stars.  The apparent ``brown dwarf desert'' that is observed for main sequence stars can be investigated for WDs, as well, and may provide information on how BDs are formed.  Large searches have found a low frequency of WD+BD systems \citep{farihi05}, but individual discoveries of WD+BD systems are proceeding \citep{farihizuck05,dayjones08,steele09,qian09,dayjones11,luhman11}, including the discovery of some BDs that are participating in a phase of mass transfer or have survived post common envelope evolution \citep{debeslopez06,maxted06}.  The nominal sensitivity of {\em WISE} to such systems will be useful for characterizing L and T dwarfs out to distances exceeding 70~pc, providing new objects for study.

Dusty WDs, in particular, provide information on the future fate of our own Solar System, as well as planetary systems around other stars.  Planetary systems can survive post-main sequence evolution and mass loss as a central star becomes a WD \citep{duncan98}, although many objects in the inner system are expected to be destroyed through engulfment or evaporation \citep{villaver07,villaver09,nordhaus10}.  Rocky planetesimals can survive gas drag and sublimation during post-main sequence evolution \citep{jura08,dong10,bonsor10}, while simple models suggest icy planetesimals should be evaporated out to a few hundred AU for most stars \citep{stern90}.  The parent bodies of main sequence star debris disks that have been observed as part of various {\em Spitzer} surveys evolve through the post-main sequence evolution of their star and can become detectable during the planetary nebula phase \citep{chu07,chu09}, but rapidly become too cold to be observed at any wavelength relative to the luminosity of the hot WD \citep{bonsor10}. 

These planetesimals may once again become detectable later in the evolution of a WD as it cools.  Roughly 25--30\% of cool, isolated WDs show metal enrichment through optical spectroscopic detection of Ca or other lines \citep{zuckerman03,koester05,zuckerman10} or show emission due to heated gas \citep{gaensicke06,gaensicke07,gaensicke08}.  Metal polluted, hydrogen atmosphere WDs have short settling times for metals \citep{koester09} and are inferred to be actively accreting metal rich material.  Eighteen of the known metal enriched WDs show IR excesses due to optically thick dust located at a radius of $\sim10$~R$_{\rm WD}$ \citep{jura03,reach05,kilic06,vonhippel07,farihi10}.  Given the short lifetimes of dust due to collisions or Poynting-Robertson drag at such distances, the optically thick disk of dust evolves on a viscous timescale, while the presence of strong silicate emission features in most of these dusty disks that have been observed with the Infrared Spectrograph on {\em Spitzer} points to an additional reservoir of optically thin material \citep{jura07b,reach05,reach09}.

The presence of dust within a region that should be devoid of any material due to post-main sequence evolution of the WD progenitor is, at face value, challenging to explain.  Surviving planetesimals must be perturbed (presumably by a planet) on timescales that range from a few Myr to a few Gyr, and be tidally disrupted by the central WD \citep{jura03,jura08}.  The perturbation of planetesimals by the post-main sequence destabilization of giant planetary systems has been proposed but, to date, no quantitative predictions for the lifetime or efficiency of this mechanism have been made \citep{debes02}.  

Nonetheless, the evidence for a link between dusty WDs and remnant planetary systems is mounting.  Order of magnitude estimates for the expected number of metal polluted WDs from \citet{debes02} are consistent with observations, and are similar to both the frequencies of giant planets in orbit around early type stars \citep{johnson} and the estimated frequencies of close-in Earth mass planets \citep{howard10}.  Post-main sequence planetary systems provide an important complementary sample to main sequence planetary systems, and will provide crucial compositional information on extrasolar planetesimals impossible to obtain with other observational techniques.  Yet, several questions about dusty and metal enriched WDs remain.  The lifetime of dust, the exact structure of the dusty disks, and their evolution are all highly uncertain.  The answers to such questions may come via large number statistics from which trends and correlations can be identified.

In this respect, the WIRED Survey is well positioned to provide a large number of new dusty WDs to help answer these and other questions.  {\em WISE}'s sensitivity in the {\em W1} and {\em W2} bands is sufficient to detect dusty disks, such as the one around G29-38 \citep{zuckerman87}, out to $\sim$140~pc.  WISE can detect bright silicate emission features like G29-38's out to $\sim$55~pc.  In this survey we discover 52 candidate disk systems.  The SDSS DR7 catalog is not the only repository of WDs-- we expect to detect a majority of the WDs in the McCook \& Sion catalog \citep{mccook99}, and find another 30--40 candidates, effectively quadrupling the total number of known dusty WDs (currently 20; \citealt{farihi_wdbook}).  

In this paper, we focus on the {\em WISE} detections of SDSS DR7 WDs \citep{kleinman11}, which we examined as part of the WIRED Survey.  Out of 1527 {\em WISE} detections, 95 WDs show excesses likely due to either BDs or dusty disks.  First, in Section \ref{s:wisefacts}, we briefly describe aspects of the {\em WISE} mission important for the WIRED Survey.  In Section \ref{s:photo}, we describe the {\em WISE} photometry dataset.  In Section \ref{s:method}, we discuss our method for finding and identifying excesses. In Section \ref{s:results}, we look at the targets found to show excesses due to M dwarfs, brown dwarfs, and dusty disks, and determine the frequency with which each population occurs.  Finally, we present our conclusions in Section \ref{s:conc}.

\section{{\em WISE} Photometry and Image Quality}
\label{s:wisefacts}

A full description of the characteristics of the {\em WISE} mission can be found in \citet{wright10}, but we summarize details relevant to the WIRED Survey here; namely, the characteristics of the four {\em WISE} imaging channels and the point spread function (PSF) within each channel.

The four {\em WISE} channels span a wavelength range of 3--25\micron. The  $W1 (\lambda_{\rm iso}=3.35\micron)$, $W2 (\lambda_{\rm iso}=4.60\micron)$, $W3 (\lambda_{\rm iso}=11.56 \micron)$, and $W4 (\lambda_{\rm iso}=22.09\micron)$ channels were partly designed for the efficient detection and characterization of field BDs, ultra-luminous galaxies, dusty AGN, and solar system asteroids.  With 99\% of the sky covered to a depth of $>$8 frames, the $5\sigma$ point source sensitivities in each channel are 0.08, 0.11, 1, and 6~mJy for $W1$ to $W4$, respectively (corresponding to Vega magnitudes of 16.5, 15.5, 11.2, and 7.9), with deeper coverage in selected regions of the sky.  

The image quality for {\em WISE} produces well defined PSFs that have FWHMs of 6\farcs1, 6\farcs4, 6\farcs5, and 12\arcsec, for $W1$ to $W4$, respectively.  This dictates the level to which the photometry can resolve blended sources.  In the {\em WISE} catalog, sources at separations $>$1.3 FWHMs (7\farcs8) were resolved through profile fitting.  For this reason, we have used the profile fitted magnitudes for all of our sources.  Sources with separations less than 7\farcs8 will be blended and may represent a possible source of contamination.  However, in many cases, contamination will not mimic the signature of a true excess and can be rejected based on a visual inspection of the spectral energy distribution (SED) of a WD, as well as by comparing images of the target field from {\em WISE} with higher resolution images from SDSS and {\em Spitzer} (when available).  A final determination, however, will require follow-up high resolution infrared imaging or infrared spectroscopy.  For this reason, all excess candidates that we detected are listed in this paper, even if there is reason to believe that the observed excess might be due to contamination from (unrelated) blended sources in the {\em WISE} PSF; we have provided notes for each target-of-interest detailing reasons for possible concern with regard to photometric contamination -- see the Appendix.  

An example of how well WISE can do with regard to moderate contamination is demonstrated by the photometry of GALEX 1931+0117, which is at a galactic latitude of -8.4$^\circ$.  High resolution $K$ and $L^\prime$ images showed dimmer sources within the WISE beam, and it was suggested that this could account for a discrepancy between the WISE and ground-based photometry \citep{melis10}.  J. Farihi has kindly provided the ISAAC $L^\prime$ images obtained on GALEX 1931+0117.  We performed simple aperture photometry on the two sources detected in ISAAC that would lie close to or interior to the WISE beam and on GALEX 1931+0117.  Source 1 was located at 2\farcs4 and source 2 was located at 5\farcs9, corresponding to 0.8 and 1.9 half-width at half maximums (HWHMs) away from GALEX 1931+0117 in the W1 beam.  While these sources would not be actively deblended in the WISE catalog, profile fitting would mitigate contamination for sources $>$1 HWHM away.  From the aperture photometry, we find that these sources have 20\% and 19\% of GALEX 1931+0117's flux and at first glance could account for the $\sim$40\% discrepancy between the ground-based and WISE photometry.  However, another test is to take the ratio of GALEX 1931+0117 to WISE J193157.89+011736.3, the bright source to the SE of GALEX 1931+0117, in both the WISE and $L^\prime$ data.  If there were significant contamination from the two sources in the WISE beam, the ratios would be different.  Instead, we find that the ratios between the two sources in each image are within the uncertainties, suggesting that the flux calibration from the ground rather than contamination in $W1$, explains the discrepancy between $W1$ and $L^\prime$.
. 
Since many of our target WDs have appreciable proper motion as defined by measured proper motion in the DR7 catalog, we calculated the offsets to their SDSS J2000 positions on 1 May 2010, the approximate mid-point of the {\em WISE} mission.  However, one can guess that most of the targets will not have moved by a distance larger than a {\em WISE} PSF -- in fact, only 14 of our targets moved a distance on the sky larger than 3\arcsec, and only two of these moved more than 5\arcsec (WIRED J104559.13+590448.3\footnote{A note on source names:\ the preliminary SDSS DR7 WD catalog, kindly provided to us by S.\ Kleinman, identifies its targets via a unique combination of plate and fiber numbers.  We have followed the SDSS naming convention in constructing names from the right ascension and declination coordinates of the targets, so expect that in the overwhelming majority of cases, our source names will match the SDSS names.  However, we are wary of the potential for small differences between the preliminary and final source positions; consequently, in order to reduce future confusion, we refer to any targets named in the text and tables of this paper as ``WIRED'', rather than ``SDSS'', sources.  The WIRED names are typically either the same as the corresponding SDSS names for targets that already have published SDSS names, or have small differences in the least significant digits of the right ascension and/or declination components of their names.  We defer the ``official'' naming of the SDSS sources presented here to the publication of the final SDSS DR7 WD catalog.}, 12\farcs3; WIRED J105612.31-000621.6, 5\farcs8).  Excluding the 11\% of the targets with proper motions of zero (i.e., either unknown or too small to have been measured), the mean proper motion offset of our targets is 0\farcs19, and the distribution of all non-zero proper motion offsets is gaussian peaked around zero with HWHM=0\farcs13.  Intrinsically, the astrometric accuracy of {\em WISE} is of the order FWHM/(2 SNR), with a worst case scenario for a SNR=5 detection of $\sim$0\farcs6.  However, there are also systematic declination errors for faint ($W1 > 13.0$) sources due to an error in the source extraction as part of the {\em WISE} reduction pipeline \citep{wisesupp}.  To reflect this, the positional uncertainties have been inflated in quadrature by 0\farcs5 in declination.  Due to this uncertainty and any proper motion uncertainties, we searched for {\em WISE} sources within 2\arcsec of the proper motion corrected WD positions.

\section{{\em WISE} Detections of DR7 WDs}
\label{s:photo}

The preliminary SDSS DR7 WD Catalog \citep{kleinman11} contains 17,955 unique and valid targets, after rejecting 1788 duplicate targets and 94 targets with unusable SDSS photometry.  We defined a target as a duplicate if its coordinates matched those of another target within 1\farcs4 (i.e., the mean FWHM of the SDSS PSF) in both right ascension and declination.  We then cross-correlated sources within a 2\arcsec\ radius around the proper motion corrected SDSS coordinates from (in order of precedence):\ the {\em WISE} Preliminary Release Catalog (p3a), the {\em WISE} Atlas Catalog (i3a), and the {\em WISE} Co-add Catalog (i3o).  The p3a catalog was made public on 14 April 2011, while the other two catalogs are currently proprietary.  The i3a catalog contains all of the p3a catalog, but also includes source detections with S/N $<7$ that were rejected from the p3a catalog.  The i3o catalog contains the full sky source list, but has not yet undergone a final calibration or photometric quality vetting.  

\begin{figure}
\plotone{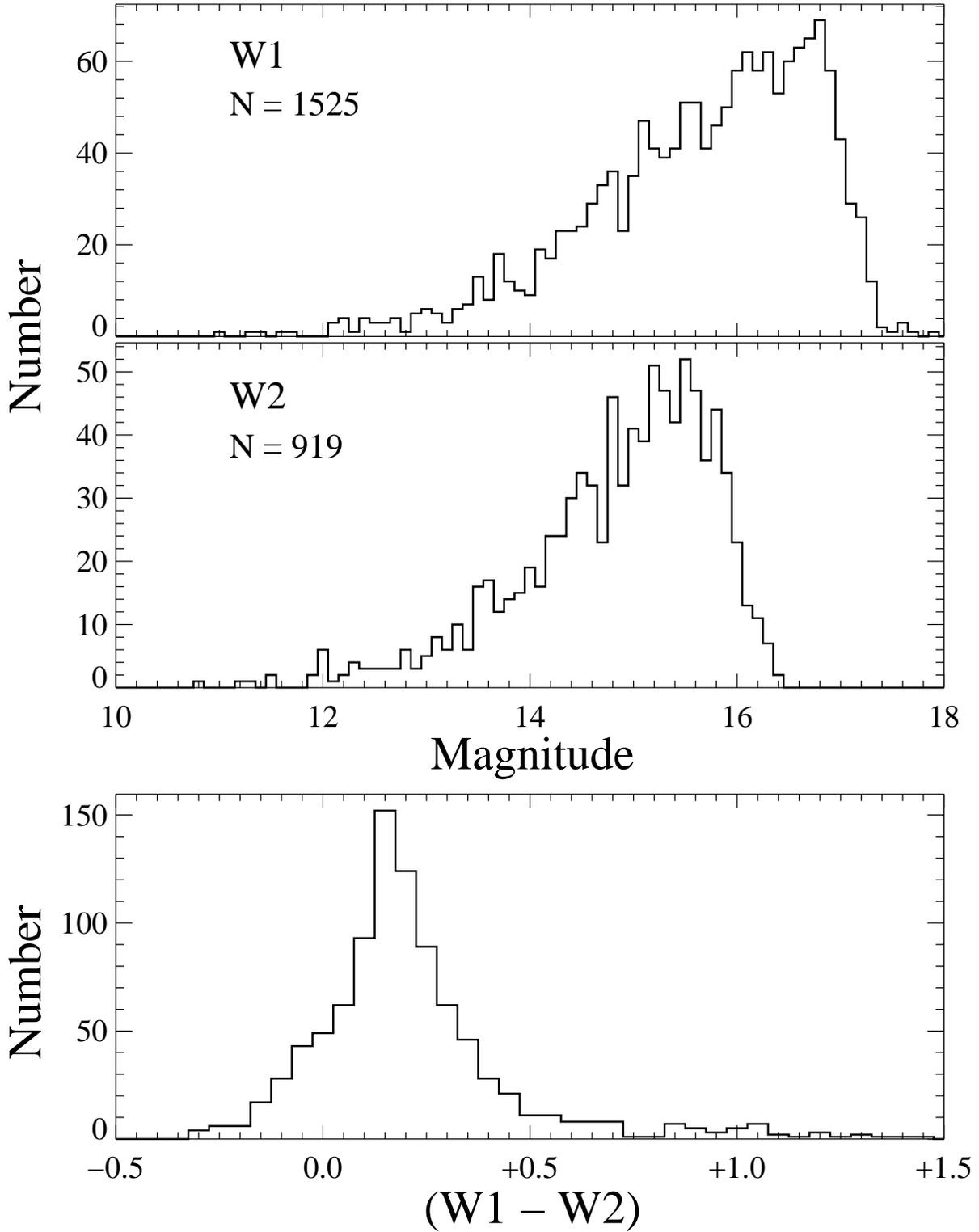}
\caption{\label{fig:hist} Top two panels show histograms of the number of detected targets as function of source brightness in the {\em WISE} $W1$ and $W2$ bands.  The bottom panel shows a histogram of the corresponding distribution of $(W1-W2)$ colors.}
\end{figure}

Of the 17,955 unique and valid SDSS targets, sources corresponding to 1858 of the targets were detected in at least one of the {\em WISE} bands in at least one of these {\em WISE} catalogs.  When multiple sources were detected within 2\arcsec of a target, only the closest source was retained.  We then rejected any detection in a particular {\em WISE} band that had S/N $<5$ (these were retained as upper limits), or had quality flags (cc\_flag) that indicated severe image artifact contamination.  Non-detections in a particular band were also retained as upper limits, provided that the target was reliably detected in at least one band.  A target that displayed any combination of rejection and/or non-detection in all four {\em WISE} bands was rejected.  The order of the {\em WISE} catalogs listed above is an order of precedence; target photometry was only retained from the highest precedence catalog, and if a target was detected in more than one catalog, but rejected in the highest precedence catalog, then the photometry from a lower precedence catalog was considered to be rejected as well.  After completing the rejection process, a total of 1527 targets remain with a reliable {\em WISE} detection in at least one band.  Of these targets, 1525, 919, 22, and 4 have a detection in the $W1$, $W2$, $W3$, and $W4$ bands, respectively.  The results for the $W1$ and $W2$ bands are summarized in Figure \ref{fig:hist}, which shows histograms of the number of detections as a function of source brightness, as well as the distribution of $(W1-W2)$ colors.

We repeated the search process (using the SDSS coordinates {\em without} proper motion correction) in the 2MASS All Sky Data Release Point Source Catalog \citep{skrutskie} and the UKIDSS\footnote{The UKIDSS project is defined in \citet{lawrence07}. UKIDSS uses the UKIRT Wide Field Camera (WFCAM; \citealt{casali07}). The photometric system is described in \citet{hewett06}, and the calibration is described in \citet{hodgkin09}. The pipeline processing and science archive are described in \citet{hambly08}.} DR5plus Point Source Catalogs from the Large Area, Galactic Plane, and Galaxy Cluster Surveys.  A total of 1010 targets from the 1527 targets with a {\em WISE} detection have a near-IR detection in at least one band ($J$, $H$, and/or $K_{\rm s}$).  In cases for which both a 2MASS and UKIDSS measurement were available in a given band, we used both in the fitting of excesses.  In almost all of the cases where both surveys were available (of order 200 sources), the photometry between surveys was consistent to within 2$\sigma$ of the uncertainties (92.5\% for $J$, 99.4\% for $H$ 98.6\% for $K_s$).  

All of the merged photometry for our targets is listed in Tables \ref{tab:phot1}, \ref{tab:phot2}, and \ref{tab:phot3}.  Because of the higher sensitivity in $W1$ and $W2$, coupled with the long wavelength faintness of most WDs in the input catalog, almost all of the WD detections are in the shorter wavelength {\em WISE} channels.  Figure \ref{fig:hist} shows the source counts as a function of magnitude for $W1$ and $W2$.  As expected, our source counts begin to drop at magnitudes fainter than the nominal $5\sigma$ sensitivity limits for the {\em WISE} mission in $W1$ and $W2$.  The fainter detected objects were observed with $>$8 {\em WISE} frames and/or in regions of the sky where source confusion or background level was minimal.

\begin{figure}
%\plotone{fig_ccd1.eps}
\plotone{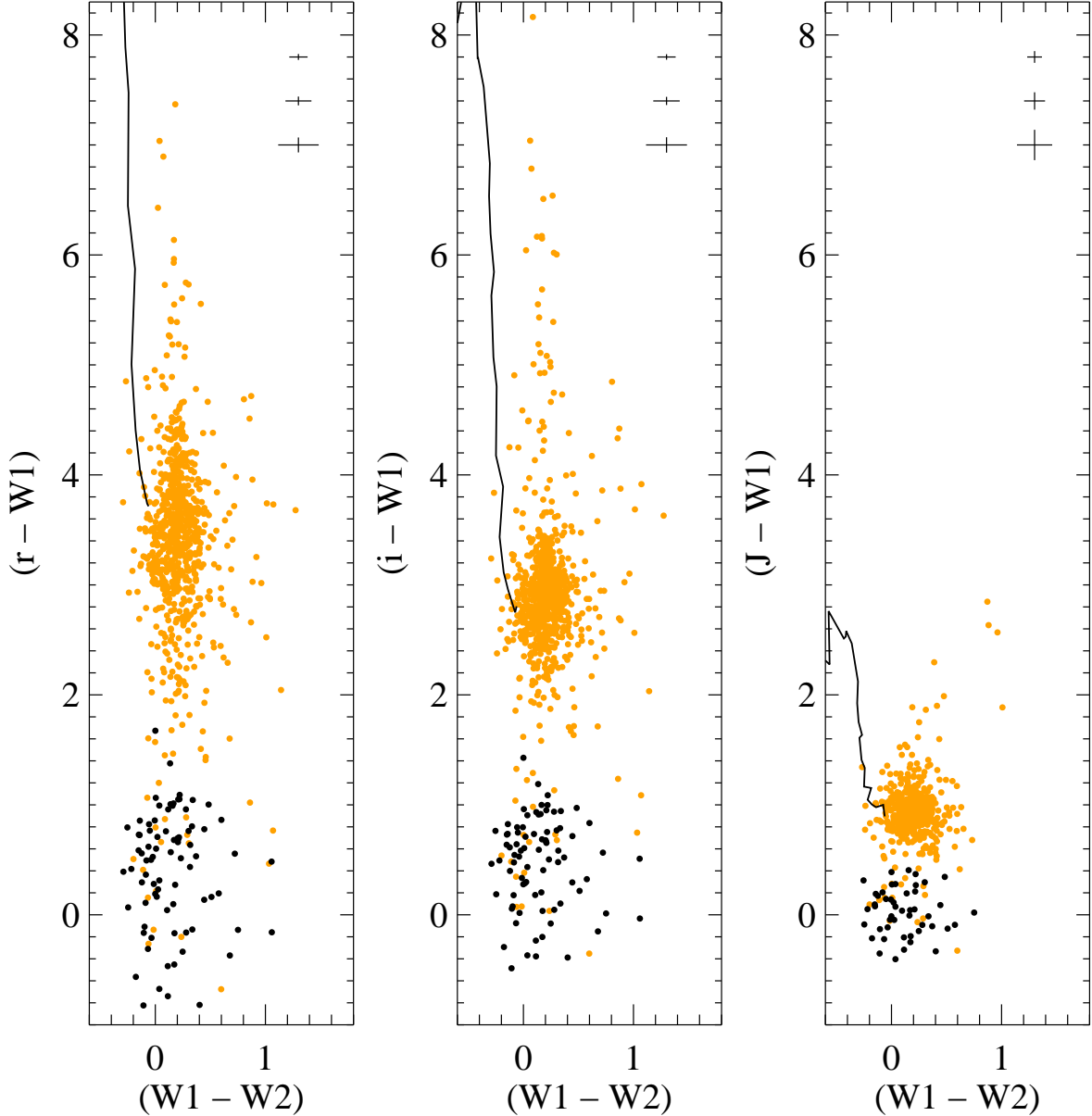}
\caption{\label{fig:ccd1} Color-color diagrams for the SDSS WDs detected by {\em WISE}.  All of the panels have the {\em WISE} color index $(W1-W2)$ on the abscissa, while the ordinates show color indices constructed by subtracting the {\em WISE} $W1$ values from (left to right):\ SDSS $i$, SDSS $r$, and $J$.  For $J$ band, UKIDSS photometry was used preferentially, or 2MASS $J$ transformed to the UKIDSS photometric system using the relations in \citet{hewett06} when a corresponding UKIDSS data point was not available for a given target.  The black points show the targets identified by our SED fitting as naked WDs, while the grey (orange) points show targets identified as unresolved WD+M dwarf binaries.  Representative error bars are shown in the upper right of each panel, and depict the total quadrature uncertainties of different selections of the plotted data:\ 50\%, 67\%, and 90\% of the targets have uncertainties smaller than the top, middle, and bottom error bars, respectively.}
\end{figure}

\begin{figure}
%\plotone{fig_ccd2.eps}
\plotone{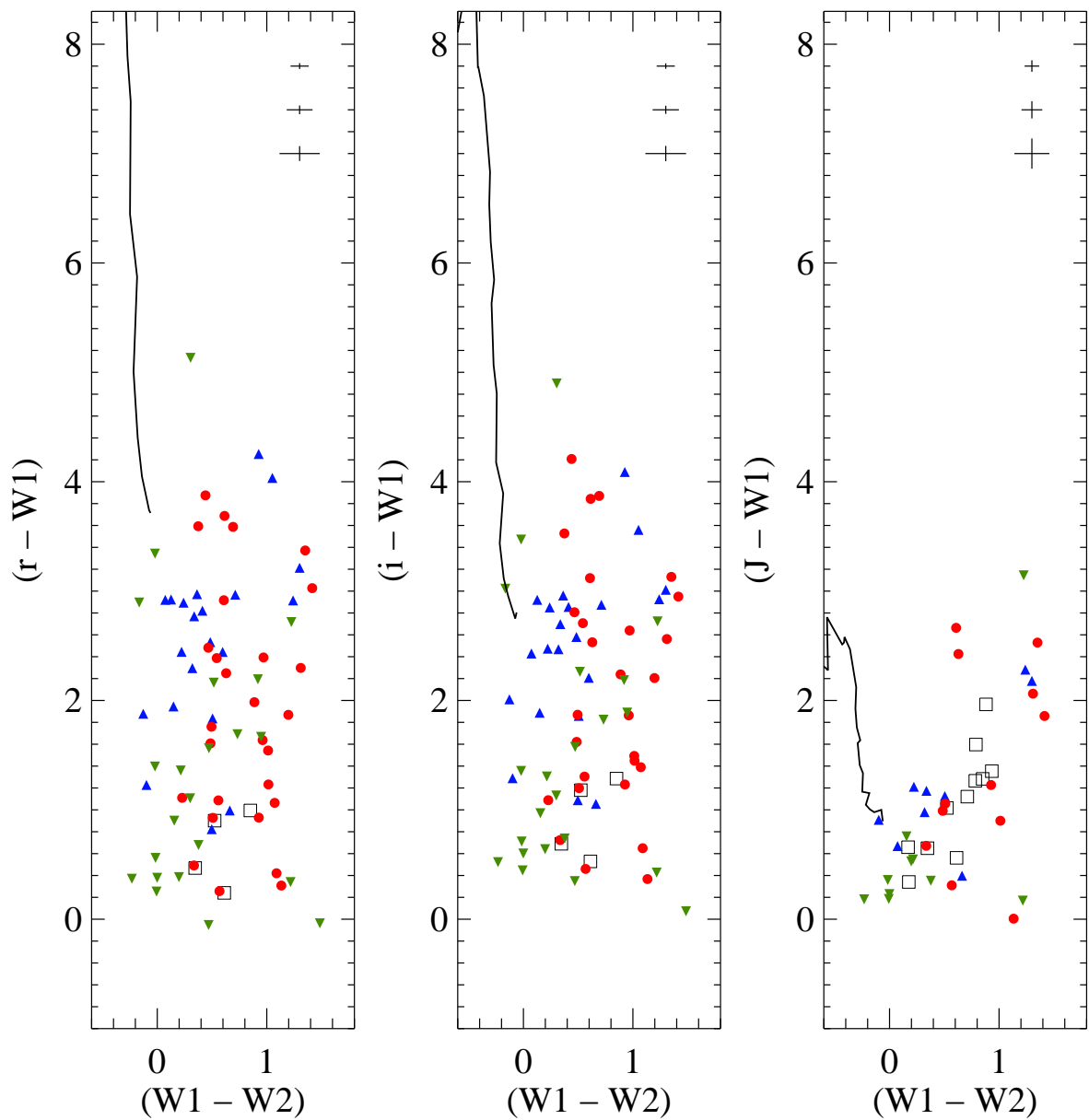}
\caption{\label{fig:ccd2} As in Figure \ref{fig:ccd1} (and using the same axis ranges), but showing only the targets of interest identified from our SED fitting, and listed in Tables \ref{tab:m}--\ref{tab:ind}.  The points are symbol-coded as follows:\ (blue) upward facing triangles = WD+BD; (green) downward facing triangles = indeterminate systems; and (red) circles = WD+dust disk.  For comparison, several previously known WD+dust disk systems drawn from the literature (but not part of our SDSS-selected sample) are plotted as large unfilled squares.}
\end{figure}

Figures \ref{fig:ccd1} and \ref{fig:ccd2} show color-color diagrams of our detected targets using the SDSS $r$ and $i$ bands, $J$ band (UKIDSS or 2MASS transformed to UKIDSS using the relations in \citet{hewett06}), and the {\em WISE} $W1$ and $W2$ bands.  For clarity, we divided the targets into two groups and plotted their color-color diagrams separately.  Figure \ref{fig:ccd1} shows the color-color diagrams for only the targets identified in our SED fitting analysis (see Section \ref{s:results}) as naked WDs or unresolved WD+M dwarf binaries.  These two groups clearly occupy different color loci.  This figure also highlights the much larger number of detected WD+M systems compared to the naked WDs.  This is a selection effect caused by the fact that a WD+M star binary might be detected by {\em WISE} due to the long wavelength contribution of the M star, even in some cases for which the WD in the system by itself would be below the {\em WISE} detection limit.   Figure \ref{fig:ccd2} shows the smaller number of targets-of-interest, along with a number of previously known WD+dust disk systems that were not part of our SDSS sample.

\section{Searching for Infrared Excesses}
\label{s:method}

In order to detect an IR excess around a WD, we require accurate photometry and accurate predictions of the WD photospheric emission in the photometric bands in which the WD has been observed.  Color-color selection is often useful for detecting excesses \citep{hoard07,wellhouse05,wachter03}, but it can fail for a target whose IR excess has colors that approximate the colors corresponding to a WD photosphere.  Similarly, BD candidates and disk candidates share similar colors in the mid-IR, making it potentially difficult to distinguish between the two populations based solely on color. 

Conversely, constructing a SED of the WD allows very small excesses to be detected.  In the case of the preliminary SDSS DR7 WD catalog, we have, at minimum, five optical photometric measures from SDSS photometry and at least one {\em WISE} measure.  In many cases, near-IR photometry in the $J$, $H$, and/or $K_{\rm s}$ bands, was also available.  The large number of photometric points in the visible provides a strong anchor with which to detect weak excesses above the expected WD photosphere in the near- to mid-IR. 

For the purposes of SED-fitting, we converted each photometric measurement from magnitude into a flux density by using zeropoints reported for each photometric band \citep{wright10,skrutskie,sloandr7}.  We then compared the SEDs of each WD to synthetic photometry based on WD cooling models including synthetic photometry in the {\em WISE} bandpasses (kindly provided by P.\ Bergeron).  Based on the reported $T_{\rm eff}$ and $\log{g}$ values from the preliminary SDSS DR7 WD catalog, the cooling models provided a model age and mass for the WD, as well as model photometry for the WD in all the bands studied.  Scaling the predicted photometry to the observed photometry in the $u$ and $g$ bands provided a provisional distance.   If no excesses were detected at longer wavelengths, as many photospheric points as possible were used to determine a photometric distance to the WD.  There were 276 WDs with $\log{g}<7$ that were forced to a $\log{g}=7$, since this was the lower bound of our cooling models.  These objects were either classified at WD+M (90\%) or were classified as not having any meaningful H or He lines and thus their $\log{g}$ values are probably suspect.  This means that, for this subset of objects, the photometric distances could be uncertain by factors of a few.  
%Finally, we set the following baseline criterion on provisionally flagging a target as having an IR excess:\ at least one photometric band had to have a significant excess (defined below) above the level of a WD photosphere.

For the purposes of the WIRED Survey, in order to confirm the presence of an IR excess, a provisionally flagged target had to fulfill one of two criteria:\ $10\sigma$ excess(es) in at least one filter at the $r$ band or longer but shortward of the $W1$ band, or $>3\sigma$ excess(es) in the $W1$ band or longward.  If {\em WISE} detected a WD, but it did not show a significant excess (as defined above), then that WD was flagged as a photospheric detection (i.e., a naked WD). We then subjected all objects {\em with} detected excesses to a second iteration of SED fitting in which several different additional components were added individually to the WD photosphere model to attempt to account for the IR excess.  These additional model components were:\ (i) a model stellar or BD companion at the photometric distance of the WD or (ii) at a best-fit distance that was allowed to be larger than the WD distance (i.e., appropriate for an unrelated contaminating source), and (iii) a simple dust disk model following Jura \citep{jura03}. The stellar and BD model companions were constructed from empirical SDSS colors as a function of spectral type for M, L, and T dwarfs with known spectral types and parallaxes \citep{hawley02}.  {\em WISE} color relations as a function of spectral type were tied to the \citet{hawley02} relations at $J$ band using objects detected in $W1$, $W2$, and $W3$ in {\em WISE} \citep{kirkpatrick11}.  White dwarfs already classified as having an M-dwarf companion in the preliminary DR7 WD catalog were not fit with disk models and were assumed to have a companion only. 

In the Jura dust disk model, a geometrically flat, optically thick and vertically isothermal disk is assumed to be passively re-radiating light from the central WD.  The inner edge corresponds to the sublimation radius of the dust, while the outer edge can extend to $\approx$100~R$_{\rm WD}$.  The flux from a constant temperature ring in such a disk can be expressed as:
\begin{equation}
F_{\nu, \rm ring} = 12\pi^{1/3}\frac{R_{\rm WD}^2 \cos i}{D^2}\left(\frac{2 k_B T_{\rm WD}}{3h\nu}\right)^{8/3} \frac{h\nu^3}{c^2}\int_{x_{min}}^{x_{max}} \frac{x^{5/3}}{e^x-1}dx,
\end{equation}
\noindent where $i$ is the inclination of the disk, $k_{\rm B}$ is Boltzmann's constant, $h$ is Planck's constant, $D$ is the distance to the white dwarf, and $x=h\nu/k_{\rm B}T_{\rm ring}$ \citep{jura03}.  The temperature of the ring in the disk, $T_{\rm ring}$, is given by:
\begin{equation}
\label{eq:tring}
T_{\rm ring} \simeq \left(\frac{2}{3\pi}\right)^{1/4}\left(\frac{R_{\rm WD}}{R_{\rm ring}}\right)^{3/4}T_{\rm WD}.
\end{equation}
\noindent If we take $T_{\rm ring}$ equal to the sublimation temperature of silicate rich dust (T$_{\rm sub}\sim$1200~K) for a WD with $T_{\rm eff}=$10,000~K, then $R_{\rm in}=10$~R$_{\rm WD}$.  For a 20,000~K WD, $R_{\rm in}=25$~R$_{\rm WD}$.

With one or two excess points, the dust disk model is subject to significant degeneracies that exist between the inner radius of a disk and its inclination.  The lack of longer wavelength photometry prohibits a unique constraint on the outer radius of the disk, as well.  For our dust disk model calculations, we assumed an outer radius of $R_{\rm out}=80$~R$_{\rm WD}$ and, initially, a face-on inclination.  For excess candidates that had two or more excess points, an inclination was also fitted.  These radii should be taken as representative, rather than specific solutions, and, for the face-on assumption, represent a lower limit to the true inner radius of the disk.

We calculated the reduced $\chi^2$ value for each of the SED fits, and selected the fit with the lowest $\chi^2$ value in order to classify the nature of the candidate.  Targets that had statistically similar fits between dust disk and companion models were classified as ``indeterminate.''  

Our fitting procedure will fail if the object is a mis-identified galaxy or quasar, the $T_{\rm eff}$ or $\log{g}$ are (very) incorrect, a strong IR excess precludes an accurate measurement of the WD photosphere in all bands, or if a companion is resolved in the SDSS photometry but unresolved in the near-IR or {\em WISE} bands.  Furthermore, contamination can occur from red galaxies that are undetected in the visible (SDSS) bands that lie within the {\em WISE} PSF.  In general, these sources cannot be plausibly fit and, thus, have anomalously high reduced $\chi^2$ values.  Objects with significantly large $\chi^2$ values ($>20$ for WD+disk candidates and $>100$ for WD+companion candidates) should be treated with extra caution.  That said, we report here every discovered excess, pending confirmation or refutation via follow-up observations, but higher weight should be given to those sources with lower $\chi^2$ values.

\section{Results}
\label{s:results}

\subsection{Target-of-Interest Vetting Process}

We searched the 1527 targets with valid {\em WISE} detections for objects with IR excess, and classified by our SED model-fitting code into groups of ``naked'' WDs (344 targets), unresolved WD+M dwarf binaries (1020 targets), unresolved WD+BD binaries (42 targets), WD+dust systems (52 targets), and targets with indeterminate IR excess (i.e., IR excess that could not be distinguished significantly between brown dwarf or dust disk models; 69 targets).  We consider the latter three categories (brown dwarf, dust, indeterminate) to be targets-of-interest (TOIs), and subjected them to greater scrutiny in order to better evaluate the possibility that additional unresolved background sources within the large {\em WISE} PSF could have resulted in spurious IR excess detections.  We purposefully did not reject any excess candidate from these categories, since they will need to be confirmed either spectroscopically or photometrically from higher angular resolution IR images.  We provide cautionary notes for each target in the Appendix.

To validate our targets, we first examined the {\em WISE} images for each TOI, especially in the $W1$ band (where the targets were generally brightest and, in many cases, {\em only} detected).  We noted targets whose {\em WISE} catalog entries included non-zero photometric quality flags (cc\_flag) values and/or nb, na values other than the default (nb=1, na=0).  

The cc\_flag values are assigned for each {\em WISE} band and indicate a spurious detection or photometric contamination from:\ a diffraction spike of a nearby bright source (D), persistence effect from a latent image left by a bright source (P), scattered light halo from a nearby bright source (H), or an optical ghost caused by a nearby bright source (O).  When the cc\_flags are given as lower case letters, it indicates that the source detection is believed to be real (i.e., non-spurious) but the photometry may still be contaminated by a nearby artifact.  A cc\_flag value of 0 (zero) indicates that the photometry in that band is unaffected by known artifacts.  We retained all photometry with cc\_flag values of 0 or any lower case letter (with a note in the Appendix), but rejected all photometry with an upper case cc\_flag.

The nb and na parameters relate to the deblending of sources in the {\em WISE} catalog.  In general, a source is considered for deblending if additional sources are located within 25\arcsec of the target.  The nb parameter denotes the number of PSF components used simultaneously in the profile-fitting for a given source. This number includes the source itself, so the minimum value of nb is 1; nb is $>1$ when the source is fit concurrently with other nearby detections (passive deblending), or when a single source is split into two components during the fitting process (active deblending).  In cases of nb $>1$, the na parameter indicates whether passive (na=0) or active (na=1) deblending was used.  We have noted in the Appendix any TOI with non-default values of nb and na.  Note that non-default nb, na values are {\em not} necessarily reasons for concern about photometric quality, since either deblending process will accurately recover the photometry of the target source.  The one caveat to this is that the photometric extraction for the {\em WISE} catalogs was set to stop searching for blended sources at radii smaller than 7\farcs8 (i.e., $1.3\times$ the $W1$ PSF FWHM).  Consequently, an important component of our vetting of the TOI {\em WISE} images was to consider whether there might be an additional source within 7\farcs8 of the target.  In some cases, this was apparent upon visual inspection, but in all cases we also measured the target FWHM using the IRAF task imexam, and compared it to the FWHM values measured for other point sources of comparable brightness in each {\em WISE} image.  TOIs with larger FWHM values are noted in the Appendix as either possibly or likely contaminated by an unresolved point source within 7\farcs8, depending on the severity of the increase in FWHM relative to other point sources.  Typically, an increase in FWHM by $<$0\farcs5 was noted as possibly contaminated, while larger increases were noted as likely contaminated.

Next, we examined the SDSS $i$-band image for each target, as well as the SDSS color jpg image delivered by the SDSS Finder Chart tool.  We noted the presence of additional sources visible within 7\farcs8 in the higher resolution SDSS images, especially the presence of (barely) resolved companion sources within 3\arcsec (i.e., within the {\em WISE} $W1$ PSF HWHM).  As a double check to this process, we also ran the TOI coordinates through the SDSS Cross-ID service to obtain all cataloged primary SDSS targets within 9\arcsec of each TOI.  In general, we found that our inspection of the images yielded consistent results with the Cross-ID search; we note that in a number of cases, blended or barely resolved sources that were identifiable due to color difference compared to the WD in the SDSS color jpg image were {\em not} reported by the Cross-ID search.  TOIs with other SDSS sources within 7\farcs8 are noted in the Appendix.  If these additional sources are stellar and very faint in the SDSS bands, then it is likely that they do not pose a significant contamination risk in the {\em WISE} bands.  Extended sources (i.e., background galaxies) seen in the SDSS images are of more concern since these tend to have red SEDs, so might pose a contamination risk in the {\em WISE} bands, even if they are relatively faint in the SDSS bands.  In numerous cases, blended or barely resolved red stellar companions (bright or faint relative to the WD) are obvious in the SDSS images.  As these coincide with the TOIs identified by our model fitting code as brown dwarf or indeterminate systems, the presence of the additional source in these cases is not necessarily reason for alarm, as its contribution to the target SED in the {\em WISE} bands is being correctly identified as due to the presence of a red stellar companion to the WD.

\subsection{Photospheric Detections}

\begin{figure}
\plottwo{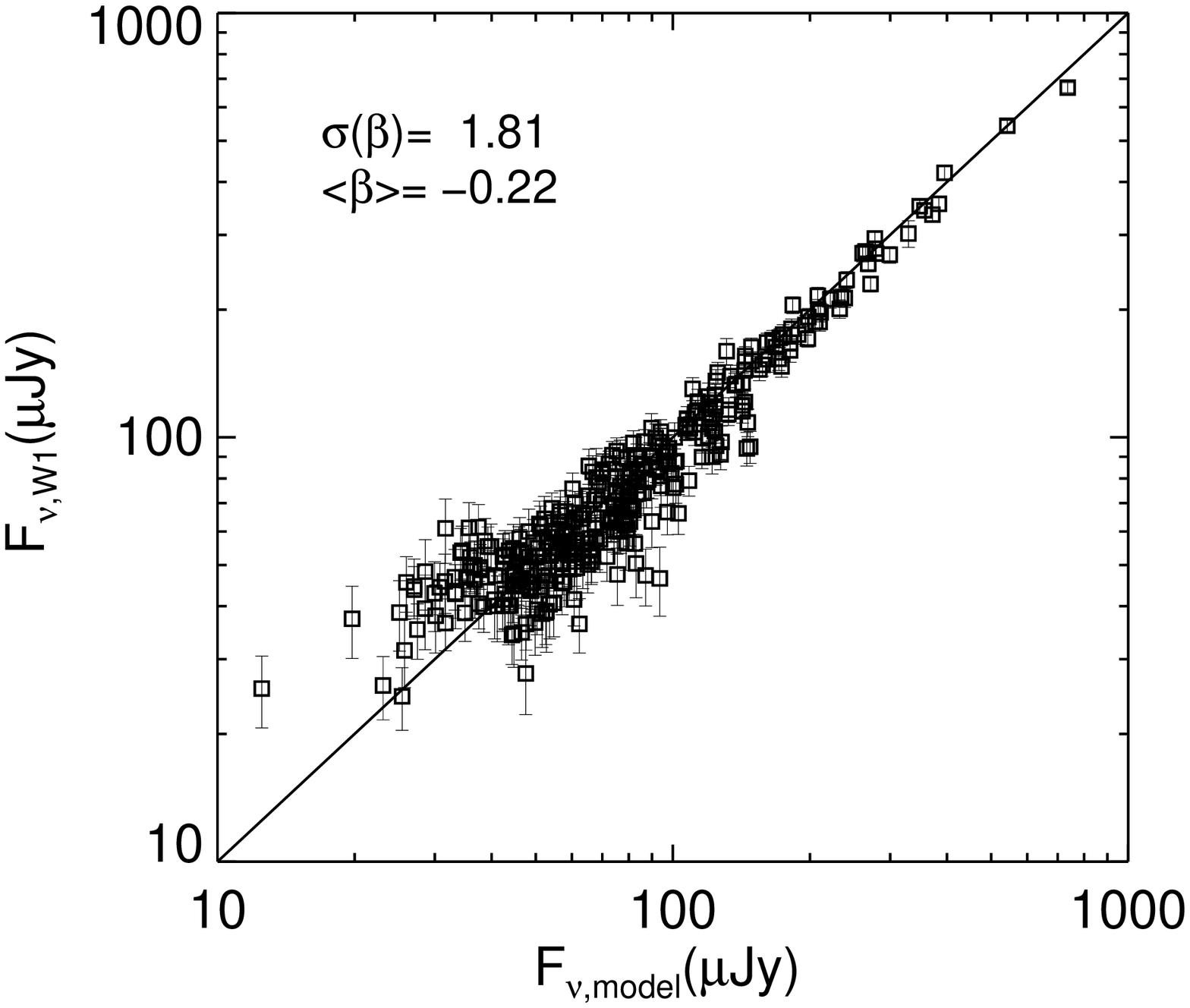}{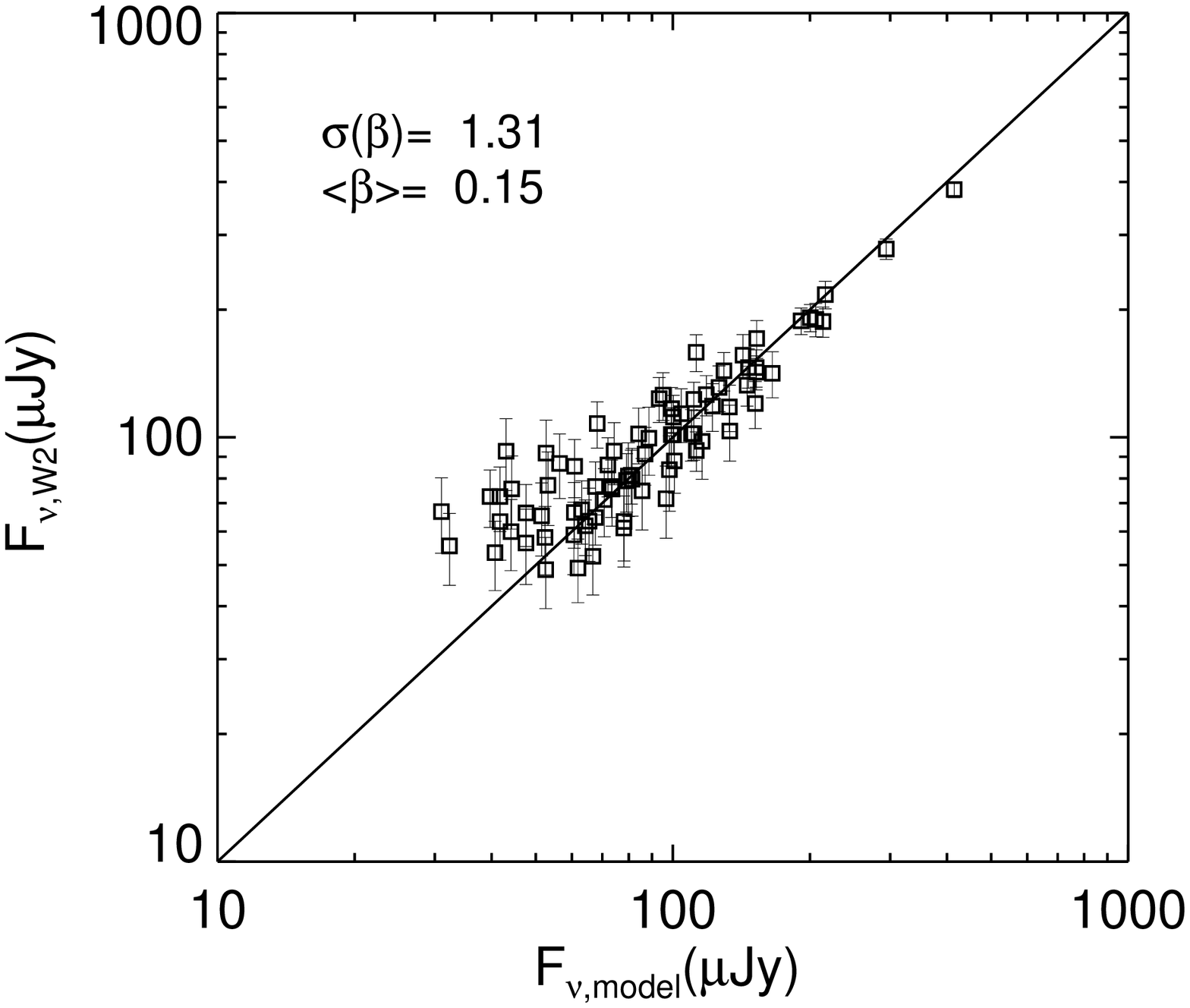}
\caption{\label{fig:photcomp} Comparison between observed {\em WISE} flux densities ($F_{\nu,W1}$, left, and $F_{\nu,W2}$, right) and predicted photospheric flux density ($F_{\nu,mod}$) for all WD photosphere detections. The symbol $\beta$ is a measure of the uncertainty weighted deviation of each observed flux density from its predicted value.  A value of $\sigma(\beta) \leq 1$ is equivalent to the entire sample matching the predicted photometry to within the uncertainties.}
\end{figure}

{\em WISE} was sensitive enough to detect 344 WD photospheres in the $W1$ band and 81 in both $W1$ and $W2$ over a significant fraction of the total distance to which it is sensitive to nearby dusty white dwarfs and brown dwarfs.  The majority of detections are of WDs with photometric distances of $<80$~pc.  The WD photospheric detections provide a useful test of both our WD cooling models and the {\em WISE} photometry.  Figure \ref{fig:photcomp} shows the observed $W1$ and $W2$ flux densities compared to the predicted flux densities in those bands for our WD photospheric detections.

\begin{figure}
\plottwo{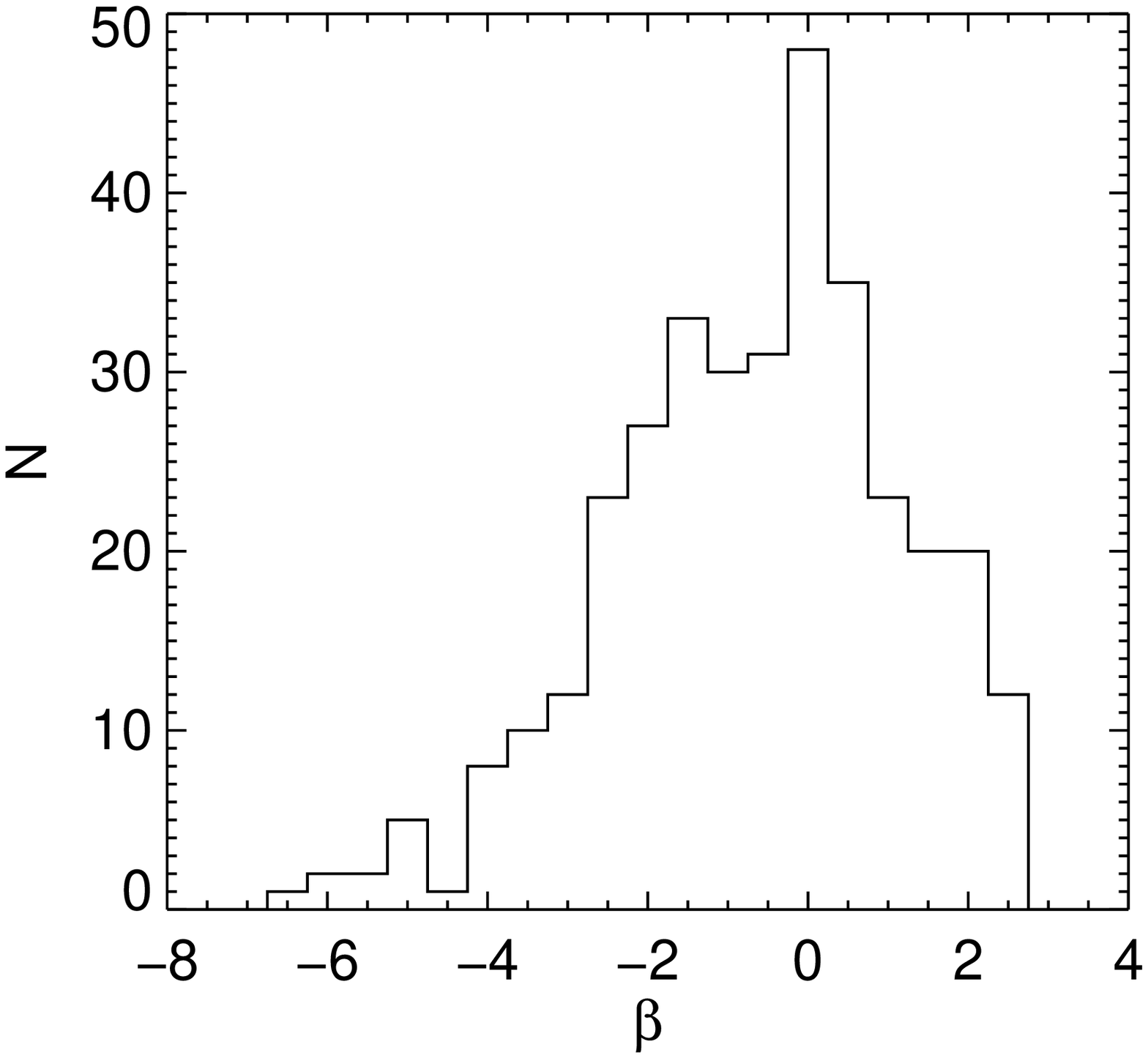}{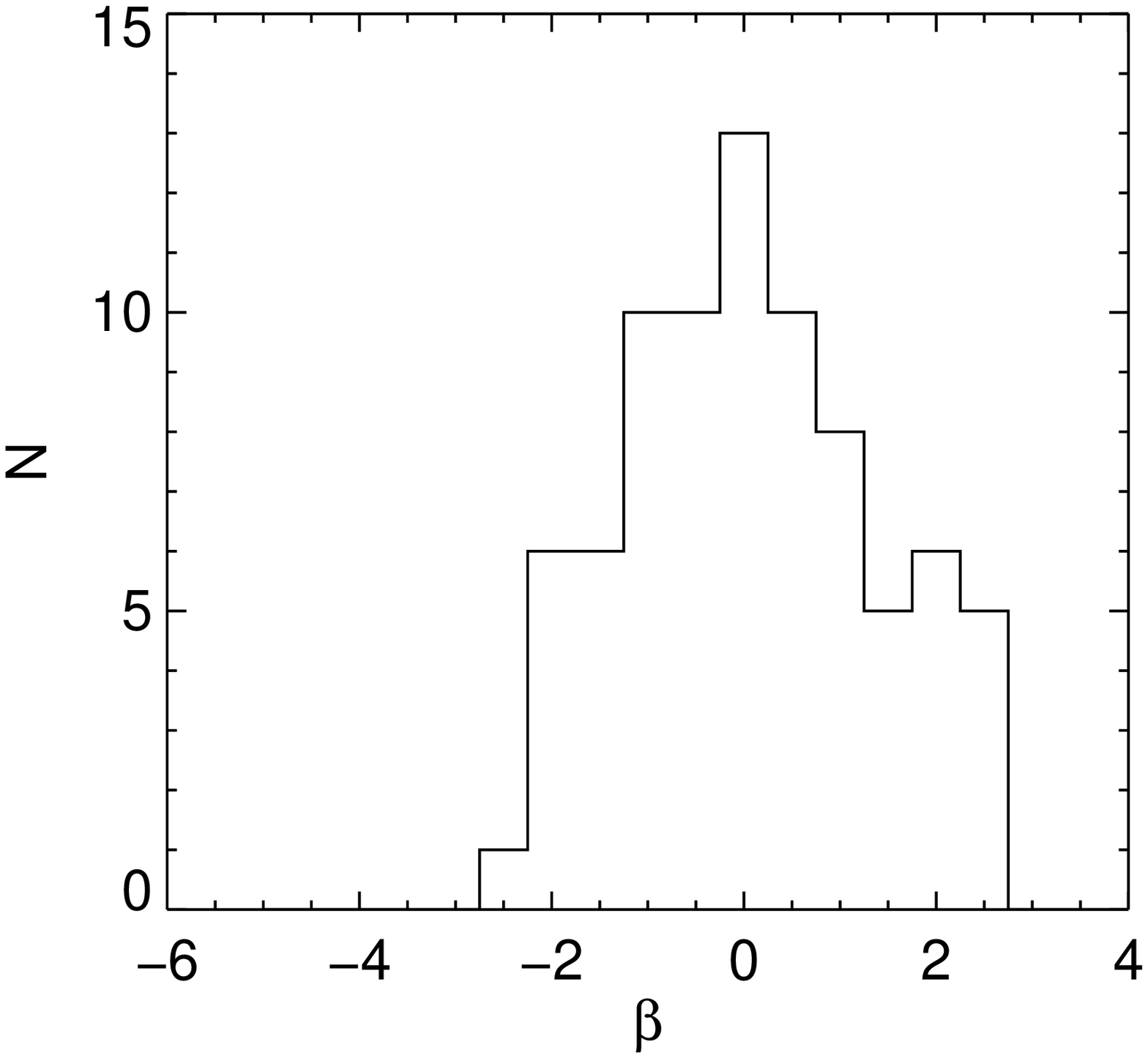}
\caption{\label{fig:phothist} Detailed distribution of $\beta$, the deviation (in units of $\sigma_{\rm obs}$) between the observed and model photometry for our sample of naked WDs detected in $W1$ (left) and $W2$ (right).}
\end{figure}

In general, the observed flux densities and the predicted model photospheres are well matched.  We can also check the distribution of the deviation of each observed WD relative to the predicted photosphere and weighted by the uncertainty, $\beta=(F_{\rm obs}-F_{\rm model}$)/$\sigma(F_{\rm obs})$.  Figure \ref{fig:phothist} shows the distribution of $\beta$ for the $W1$ and $W2$ channels.  Most sources fall within $\pm3\sigma$, and show a reasonable degree of symmetry about zero.  We find 28 targets with deviations of $< -3\sigma$ for $W1$.  A large number of these targets (21) are from the i3o catalog, which has not undergone a final calibration or formal quality vetting process. Nonetheless, seven of these targets have $W2$ photometry that matches our models.  At worst, a similar number of our excess candidates may be due to the greater photometric uncertainty from the i3o catalog.  It is difficult, then, to predict how many marginal excesses (those $<$8$\sigma$ above the predicted photosphere) may be contaminated by the positive tail of this i3o uncertainty.

One of the targets detected by {\em WISE}, WIRED J230645.72+212859.3, is labeled as a DAZCOOL star in the DR7 catalog, but is listed as having an effective temperature of 93,300~K.  It was initially flagged as a WD+disk candidate by our fitting algorithm, but it is most likely {\em not} a disk candidate and, instead, suffers from a severe mis-match in its predicted $T_{\rm eff}$, as well as an atypical photosphereic SED that is not reproduced adequately by ``normal'' DA WD models.  Given that the SED of the WD in this target peaks closer to 1~$\mu$m, its T$_{\rm eff}$ must be closer to 3000--4000~K.  We manually reclassified it as a naked WD.

\subsection{Infrared-excess WDs with Previous {\em Spitzer} Photometry}

\begin{figure}
\epsscale{0.8}
\plotone{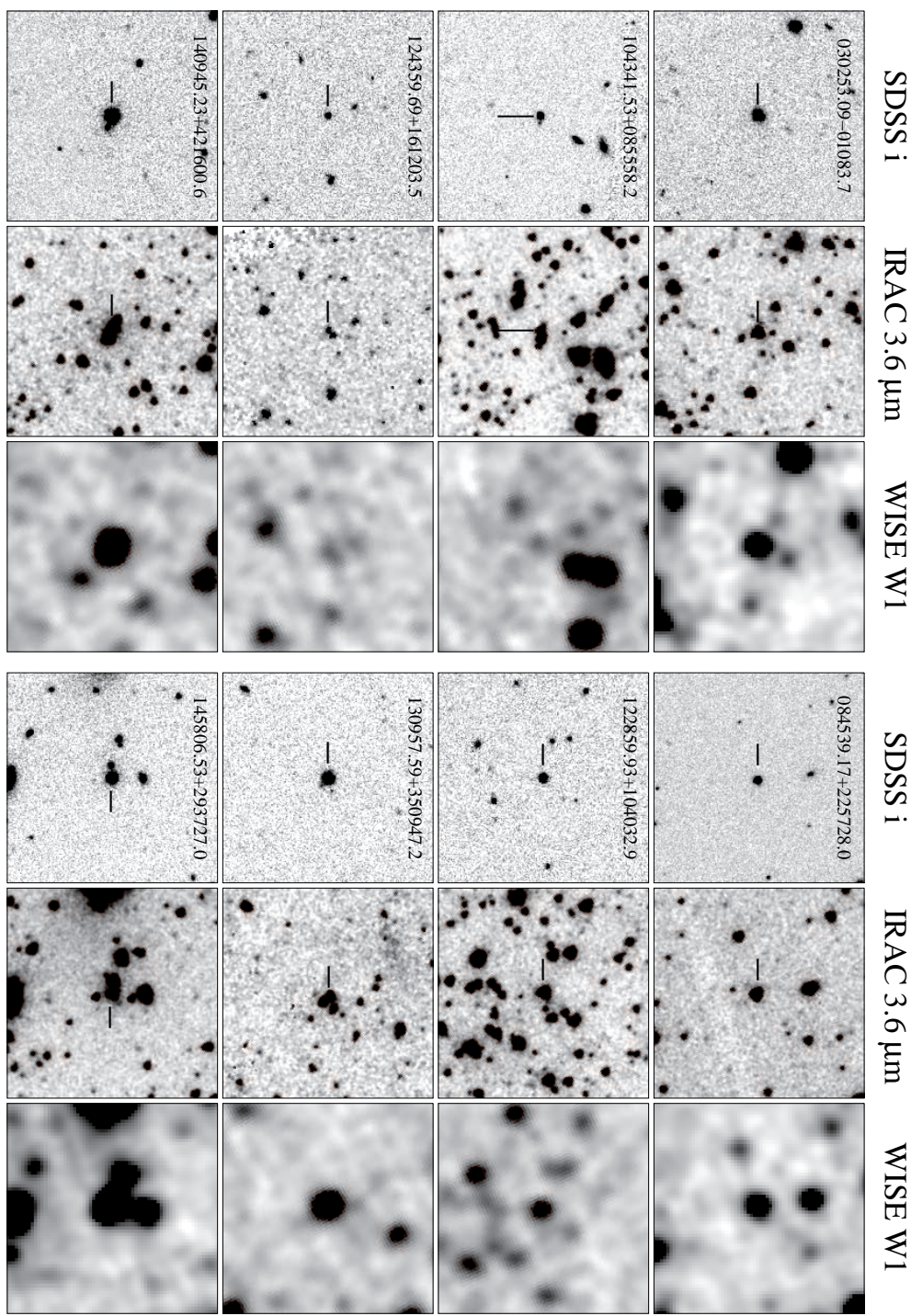}
\epsscale{1.0}
\caption{\label{fig:spitzcomp} The SDSS $i$, {\em Spitzer} IRAC 3.6$\mu$m, and {\em WISE} $W1$ (3.4$\mu$m) images for our TOIs with {\em Spitzer} IRAC photometric data. A 90\arcsec $\times$ 90\arcsec field-of-view is shown for each target. Note that for WIRED J084539.17+225728.0 and WIRED J124359.69+161203.5, the IRAC 4.5$\mu$m images are displayed since neither source has been observed with IRAC at 3.6$\mu$m. WIRED J161717.04+162022.3 is not included because the {\em Spitzer} data for this target are still proprietary.}
\end{figure}

Due to the limited spatial resolution of the {\em WISE} images, all of our IR-excess candidates will ultimately have to be confirmed through observations at higher angular resolution (e.g., with {\em Spitzer}). We searched the {\em Spitzer} Heritage Archive for data covering our TOIs, and found IRAC observations for 9 of our 164 TOIs listed in Tables \ref{tab:bd}, \ref{tab:disk}, and \ref{tab:ind}.  We retrieved published {\em Spitzer} flux density measurements for these targets from the literature or performed our own photometry using the method described in \citet{hoard09}.  A comparison between the resulting {\em WISE} and {\em Spitzer} photometry is compiled in Table~\ref{tab:spitzcomp}. We also show the SDSS $i$, {\em Spitzer} IRAC 3.6$\mu$m, and {\em WISE} $W1$ (3.4$\mu$m) images for these targets in Figure~\ref{fig:spitzcomp}, with the exception of WIRED J161717.04+162022.3, for which the {\em Spitzer} data are still proprietary. 

For all of the targets that exhibit large discrepancies between the {\em WISE} and {\em Spitzer} photometry ($>15\%$), Figure~\ref{fig:spitzcomp} indicates that the {\em WISE} PSF is contaminated by nearby sources. This accounts for their higher flux densities from the {\em WISE} photometry, above that reported based on {\em Spitzer} data. The only exception is WIRED J122859.9+104032.9, which shows a significant decrease in flux density in $W1$ compared to its {\em Spitzer} photometry in IRAC-1, yet appears to be a relatively isolated source.  Its $W2$ and IRAC-2 photometry, on the other hand, agree at a level consistent with the photometric uncertainties.  This suggests that the dust disk emission from this target could possibly be variable.

Table~\ref{tab:spitzcomp} contains three objects classified as Indeterminate IR excess sources (WIRED J124359.69+161203.5, WIRED J130957.59+350947.2, and WIRED J140945.23+421600.6), while the remaining six were classified in our group of WD + dust disk candidates. WIRED J130957.59+350947.2 and WIRED J140945.23+421600.6 have already been shown to be single WDs based on their {\em Spitzer} photometry. {\em All} of our TOIs with {\em Spitzer} coverage that belong to the WD + dust disk candidates category have been previously confirmed as bona fide dust disk systems via {\em Spitzer} photometry. We successfully detect their IR excess with {\em WISE}, also, and classify these targets correctly, despite the contaminated photometry for WIRED J104341.53+085558.2 and WIRED J145806.53+293727.0. In addition, we investigated whether our excess detection methodology might have missed any known WD + dust disk systems.  Five of the 20 known dusty WDs (as of late 2010; \citealt{farihi_wdbook}) are contained in the preliminary SDSS DR7 WD catalog, and all five are recovered in the WIRED Survey and classified by our model fitting algorithm as WD + dust disk systems.

\subsection{Infrared-excess WDs discovered through recent surveys using UKIDSS photometry}

We note that several WD IR excess candidates have been compiled in the UKIDSS surveys of \citet{girven11} and \citet{steele11}.  If any previously unknown objects were also detected in our survey, we mark that in Table \ref{tab:phot1} as well as in the individual notes on each object.  In general, the objects found in these two surveys are WD+M binaries, with a few objects confirmed as being contaminated by background or foreground objects.  Three objects from WIRED with classifications of WD+BD, WD+disk, or indeterminate that coincide with \citet{girven11} are WIRED J013532.97+144555.9 (WD+L6, compared to our fit of WD+L5$\pm$3), WIRED J133100.61+004033.5 (WD+?, compared to our classification as indeterminate), and WIRED J141448.24+021257.7 (WD+?, compared to our classification as indeterminate).  There are five similar overlapping objects from \citet{steele11}, WIRED J013532.97+144555.9 (WD+L5, compared to WD+L6 from \citet{girven11} and our classification of WD+L5$\pm$3) WIRED J093821.34+342035.6 (WD+L5, compared to our classification of WD+L3$^{+1}_{-4}$), WIRED J 124455.15+040220.6 (WD+contamination, listed as WD+disk, but with a poor model fit), and WIRED J141448.24+021257.7 (WD+disk, indeterminate from our classification and from \citet{girven11}).

\subsection{White Dwarf + M Dwarf Candidates}
\label{s:wd_mdwarf}

\begin{figure}
\plotone{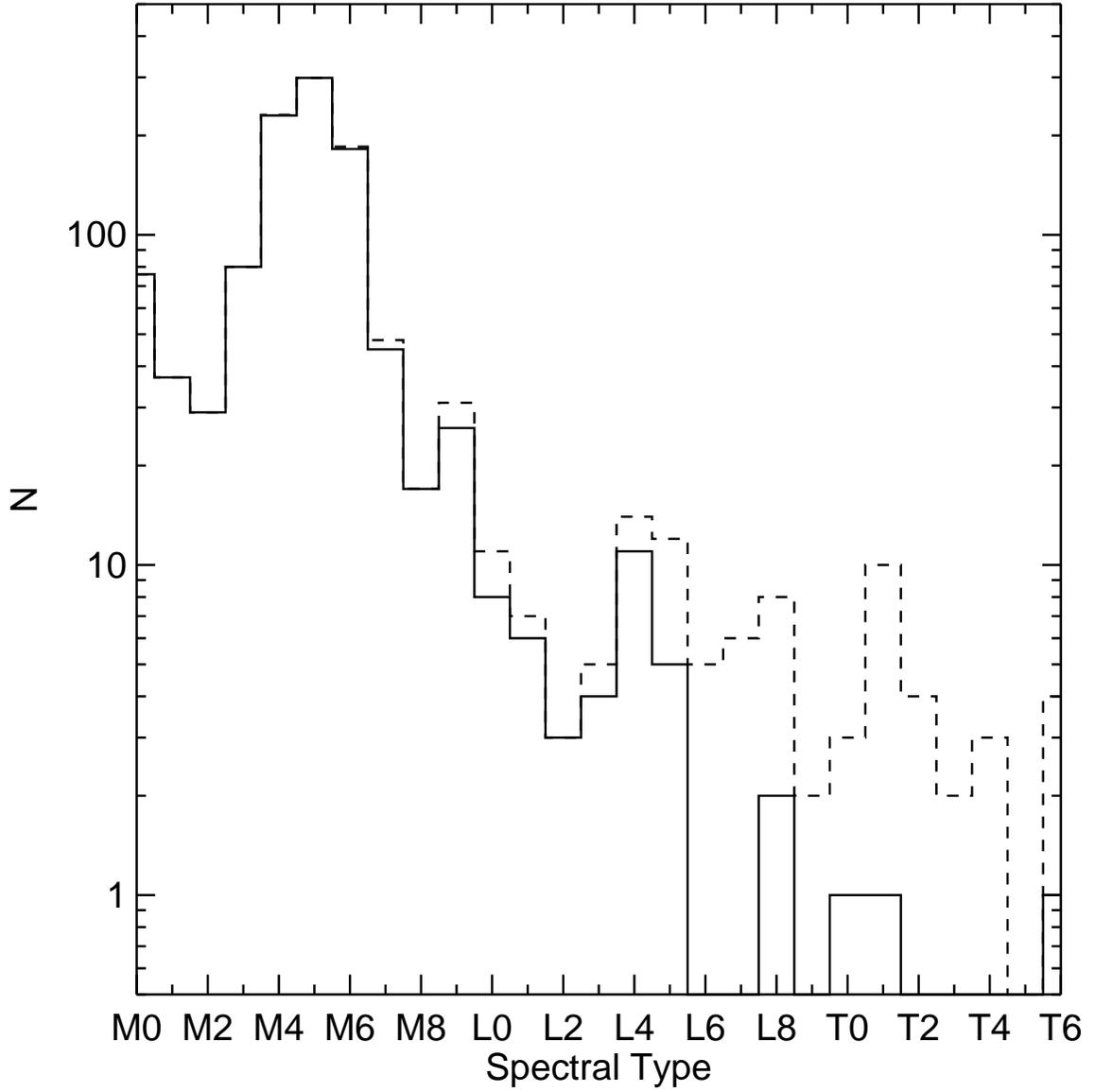}
\caption{\label{fig:compdist} Distribution of fitted companion star spectral types.  The solid line shows all WD+M star and WD+BD candidates, while the dashed line also includes the indeterminate excess targets.}
\end{figure}

Table \ref{tab:m} lists the targets detected by {\em WISE} that we classified as WD+M star binary candidates. Figure \ref{fig:compdist} shows the distribution of all inferred companion star spectral types from our fitting routine.  These spectral types should be viewed with some caution -- while they should be reasonable matches to within $\pm$1--2 spectral types, the fitting algorithm can give spurious results if a companion was resolved in the SDSS photometry but not at longer wavelengths, or if the $T_{\rm eff}$ of the WD provided in the preliminary SDSS DR7 WD Catalog was incorrect -- the targets in Table \ref{tab:m} with very large $\chi^2$ values ($\gtrsim$10,000) are exemplars of these problems.  As can be seen in Figure \ref{fig:compdist}, there is an excess of fitted M0 spectral types, most likely due to the above issues.  Severe mismatches in the photometric distance of the companion and candidate M dwarf are most likely a sign of some significant issue, and will be the result of further study to ensure no weak excesses were actually fit with a spurious M dwarf companion at $>1$ kpc.

WD+M systems make up a significant fraction (2/3) of our {\em WISE} detections, and we can measure the frequency of occurrence amongst all WDs.  In order to determine such a frequency, we must first determine how efficiently we detected WDs in the SDSS sample and to what flux density level.  The total number of individual WDs in the preliminary DR7 sample is $\approx$18,000, with a majority having expected photospheric flux fainter than the nominal detection limits of {\em WISE}.  Using the full sample's optical photometry, we estimated the expected $W1$ flux density for each DR7 WD and compared the total number of WDs in logarithmic flux bins against the predicted photospheric flux densities for all WDs detected by {\em WISE}.  Of the 1527 SDSS DR7 targets detected by {\em WISE}, a total of 395 targets have $W1$ flux densities brighter than a minimum predicted photospheric flux density level of 50~$\mu$Jy (equal to a $W1$ magnitude of 17.0, which is the peak of the detection histogram of our targets in $W1$ -- see Figure \ref{fig:hist}).  Based on the total sample of 18,000 WDs, we expected to detect 533 to that flux level, meaning that our WIRED survey is 74\% complete to 50$\mu$Jy.  Of the 395 detected WDs, 111 were classified as WD+M systems.  Assuming a Poissonian probability distribution (given the relatively large sample size) for calculation of the uncertainty, the frequency of M dwarf companions to WDs is 28$\pm$3\%.

\subsection{White Dwarf + Brown Dwarf Candidates}

\begin{figure}
\plotone{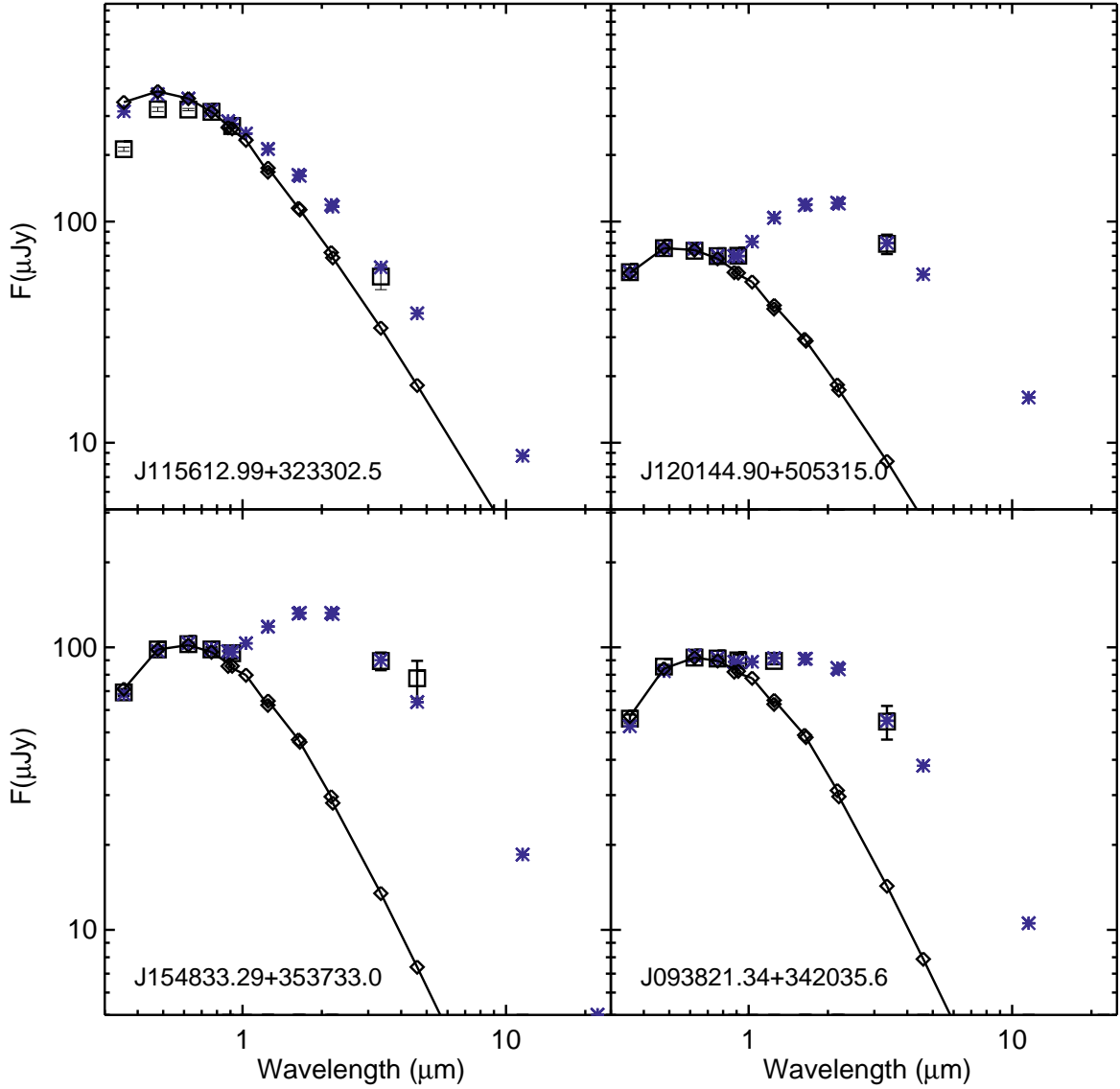}
\caption{\label{fig:wd+bd} Sample SEDs of WD+BD candidates from the WIRED Survey. Black squares are the observed photometry, (blue) asterisks are the best fitting WD+BD models, and (black) diamonds are the corresponding model WD photosphere.  In a significant number of objects, near-IR photometry was not available.}
\end{figure}

Table \ref{tab:bd} lists the targets detected by {\em WISE} that we classified as WD+BD binary candidates.  Figure \ref{fig:compdist} shows the distribution of the inferred BD spectral types from our fitting routine.  This figure shows that the companion distribution appears smooth beyond the M spectral type into early to mid-L types.  Figure \ref{fig:wd+bd} shows sample SEDs (observed and modelled) of WD+BD candidates with the lowest $\chi^2$ fits.  One previously known WD+BD system is part of our sample, WIRED J222030.69-004107.3, also known as PHL~5038 \citep{steele09}.  It possesses an L8 companion at a separation of 0\farcs94.  It was mis-classified by our automated fitting program as a WD+Disk system.  

As we did for the WD+M dwarf systems in Section \ref{s:wd_mdwarf}, we can estimate a frequency of WD+BD systems among the general population of WDs.  The expected photosphere of PHL~5038 is above our 50~$\mu$Jy flux density cut-off, so it is included in our estimated rate of BD companions.  With 8 of the 395 detected targets brighter than the predicted WD photosphere flux density cut-off being classified as WD+BD, this corresponds to 2$\pm$0.7\% of WDs possessing a BD companion.  This is a factor of five higher than previously determined values of such a frequency from near-IR surveys, $f_{\rm BD}$=0.4\% \citep{farihi05}.  

Given the higher contrast between a BD companion and its WD host in the mid-IR, one would expect that a likely explanation for this higher value must include the greater sensitivity of {\em WISE} to BD companions.  However, the higher BD frequency found here could also contain a contribution  from contamination from background sources or mis-classification of the sources of the IR excesses for some targets.  In fact, inspection of the SDSS images and optical--mid-IR SEDs of all WD+BD candidates shows some contamination of the BD candidates with galaxies mis-identified as WDs by the preliminary DR7 WD catalog or, in some cases, our SED fitting algorithm mis-identified an M dwarf companion as a BD, or it failed if the $T_{\rm eff}$ given by the DR7 autofit routine was incorrect.  Notes that describe such contaminants are given in the Appendix.  Taking these into account, of the eight WD+BD candidates, three showed such issues, reducing our observed frequency to 1.3$\pm$0.6\%, which brings it more in line with previous studies.  However, this could be a lower limit, considering that 14 candidate excess sources brighter than the 50~$\mu$Jy flux density cut-off are classified as indeterminate and some of these could be WD+BD systems.  The maximum frequency (if all of the indeterminate systems are really WD+BD systems) would then be almost 5\%.

\subsection{White Dwarf + Dust Disk Candidates}

\begin{figure}
\plotone{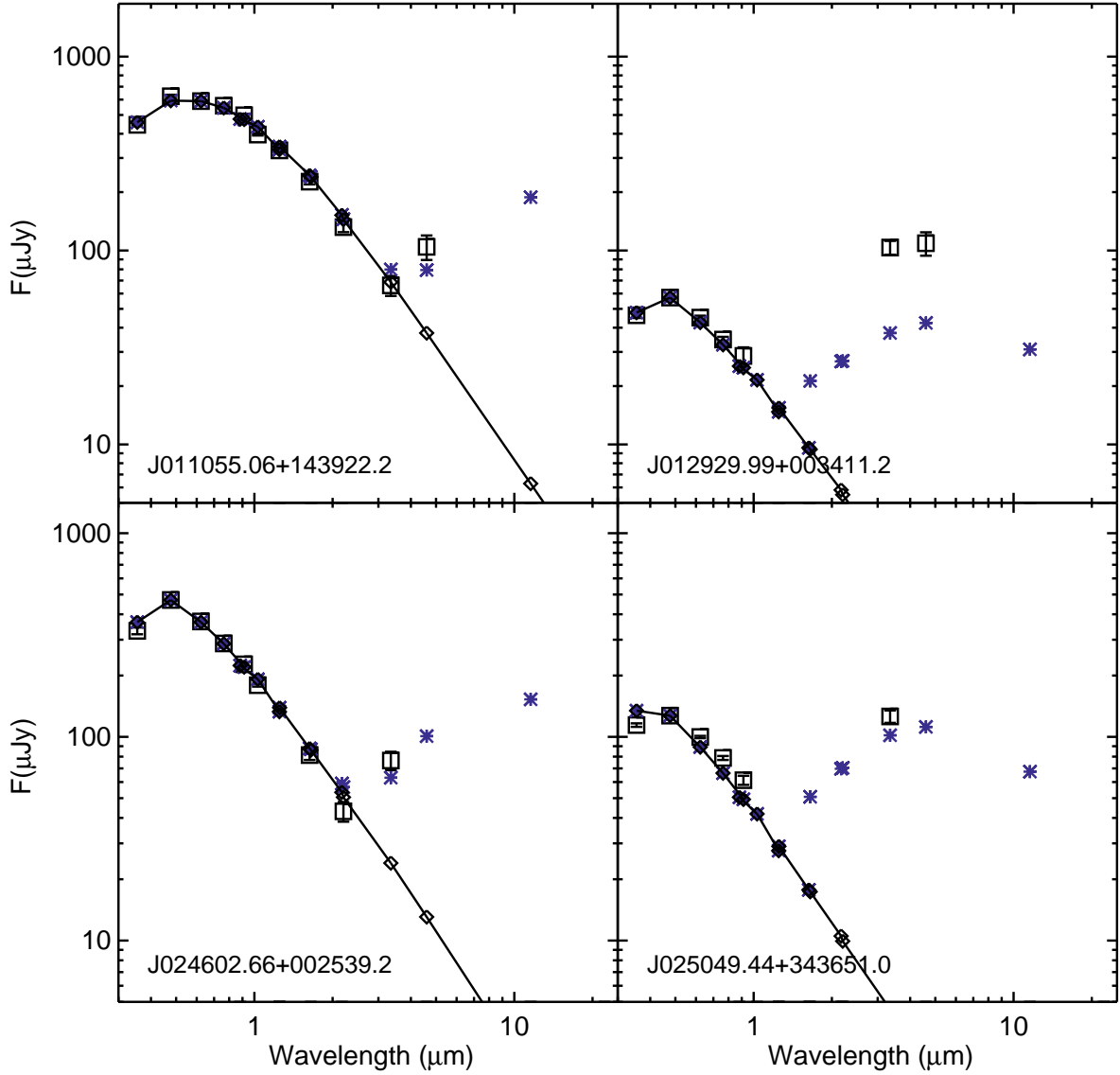}
\caption{\label{fig:d1} SEDs of WD+disk candidates. Symbols are the same as in \ref{fig:wd+bd}, with the exception that the (blue) asterisks now represent the best fitting WD+disk model.  Poor fits to the data can indicate possible contamination from unresolved sources or nearby sources that are bright in the {\em WISE} bands.}
\end{figure}

\begin{figure}
\plotone{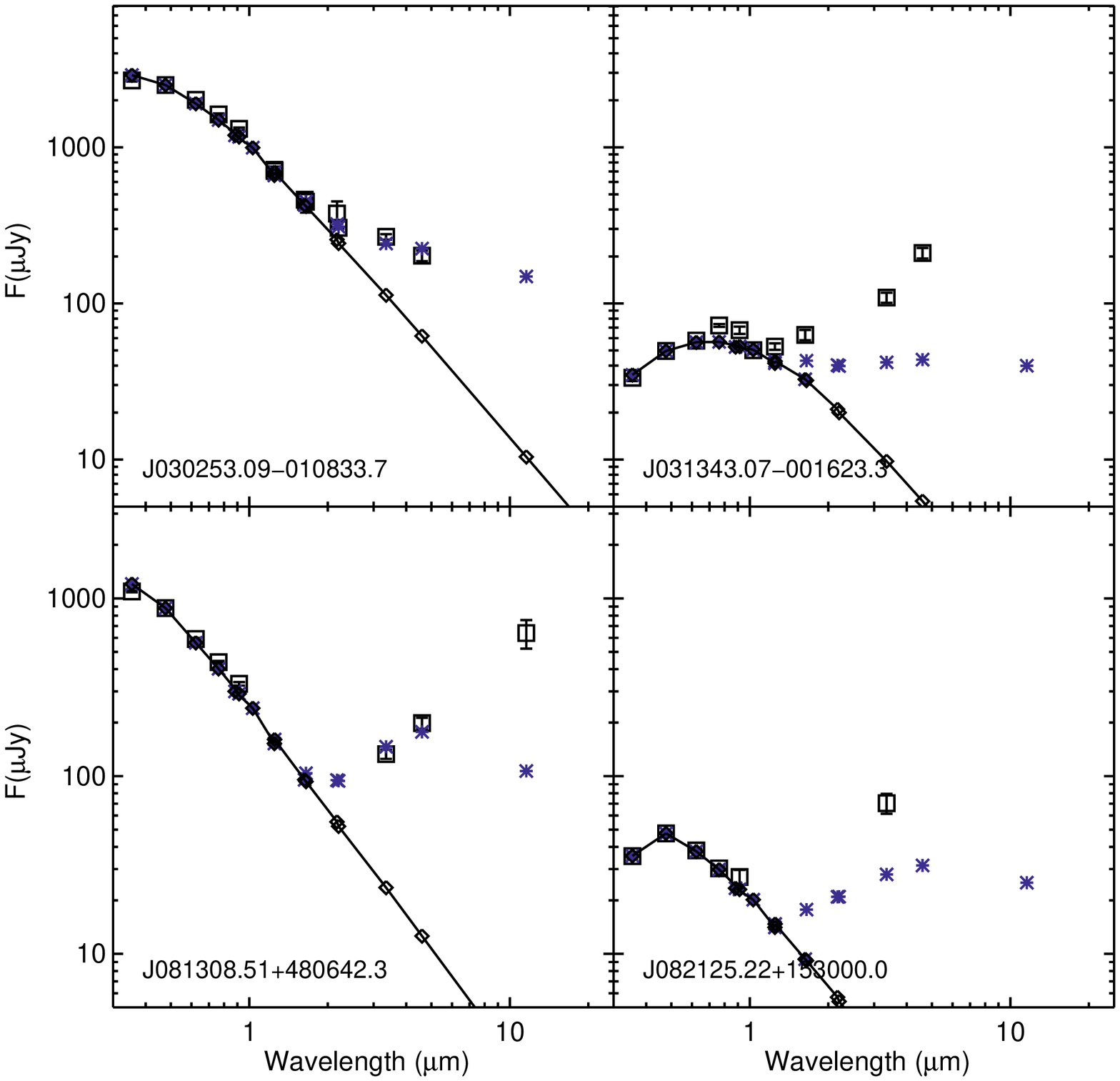}
\caption{\label{fig:d2}SEDs of WD+disk candidates (cont'd).}
\end{figure}

\begin{figure}
\plotone{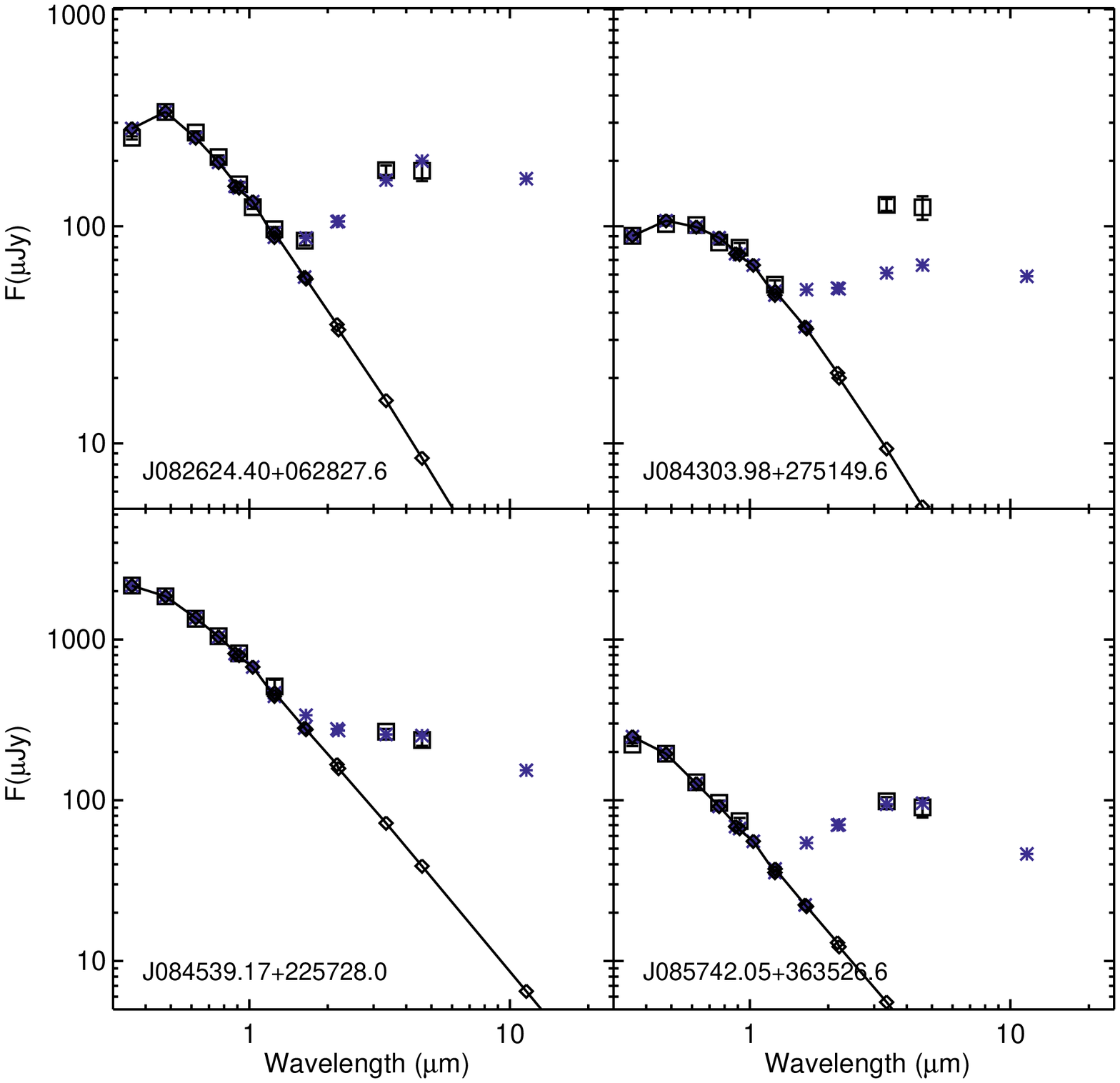}
\caption{\label{fig:d3}SEDs of WD+disk candidates (cont'd).}
\end{figure}

\begin{figure}
\plotone{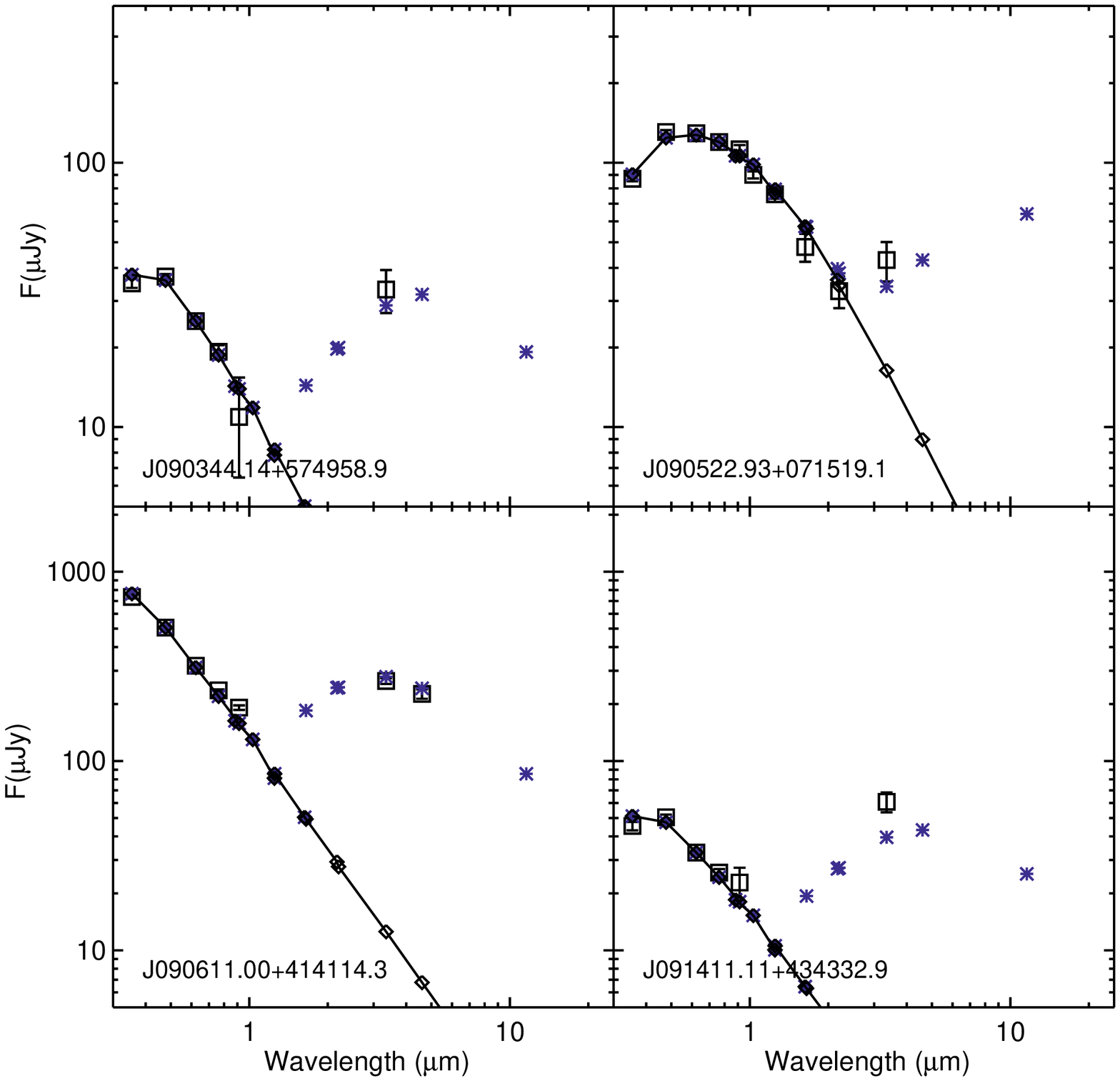}
\caption{\label{fig:d4}SEDs of WD+disk candidates (cont'd).}
\end{figure}

\begin{figure}
\plotone{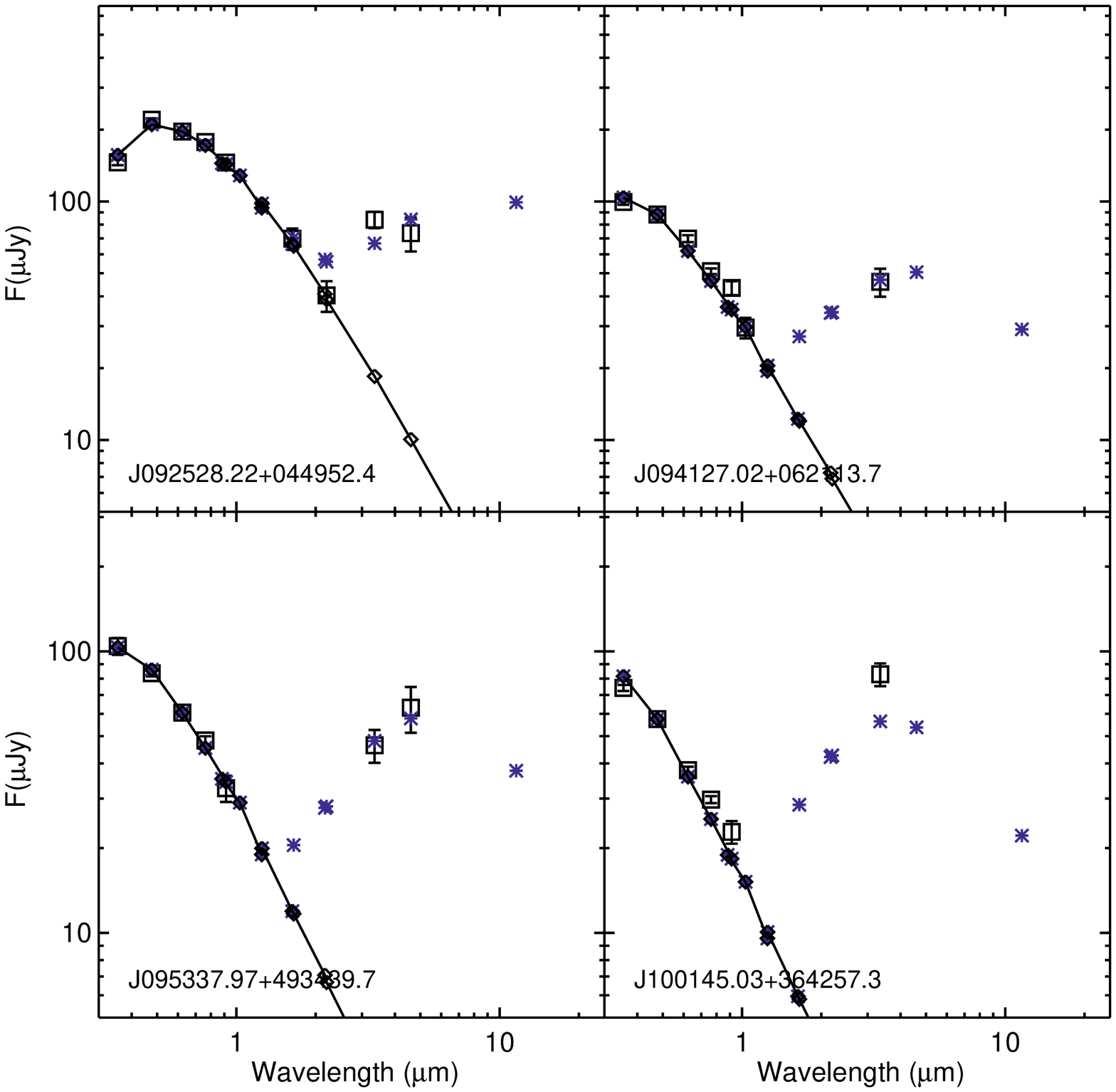}
\caption{\label{fig:d5}SEDs of WD+disk candidates (cont'd).}
\end{figure}

\begin{figure}
\plotone{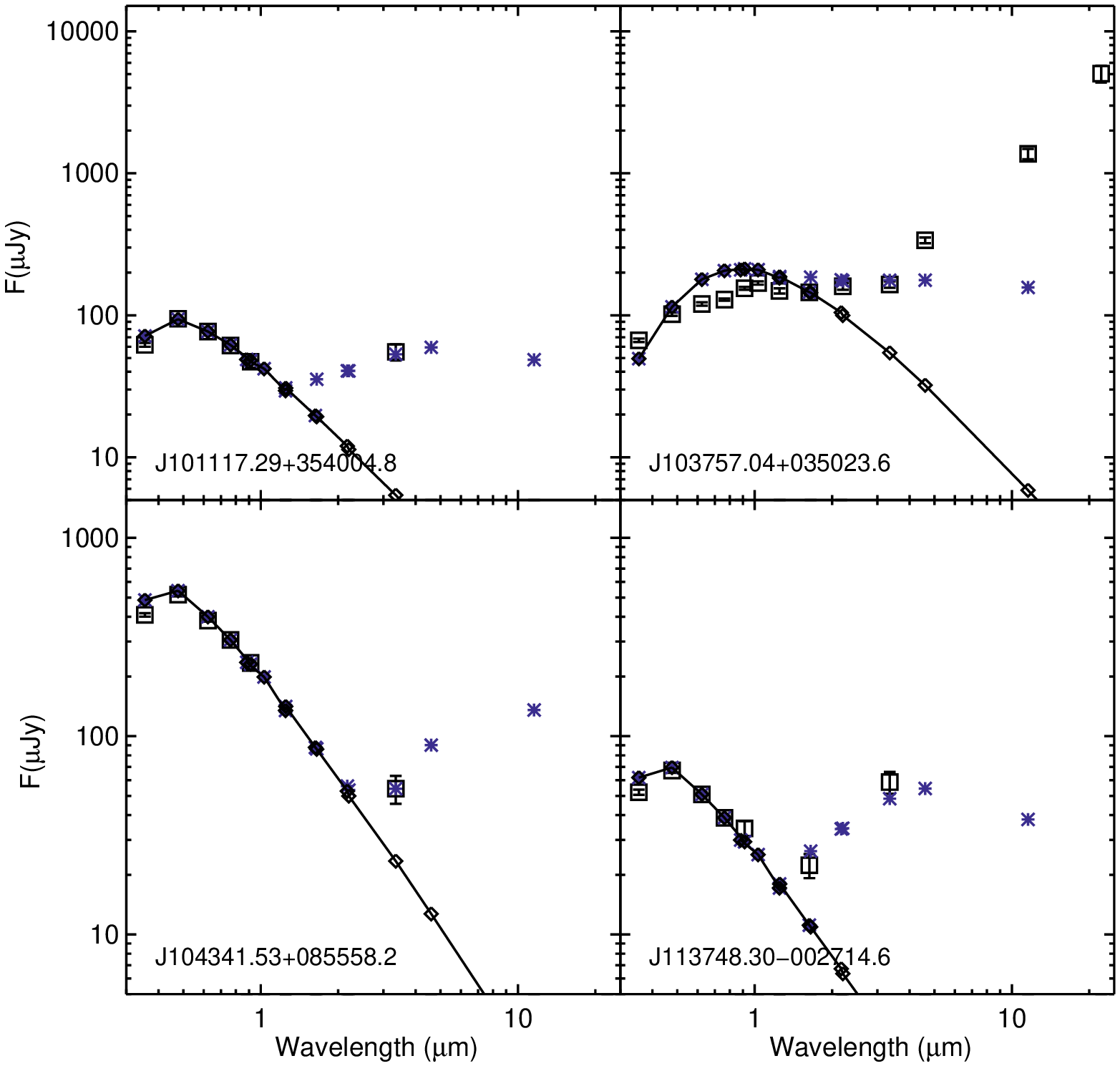}
\caption{\label{fig:d6}SEDs of WD+disk candidates (cont'd).}
\end{figure}

\begin{figure}
\plotone{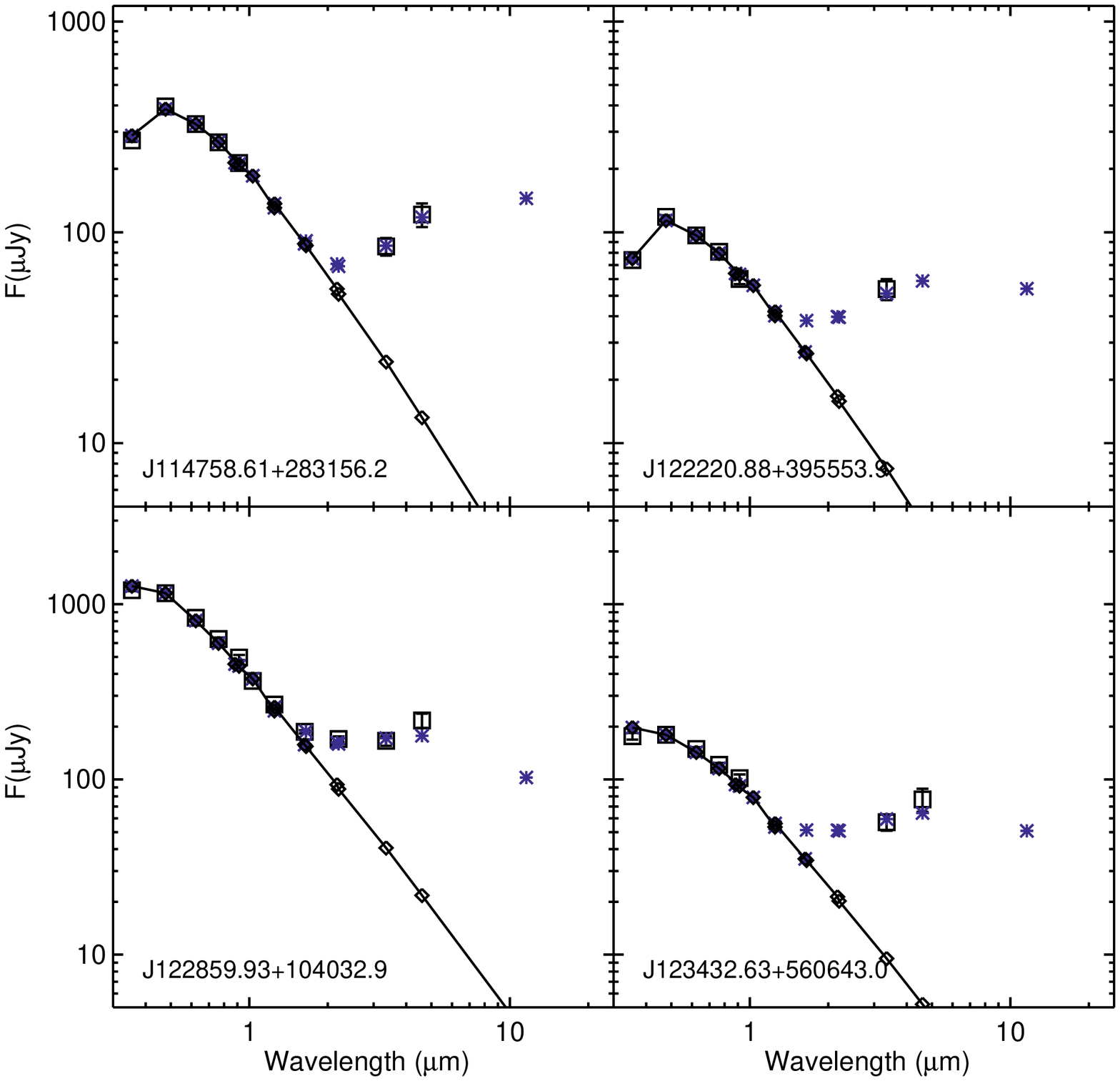}
\caption{\label{fig:d7}SEDs of WD+disk candidates (cont'd).}
\end{figure}

\begin{figure}
\plotone{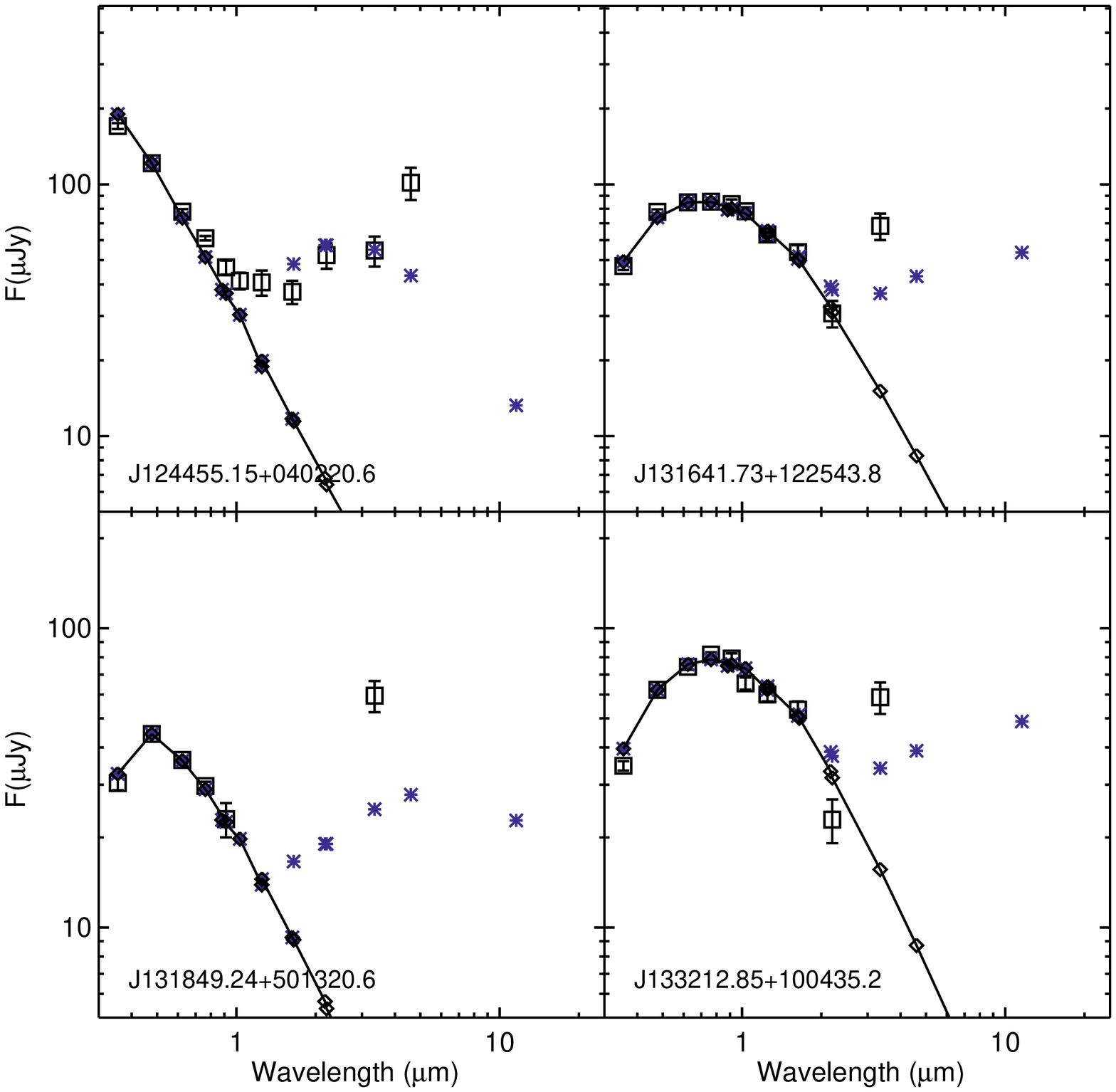}
\caption{\label{fig:d8}SEDs of WD+disk candidates (cont'd).}
\end{figure}

\begin{figure}
\plotone{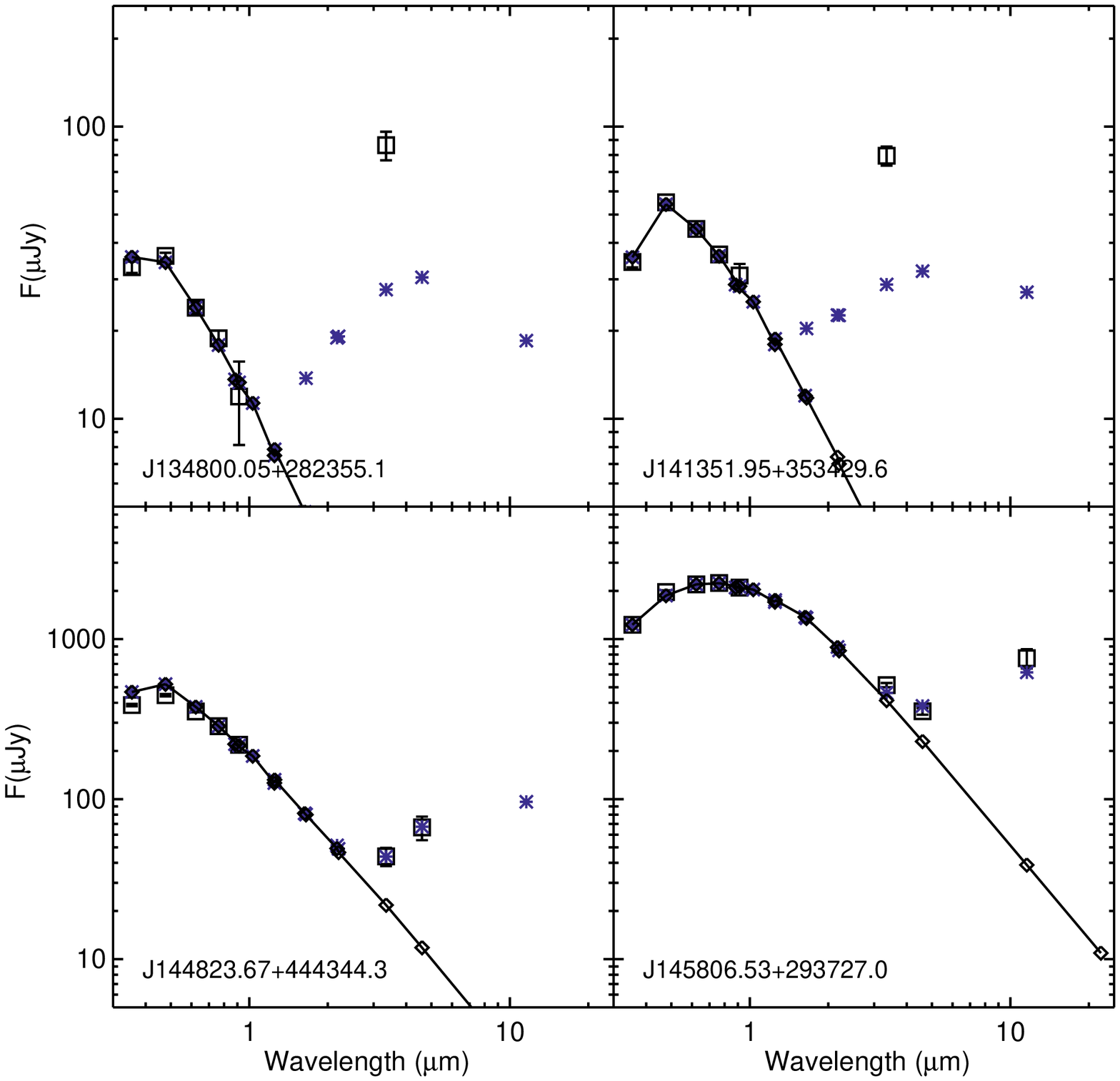}
\caption{\label{fig:d9}SEDs of WD+disk candidates (cont'd).}
\end{figure}

\begin{figure}
\plotone{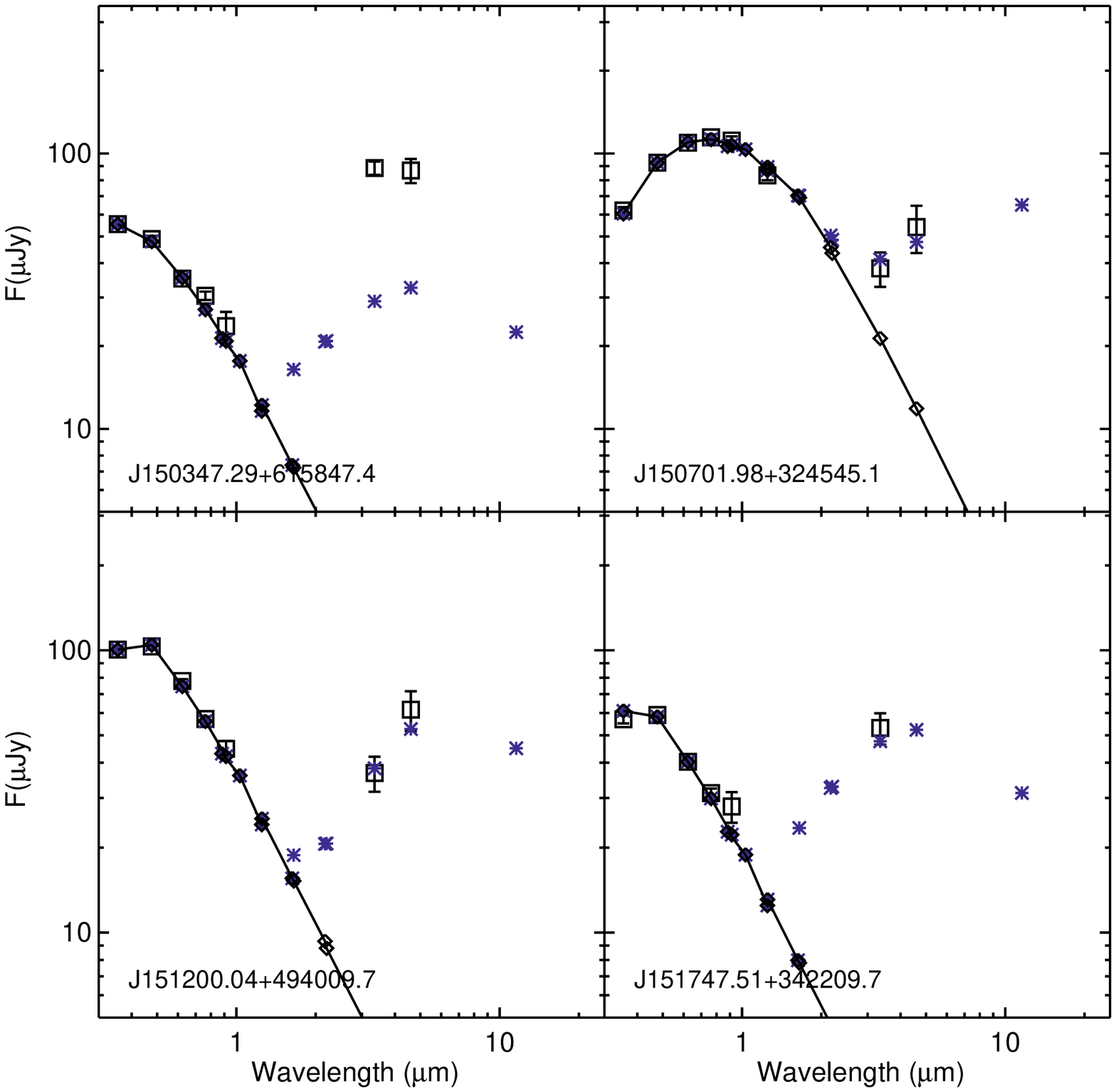}
\caption{\label{fig:d10}SEDs of WD+disk candidates (cont'd).}
\end{figure}

\begin{figure}
\plotone{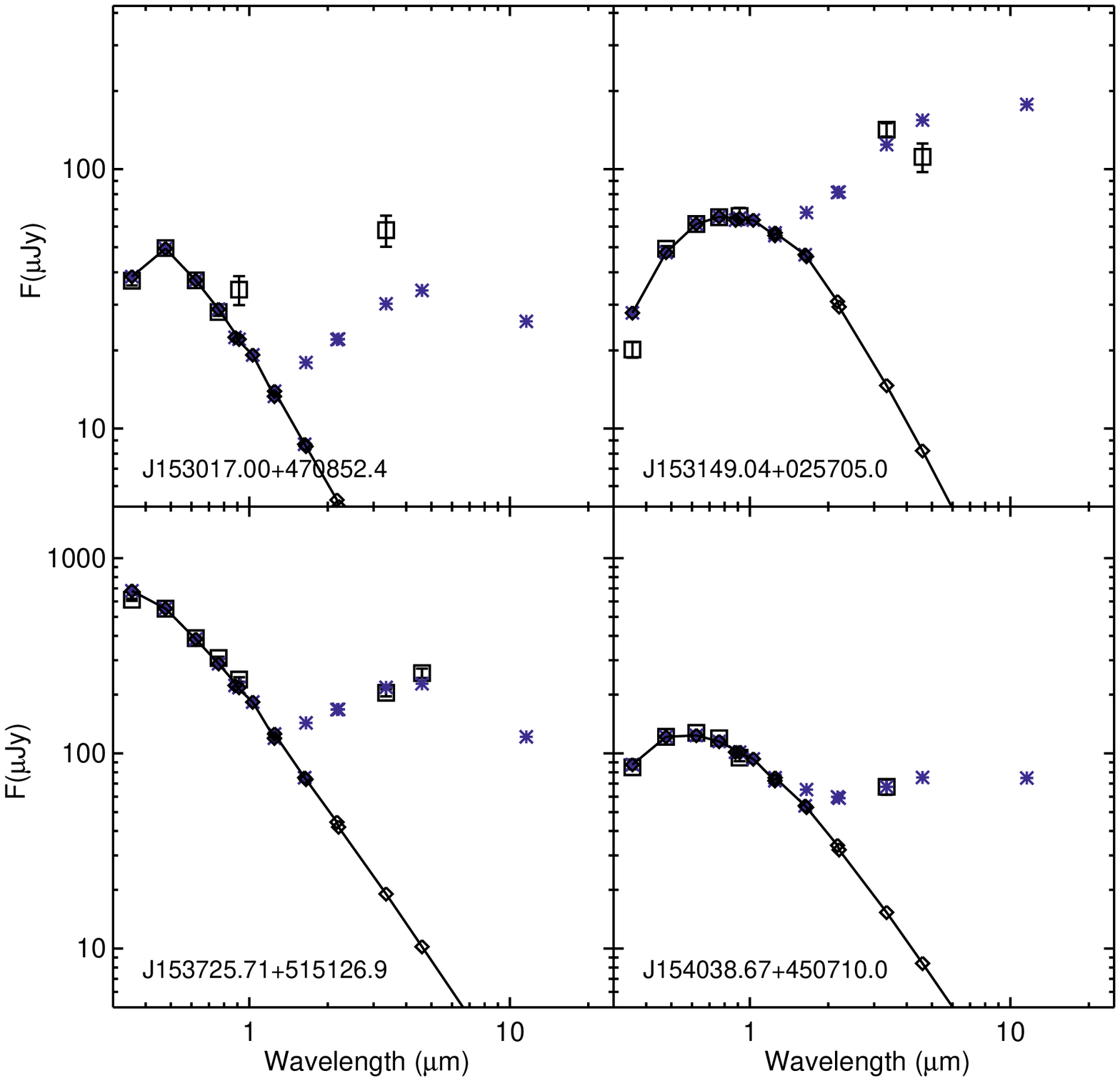}
\caption{\label{fig:d11}SEDs of WD+disk candidates (cont'd).}
\end{figure}

\begin{figure}
\plotone{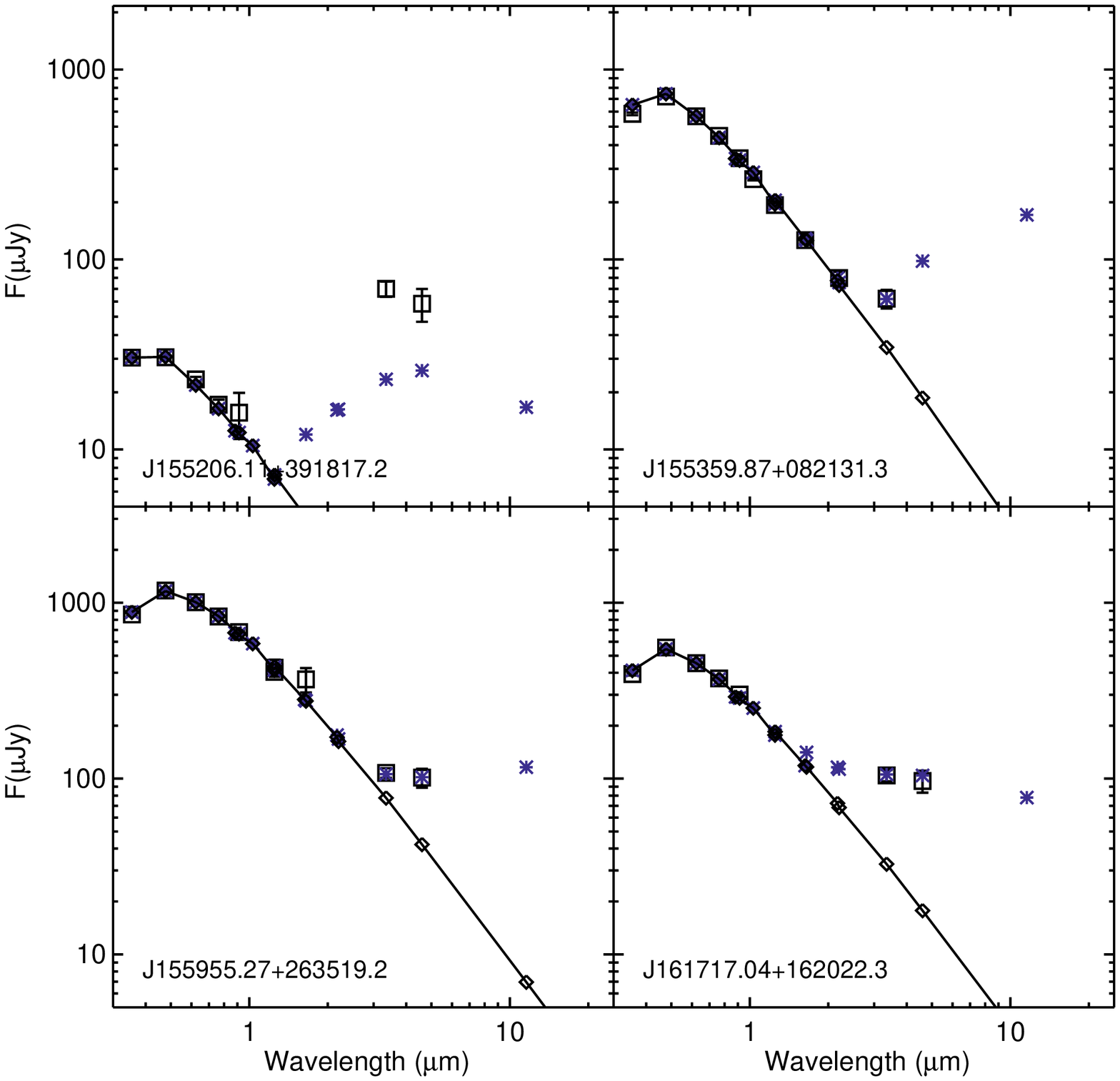}
\caption{\label{fig:d12}SEDs of WD+disk candidates (cont'd).}
\end{figure}

\begin{figure}
\plotone{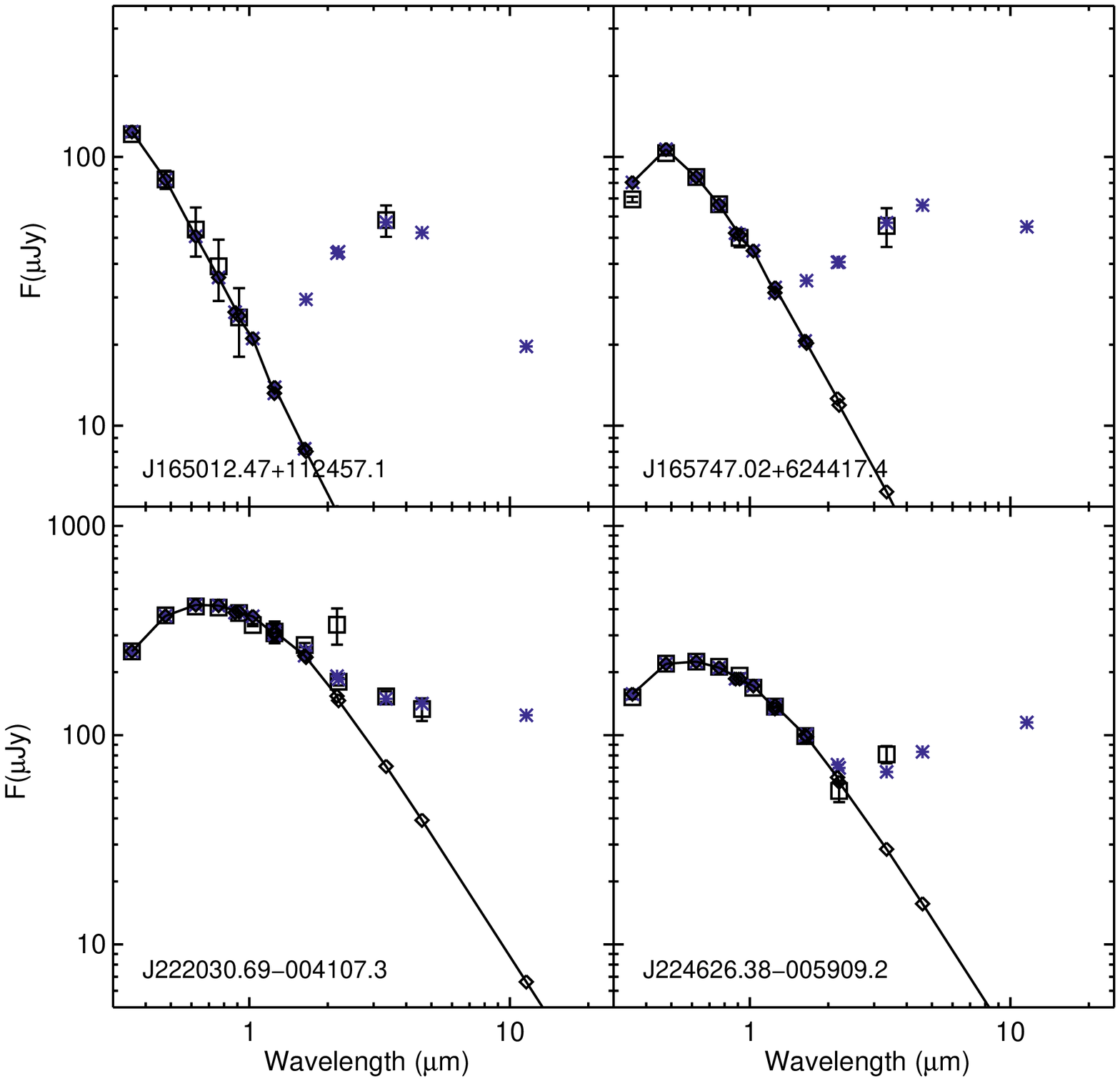}
\caption{\label{fig:d13}SEDs of WD+disk candidates (cont'd).}
\end{figure}

The 52 WD+disk candidates are listed in Table \ref{tab:disk}, with the inferred WD mass, photometric distance, age, and disk properties.  Figures \ref{fig:d1}--\ref{fig:d13} show the observed SEDs of each disk candidate, along with the best fitting WD+disk model.  As with the BD and M dwarf companions, we can estimate the overall frequency of WD+disk systems based on those candidates that had predicted WD photospheres brighter than our flux density cut-off (50~$\mu$Jy in $W1$).  This results in seven systems out of 395 (including previously known dusty white dwarfs).  WIRED J145806.53+293727 (G166-58) \citep{farihi08b} and WIRED J1043341.53+085558.2\citep{gaensicke07}, both previously confirmed as having dusty disks, were initially classified by our fitting algorithm as indeterminate -- we subsequently forced them to be assigned (and fit) as disk candidates.  G166-58 has a background galaxy located $\sim$5\arcsec away, which contaminates the {\em WISE} beam and produces spurious flux density measurements in $W1$, $W2$, and $W3$. Nonetheless, G166-58 has a predicted WD photospheric flux density brighter than our flux cut-off, so we include it in our disk frequency estimate, but the fit to its disk using the {\em WISE} photometry is not accurate.  One of the other targets, WIRED J103757.04+0354004.8 is a QSO mis-identified as WD.  The SED is clearly one of a rising IR spectrum characteristic of a galaxy or QSO.  The remaining six candidates brighter than the flux density cut-off (three of which are previously confirmed dusty disks) show no problems of possible contamination, resulting in a disk frequency of 1.5$\pm$0.6\%.  This is comparable to the frequency determined by \citet{farihi09}, but in this case we have a more homogeneous sample of WDs and a larger total sample of observed systems.  As noted for the WD+BD systems, this frequency is a lower limit, because 14 indeterminate candidates are also brighter than our flux cut-off, and some of these could be additional WD+disk systems.  The maximum frequency of disk systems would then be $\approx$5\% (if all of the indeterminate systems were actually WD+disk systems).  

\begin{figure}
\plotone{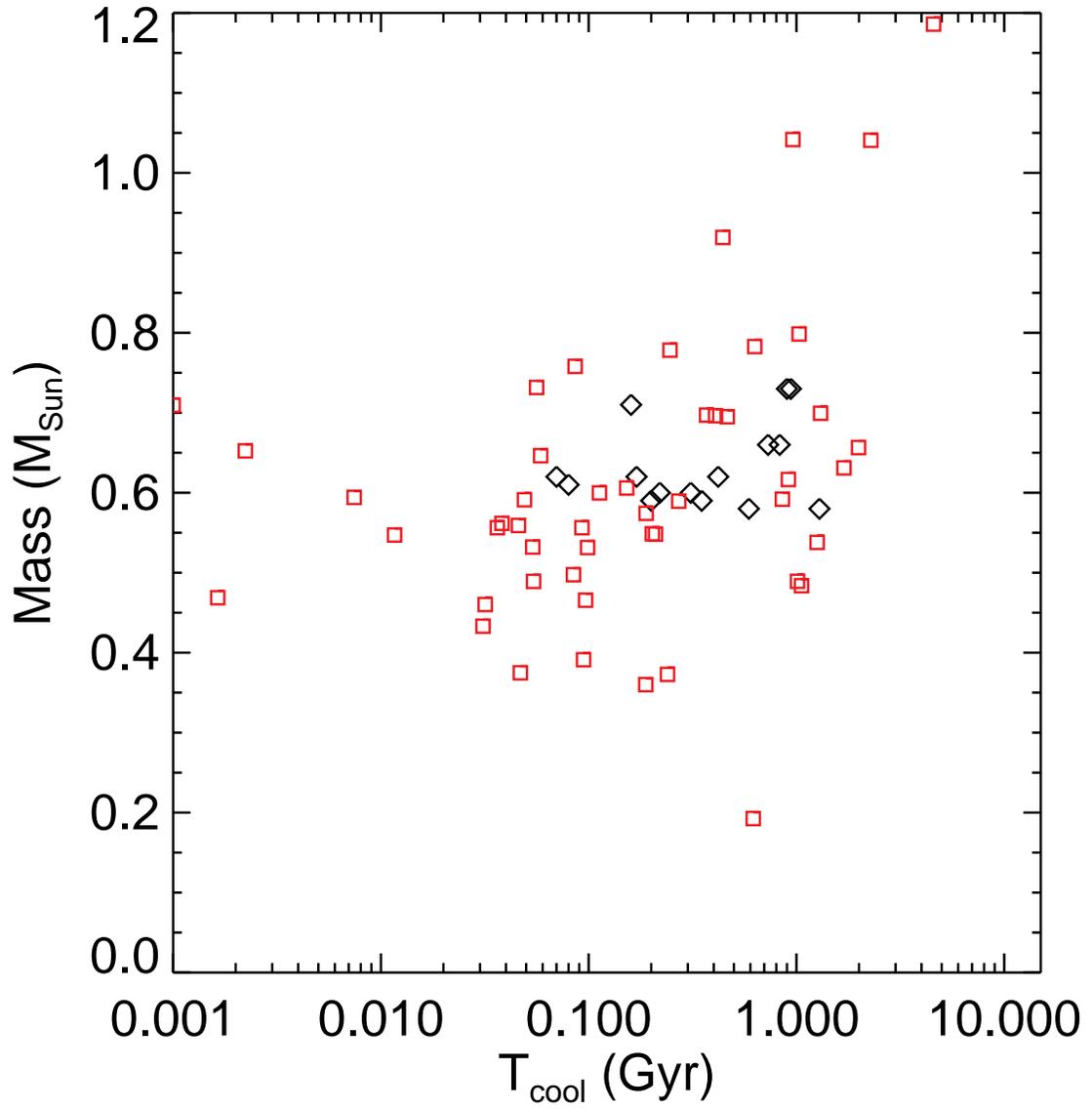}
\caption{\label{fig:massvsage} Mass vs.\ cooling age for previously known dusty WDs (diamonds) compared to candidates discovered in the WIRED Survey sample (squares).}
\end{figure}

We can also compare the distribution of previously known dusty WDs with our own sample of candidates.  In Figure \ref{fig:massvsage} we have plotted the WD mass of our candidate disk systems vs.\ their inferred WD cooling ages.   We have observed a significant number of younger WDs with disks, possibly a selection effect since these WDs are more easily detected with {\em WISE}.  Additionally, we extend our disk candidates to smaller WD masses, later cooling ages, and higher WD masses.  This more diverse sample will be useful for better constraining the origin and evolution of these disks.

\begin{figure}
\plotone{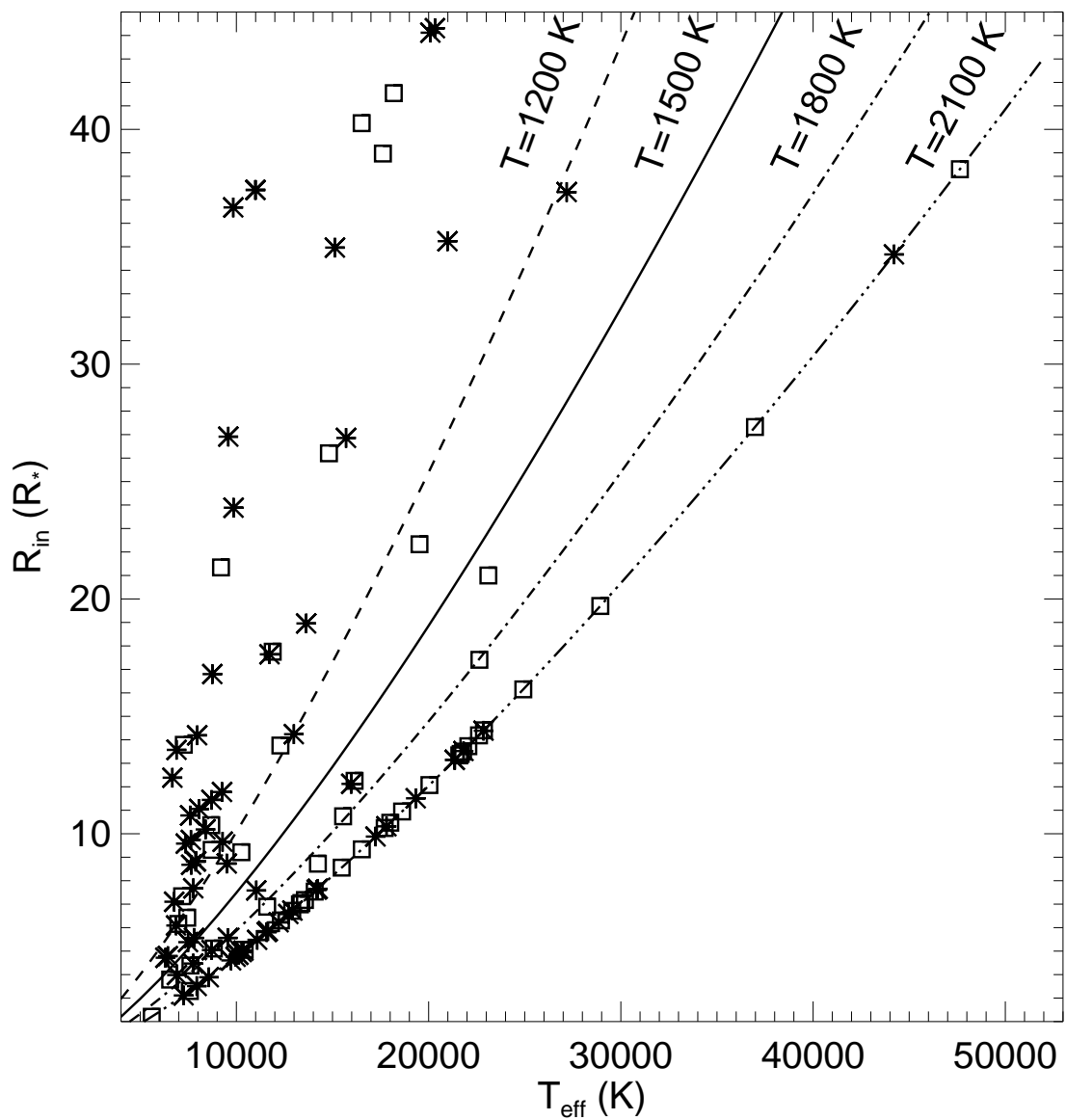}
\caption{\label{fig:rinvstemp} Inferred inner disk radius (R$_{\rm in}$) vs.\ WD effective temperature (T$_{\rm eff}$) for firm disk candidates (squares) and indeterminate systems (asterisks).
The four curves correspond to constant disk inner edge temperature for a given (T$_{\rm eff}$, R$_{\rm in}$) combinations, as follows:\ 2100~K (dash-triple dotted line), 1800~K (dash-dotted line), 1500~K (solid line), and 1200~K (dashed line).}
\end{figure}

Our inferred $R_{\rm in}$ value for each disk candidate corresponds to a particular disk inner edge temperature that depends, for example, on the temperature of the central WD.  Since we selected a minimum inner radius for each disk candidate that corresponded to a temperature of 2100~K, the approximate sublimation temperature of pure carbon dust, we can look at our disk and indeterminate excess candidates to understand where most disk radii fall in terms of radii of constant temperature.  Figure \ref{fig:rinvstemp} shows our disk candidates and our indeterminate candidates as a function of WD effective temperature.  In general, most of the disks have radii consistent with sublimation of silicate dust or even inner holes much larger than the radius corresponding to the dust sublimation temperature.  Some objects, however, are consistent with the hottest inner radii.  If these objects are confirmed, it may suggest either highly refractory dust, or inner structures that deviate from the simple vertically isothermal model that we have used, such as a ``puffy'' inner disk wall.  They should also be treated as the targets most likely to be affected by some form of photometric contamination in the {\em WISE} data, because they represent the absolute maximum brightness such a disk could obtain.  Any contamination source significantly brighter than the WD photosphere would be preferentially chosen by the hottest inner disk radii.  Of our disk candidates, ten show {\em WISE} flux densities above the hottest possible disks.  

One previously known WD with a disk, SDSS J084539.17+225728.0, was initially mis-classified by our SED fitting algorithm as a WD+M system, mainly because the best fitting T$_{\rm eff}$ for this WD is 7000~K cooler than the SDSS autofit temperature given in the preliminary DR7 WD catalog \citep{brinkworth11}.  This had the effect of creating a spurious IR excess at shorter wavelengths, mimicking a late-type M dwarf.  When this temperature change is taken into account, a disk model is marginally preferred for this target based on the $\chi^2$ criterion, and, based only on our SED model-fitting process, it is classified as indeterminate.  Its predicted photosphere in $W1$ is $>50$~$\mu$Jy, our flux density lower limit for inclusion in our calculation of the total dust disk frequency among WDs.

\subsection{Indeterminate Infrared Excess Sources} 

We found 69 indeterminate objects that met our IR excess detection criteria, but had indistinguishably good fits to both BD companion and dust disk models. These are listed in Table \ref{tab:ind}, which provides the properties of both the companion and disk as inferred from separate model fits.  For many of these targets, there is no near-IR photometry to better constrain the nature of the excess, nor is there a large excess in the {\em WISE} bands.  As mentioned previously, 14 indeterminate targets have $W1$ flux densities in excess of 50~$\mu$Jy and are included in the flux limited sample utilized above.  Of these 14 targets, several are affected by likely contamination from nearby sources which could account for both the presence of an IR excess and the ambiguity of its origin.

\section{Conclusions}
\label{s:conc}

We have cross-correlated the preliminary SDSS DR7 WD catalog with photometry in the 2MASS, UKIDSS, and {\em WISE} point source catalogs in order to discover new WDs with IR excesses.  A total of 1184 WDs show some sort of excess, the majority of which are candidate WD+M dwarf systems. A smaller percentage of these sources show excesses due to possible BD companions and dusty disks.  We find that $\approx$1--5\% of WDs detected by {\em WISE} down to a predicted photospheric brightness cut-off of 50~$\mu$Jy in the $W1$ band show excesses due to possible dust disks and another $\approx$1--5\% show excesses due to possible BD companions (in both cases, the upper end of the range assumes that all of our indeterminate excess sources are actually sources of the indicated types).  {\em WISE} is quite sensitive to WD photospheres in its $W1$ and $W2$ bands, and as the sample of WDs is completed to beyond 20~pc, a wealth of new IR excess systems may be discovered.  A large confirmed sample of WD+disk and WD+BD systems can lead to important statistical insights into the formation and evolution of dusty disks around WDs and their possible links to planetary systems, and to the formation and evolution of BDs during their parent star's lifetime and death.

\acknowledgements

This research was supported by an appointment to the NASA Postdoctoral Program at the Goddard Space Flight Center, administered by Oak Ridge Associated Universities through a contract with NASA.    
This work is based on data obtained from:\ 
(a) the Wide-field Infrared Survey Explorer, which is a joint project of the University of California, Los Angeles, and the Jet Propulsion Laboratory (JPL), California Institute of Technology (Caltech), funded by the National Aeronautics and Space Administration (NASA); 
(b) the Two Micron All Sky Survey (2MASS), a joint project of the University of Massachusetts and the Infrared Processing and Analysis Center (IPAC)/Caltech, funded by NASA and the National Science Foundation; 
(c) the UKIRT Infrared Deep Sky Survey (UKIDSS);
(d) the Sloan Digital Sky Survey (SDSS). Funding for the SDSS and SDSS-II has been provided by the Alfred P.\ Sloan Foundation, the Participating Institutions, the National Science Foundation, the U.S. Department of Energy, the National Aeronautics and Space Administration, the Japanese Monbukagakusho, the Max Planck Society, and the Higher Education Funding Council for England. The SDSS Web Site is http://www.sdss.org/.  The SDSS is managed by the Astrophysical Research Consortium for the Participating Institutions. The Participating Institutions are the American Museum of Natural History, Astrophysical Institute Potsdam, University of Basel, University of Cambridge, Case Western Reserve University, University of Chicago, Drexel University, Fermilab, the Institute for Advanced Study, the Japan Participation Group, Johns Hopkins University, the Joint Institute for Nuclear Astrophysics, the Kavli Institute for Particle Astrophysics and Cosmology, the Korean Scientist Group, the Chinese Academy of Sciences (LAMOST), Los Alamos National Laboratory, the Max-Planck-Institute for Astronomy (MPIA), the Max-Planck-Institute for Astrophysics (MPA), New Mexico State University, Ohio State University, University of Pittsburgh, University of Portsmouth, Princeton University, the United States Naval Observatory, and the University of Washington;
(e) the SIMBAD database, operated at CDS, Strasbourg, France; and 
(f) the NASA/IPAC Infrared Science Archive, which is operated by JPL, Caltech, under a contract with NASA.  
M.C. thanks NASA for supporting his participation in this work through UCLA Sub-Award 1000-S-MA756 with a UCLA FAU 26311 to MIRA.

\bibliography{wired}
\bibliographystyle{apj}

\appendix

\section{Notes on Individual Objects}
\label{s:app1}

In this section we provide notes on each target of interest (TOI) that was identified as a WD+disk candidate, WD+BD candidate, or WD+indeterminate IR excess source.  As explained in Section \ref{s:results}, we inspected SDSS (i-band and color-composite) and {\em WISE} (all bands) images of each TOI for irregularities or additional sources within $\sim$7\farcs8 (i.e., the separation below which the {\em WISE} photometry algorithm does not distinguish multiple sources).  When SDSS spectra were available for a given TOI, we inspected them for obvious non-WD spectral features or irregularities.  We also conducted a CROSSID search in the SDSS DR7 catalog for any identified objects within 9\arcsec of the targets.  In many cases, when an object is noted via visual inspection of the SDSS images close to the target, it is confirmed by the CROSSID search results.  In some cases, nearby sources might be seen in the SDSS images, but not the {\em WISE} images (or vice-versa), which is likely due to the relative sensitivities of the two surveys and/or the spectral energy distirbutions of the objects, and offers information about the potential of any nearby object for contaminating the {\em WISE} photometry (e.g., nearby objects that are blue and/or faint in the SDSS images are less likely to be detected as significant contaminating sources in the {\em WISE} images).  We individually inspected the SEDs generated by our fitting algorithm for any obvious problems in the corresponding TOI classifications.  Finally, we cite relevant results already published in the literature.  

We provide notes for each target in up to four categories, as described above, corresponding to the SDSS and {\em WISE} image and photometry quality, the CROSSID results, and the literature search results.  The absence of a particular category ({\em WISE}, SDSS, CROSSID, NOTES) for any TOI indicates that there was nothing to note with respect to that category; that is, ``no news is good news'' (e.g., no ``{\em WISE}'' entry in the notes for a TOI indicates there were no irregularities or concerns about the {\em WISE} photometry or image quality).

\subsection{Dusty WD Candidates}

\begin{description}

\item[011055.06+143922.2 --]
  %WISE: ...  
  SDSS: blue; %no additional sources within 7\farcs8  
  %CROSSID: ...  
  NOTES: WD0108+143 (G33-45), {\em not} listed as a WD+M binary in \citet{koester09}. 
 
\item[012929.99+003411.2 --]
  %WISE: ...  
  SDSS: blue;  
  %CROSSID: ...  
  NOTES: DA spectroscopic classification \citep{eisenstein06}.
 
\item[024602.66+002539.2 --]
  WISE: photometry possibly contaminated by faint blended source(s) within 7\farcs8;
  SDSS: blue; %no additional sources within 7\farcs8  
  %CROSSID: ...  
  NOTES: WD0243+002.
 
\item[025049.44+343651.0 --]
  %WISE: ...  
  SDSS: blue; possible slight pink extension (?) at $\approx$1\farcs7;
  CROSSID: galaxy at 1\farcs7 ($g=22.11$, $i=22.46$).
  %NOTES: ... 
 
\item[030253.09-010833.7 --]
  %WISE: ...  
  SDSS: blue; %no additional sources within 7\farcs8  
  %CROSSID: ...  
  NOTES: GD40, a known dusty WD (disk and metal-contamination; e.g., \citealt{klein10}).
 
\item[031343.07-001623.3 --]
  %WISE: ...  
  SDSS: blue-white;
  %CROSSID: ...  
  NOTES: quasar misidentified as a WD \citep{schneider10}.
 
\item[081308.51+480642.3 --]
  WISE: several sources at 20\arcsec, passive deblending applied (nb=2, na=0), but no indication of photometric contamination;
  SDSS: blue; %no additional sources within 7\farcs8  
  %CROSSID: ...  
  NOTES: WD0809+482.
 
\item[082125.22+153000.0 --]
  WISE: photometry possibly contaminated by blended source;
  SDSS: blue;  
  CROSSID: faint galaxy at 7\farcs9 ($g=23.49$, $i=22.62$).
  %NOTES: ... 
 
\item[082624.40+062827.6 --]
  %WISE: ...  
  SDSS: blue; faint, red source at $\approx$3\arcsec;
  CROSSID: galaxy at 8\farcs8 ($g=23.78$, $i=21.77$).
  %NOTES: ... 
 
\item[084303.98+275149.6 --]
  %WISE: ...  
  SDSS: blue-white; orange source at $\approx$3\farcs2;
  CROSSID: galaxy at 3\farcs2 ($g=22.48$, $i=21.03$);
  NOTES: EG Cnc; faint, short orbital period (1.4~h) cataclysmic variable, dwarf nova (TOAD) type; likely appears as an unresolved WD+M binary during quiescence (e.g., \citealt{m98}).
 
\item[084539.17+225728.0 --]
  %WISE: ...  
  SDSS: blue;  
  %CROSSID: ...  
  NOTES: WD0842+231, Ton 345; known DBZ Emission line WD disk star with IR excess \citep{gaensicke08,melis10}.
 
\item[085742.05+363526.6 --]
  %WISE: ...  
  SDSS: blue, slightly extended, possibly superimposed on background galaxy.
  %CROSSID: ...  
  %NOTES: ... 
 
\item[090344.14+574958.9 --]
  WISE: photometry possibly contaminated by blended source within 7\farcs8;
  SDSS: blue; blended WISE source is not visible;
  %CROSSID: ...  
  NOTES: WD0859+580.
 
\item[090522.93+071519.1 --]
  %WISE: ...  
  SDSS: blue-white. %no additional sources within 7\farcs8  
  %CROSSID: ...  
  %NOTES: ... 
 
\item[090611.00+414114.3 --]
  %WISE: ...  
  SDSS: slightly extended blue/pink blend;
  CROSSID: galaxy at 5\farcs5 ($g=23.82$, $i=21.71$);
  NOTES: WD0902+418.
 
\item[091411.11+434332.9 --]
  WISE: photometry likely contaminated by blended source within 7\farcs8;
  SDSS: blue. %no additional sources within 7\farcs8  
  %CROSSID: ...  
  %NOTES: ... 
 
\item[092528.22+044952.4 --]
  %WISE: ...  
  SDSS: blue; yellow source at $\approx$3\arcsec.
  %CROSSID: ...  
  %NOTES: ... 
 
\item[094127.02+062113.7 --]
  %WISE: ...  
  SDSS: blue. %no additional sources within 7\farcs8  
  %CROSSID: ...  
  %NOTES: ... 
 
\item[095337.97+493439.7 --]
  WISE: photometry possibly contaminated by faint blended source within 7\farcs8;
  SDSS: blue; two resolved, faint, yellow sources within 7\farcs8;
  CROSSID: star at 6\farcs7 ($g=23.76$, $i=22.18$);
  NOTES: a DB WD \citep{eisenstein06}.
 
\item[100145.03+364257.3 --]
  WISE: potential contamination from nearby bright star diffraction spike, but cc\_flag information is not available for i3o catalog; target FWHM consistent with other point sources;
  SDSS: blue; faint, red source at $\approx$1\farcs5;
  CROSSID: galaxy at 6\farcs8 ($g=24.25$, $i=21.57$).
  %NOTES: ... 
 
\item[101117.29+354004.8 --]
  WISE: target FWHM slightly extended compared to point sources in the field;
  SDSS: blue. %no additional sources within 7\farcs8  
  %CROSSID: ...  
  %NOTES: ... 
 
\item[103757.04+035023.6 --]
  WISE: photometry likely contaminated by blended source within 7\farcs8;
  SDSS: blue-white; %no additional sources within 7\farcs8  
  %CROSSID: ...  
  NOTES: currently identified as a QSO in SIMBAD.
 
\item[104341.53+085558.2 --]
  %WISE: ...  
  SDSS: blue; very faint extended source within 7\farcs8 (blended);
  %CROSSID: ...  
  NOTES: WD1041+091, known dusty WD with gaseous and dusty disk (e.g., \citealt{farihi10}).
 
\item[113748.30-002714.6 --]
  WISE: photometry likely contaminated by blended source within 7\farcs8;
  SDSS: blue; %no additional sources within 7\farcs8  
  %CROSSID: ...  
  NOTES: WD1135-001B.
 
\item[114758.61+283156.2 --]
  %WISE: ...  
  SDSS: blue; %no additional sources within 7\farcs8  
  %CROSSID: ...  
  NOTES: WD1145+288.
 
\item[122220.88+395553.9 --]
  WISE: photometry likely contaminated by blended source within 7\farcs8 (target is elongated compared to point sources in the field);
  SDSS: blue; %no additional sources within 7\farcs8  
  %CROSSID: ...  
  NOTES: currently identified as a possible QSO in SIMBAD.
 
\item[122859.93+104032.9 --]
  %WISE: ...  
  SDSS: blue; %no additional sources within 7\farcs8  
  %CROSSID: ...  
  NOTES: WD1226+110, a known WD with a gaseous and dusty disk \citep{brinkworth09}.
 
\item[123432.63+560643.0 --]
  %WISE: ...  
  SDSS: blue. %no additional sources within 7\farcs8  
  %CROSSID: ...  
  %NOTES: ... 
 
\item[124455.15+040220.6 --]
  %WISE: ...  
  SDSS: blue; possible faint, resolved object at $\lesssim$2\arcsec;
  CROSSID: faint star at 4\farcs4 ($g=23.95$, $i=22.27$);
  NOTES: DA spectroscopic classification \citep{eisenstein06}.  Listed as having foreground/background contamination in \citet{steele11}.
 
\item[131641.73+122543.8 --]
  WISE: target is in proximity of very bright source and could be affected by an associated image artifact (i3o detection only, so cc\_flag information is not available);
  SDSS: blue-white; faint sources at $\approx$1\farcs5 (white) and $\approx$6\farcs8 (red);
  CROSSID: star at 6\farcs6 ($g=22.91$, $i=20.20$).
  %NOTES: ... 
 
\item[131849.24+501320.6 --]
  %WISE: ...  
  SDSS: blue-white;  
  %CROSSID: ...  
  NOTES: DA spectroscopic classification \citep{eisenstein06}.

\item[133212.85+100435.2 --]
  WISE: photometry likely contaminated by blended source within 7\farcs8;
  SDSS: blue-white; %no additional sources within 7\farcs8  
  CROSSID: galaxy at 8\farcs6 ($g=24.20$, $i=23.09$).
  %NOTES: ... 
 
\item[134800.05+282355.1 --]
  WISE: closest source is point-like, but coordinate offset ($\approx$2\arcsec) suggests that it might be the red source seen at shorter SDSS wavelengths, not the WD; at best, photometry is likely contaminated by blended source within 7\farcs8;
  SDSS: blue; red source at $\approx$2\farcs5;
  CROSSID: star at 2\farcs4 ($g=21.94$, $i=19.26$); galaxy at 7\farcs3 ($g=22.88$, $i=21.73$).
  %NOTES: ... 
 
\item[141351.95+353429.6 --]
  %WISE: ...  
  SDSS: blue; very faint, extended, red emission at $\approx$2\arcsec;  
  CROSSID: star at 8\arcsec ($g=23.47$, $i=23.21$).
  %NOTES: possible contamination from nearby star?  
 
\item[144823.67+444344.3 --]
  %WISE: ...  
  SDSS: blue; possible very faint source at $\approx$3\arcsec;
  %CROSSID: ...  
  NOTES: WD1446+449.
 
\item[145806.53+293727.0 --]
  WISE: photometry likely contaminated by blended source within 7\farcs8;
  SDSS: white-blue; yellow, extended source at $\approx$5\farcs9;
  CROSSID: galaxy at 5\farcs5 ($g=20.80$, $i=19.06$);
  NOTES: WD1455+298 (EGGR298, G166-58), known dusty WD with Spitzer-IRAC data (e.g., \citealt{farihi09}).
 
\item[150347.29+615847.4 --]
  %WISE: ...  
  SDSS: blue;  
  %CROSSID: ...  
  NOTES: DB spectroscopic classification \citep{eisenstein06}; WISE photometry significantly brighter than best fitting disk model, possible contamination?
 
\item[150701.98+324545.1 --]
  %WISE: ...  
  SDSS: blue-white. %no additional sources within 7\farcs8  
  %CROSSID: ...  
  %NOTES: ... 
 
\item[151200.04+494009.7 --]
  %WISE: ...  
  SDSS: blue; possible very faint source at $\approx$3\arcsec.
  %CROSSID: ...  
  %NOTES: ... 
 
\item[151747.51+342209.7 --]
  %WISE: ...  
  SDSS: blue; very faint, possibly extended, red source at $\approx$2\farcs4;
  CROSSID: galaxy at 9\arcsec ($g=23.48$, $i=20.86$).
  %NOTES: ... 
 
\item[153017.00+470852.4 --]
  WISE: photometry very likely contaminated by diffraction spike from nearby very bright source; i3o detection only, so cc\_flag information is not available;
  SDSS: blue. %no additional sources within 7\farcs8  
  %CROSSID: ...  
  %NOTES: ... 
 
\item[153149.04+025705.0 --]
  %WISE: ...  
  SDSS: white-blue;  
  %CROSSID: ...  
  NOTES: FIRST radio source within 1\arcsec; classified as DAH in Preliminary DR7 WD Catalog.
 
\item[153725.71+515126.9 --]
  %WISE: ...  
  SDSS: blue; %no additional sources within 7\farcs8  
  %CROSSID: ...  
  NOTES: WD1536+520.
 
\item[154038.67+450710.0 --]
  %WISE: ...  
  SDSS: blue-white; very faint source at $\approx$3\farcs7;
  CROSSID: galaxy at 3\farcs8 ($g=24.80$, $i=22.05$).
  %NOTES: ... 
 
\item[155206.11+391817.2 --]
  WISE: photometry possibly contaminated by faint blended source;
  SDSS: blue;  
  CROSSID: faint galaxy at 5\farcs9 ($g=24.37$, $i=22.17$); faint star at 6\farcs3 ($g=24.34$, $i=22.08$);
  NOTES: DA spectroscopic classification \citep{eisenstein06}.
 
\item[155359.87+082131.3 --]
  %WISE: ...  
  SDSS: blue; faint, red source at $\approx$2\farcs7.
  %CROSSID: ...  
  %NOTES: ... 
 
\item[155955.27+263519.2 --]
  WISE: faint source just outside the 7\farcs8 radius, active deblending used (nb=1, na=1);
  SDSS: blue. %no additional sources within 7\farcs8  
  %CROSSID: ...  
  %NOTES: ... 
 
\item[161717.04+162022.3 --]
  %WISE: ...  
  SDSS: blue; %no additional sources within 7\farcs8  
  %CROSSID: ...  
  NOTES: known IR excess indicative of gaseous and dusty disk \citep{brinkworth11}.
 
\item[165012.47+112457.1 --]
  %WISE: ...  
  SDSS: blue; brighter, white source at $\approx$2\farcs7;
  CROSSID: galaxy at 2\farcs5 ($g=18.57$, $i=17.96$); star at 8\farcs7 ($g=24.44$, $i=21.98$).
  %NOTES: ... 
 
\item[165747.02+624417.4 --]
  %WISE: ...  
  SDSS: blue; %no additional sources within 7\farcs8  
  %CROSSID: ...  
  NOTES: WD1657+628.
 
\item[222030.69-004107.3 --]
  %WISE: ...  
  SDSS: blue; very faint source at $\approx$7\arcsec;
  CROSSID: star at 6\farcs7 ($g=22.44$, $i=21.52$);
  NOTES: PHL5038, a $T_{\rm eff}=8000$~K DA WD with an L8 companion at a separation of 0\farcs94 \citep{steele09}.
 
\item[224626.38-005909.2 --]
  %WISE: ...  
  SDSS: blue; possible barely resolved, very faint, red additional source.
  %CROSSID: ...  
  %NOTES: ... 

% ultracool WD, moved to "naked WD" classification 
%\item[230645.72+212859.3 --]
%  %WISE: ... 
%  SDSS: images not available. 
%  %CROSSID: ... 
%  %NOTES: ... 

\end{description}

\subsection{WD+BD Candidates}

\begin{description}

\item[012532.02+135403.6 --]
 %WISE: ...  
 SDSS: blue-white; %no additional sources within 7\farcs8
 %CROSSID: ...  
 NOTESS: WD0122+136, a DC WD \citep{eisenstein06}.

\item[013532.97+144555.9 --]
 %WISE: ...  
 SDSS: blue; %no additional sources within 7\farcs8
 %CROSSID: ...  
 NOTES: WD0132+145, reported WD+L5 in \citet{steele11} and WD+L6 in \citet{girven11}.

\item[013553.72+132209.2 --]
 %WISE: ...  
 SDSS: blue-white; %no additional sources within 7\farcs8  
 %CROSSID: ...  
 NOTES: currently identified as a QSO in SIMBAD. 

\item[033444.86-011253.8 --]
 %WISE: ...  
 SDSS: blue-white; %no additional sources within 7\farcs8  
 %CROSSID: ...  
 NOTES: WD0332-013, a DC WD \citep{eisenstein06}.

\item[064607.86+280510.1 --]
 %WISE: ...  
 SDSS: images not available; 
 CROSSID: galaxy at 2\farcs1 ($g=21.91$, $i=19.25$); stars at 4\farcs4 ($g=24.03$, $i=22.53$), 6\farcs8 ($g=23.43$, $i=21.39$), and 7\farcs9 ($g=23.98$, $i=22.39$).
 %NOTES: ... 

\item[081113.73+144150.6 --]
 WISE: photometry possibly contaminated by faint blended source within 7\farcs8; 
 SDSS: blue-white/red-orange blend at separation of $\approx$1\farcs7.  
 %CROSSID: ...  
 %NOTES: ... 

\item[082412.27+175155.8 --]
 %WISE: ...  
 SDSS: white;
 %CROSSID: ...  
 NOTES: possible galaxy; SED is poor fit to companion (BD or M star) model.

\item[083038.79+470247.0 --]
 %WISE: ... 
 SDSS: pink-white; 
 %CROSSID: ...  
 NOTES: known DA+M spectroscopic binary \citep{eisenstein06}. % \citep{2006ApJS..167...40E}.

\item[083254.38+313904.2 --]
 %WISE: ...  
 SDSS: blue-white; faint, yellow source at $\approx$1\farcs8; 
 CROSSID: galaxy at 1\farcs9 ($g=21.46$, $i=19.36$), star at 8\farcs7 ($g=24.74$, $i=22.24$); 
 NOTES: LPP311-32.

\item[085930.41+103241.1 --]
 WISE: photometry possibly contaminated by fainter blended source within 7\farcs8; 
 SDSS: blue;  
 CROSSID: red star at 8\farcs9 ($g=23.41$, $i=20.58$);  
 NOTES: classified as DA+M in Preliminary DR 7 WD Catalog, DA in \citep{eisenstein06}. 

\item[092233.13+050640.0 --]
 %WISE: ... 
 SDSS: blue.
 %CROSSID: ...  
 %NOTES: ... 

\item[093821.34+342035.6 --]
 WISE: photometry possibly contaminated by nearby source at $>$7\farcs8; 
 SDSS: blue-white; faint, red source at $\approx$10\arcsec. 
 %CROSSID: ...  
 NOTES: reported as having unresolved companion in \citet{steele11}. 

\item[100128.30+415001.6 --]
 WISE: photometry likely contaminated by blended source within 7\farcs8; 
 SDSS: blue-white (slightly pink at one end, possible unresolved double); additional faint, red source at $\approx$6\farcs0; 
 CROSSID: galaxy at 6\farcs1 ($g=23.37$, $i=21.09$).
 %NOTES: ... 

\item[100646.07+413306.5 --]
 %WISE: ... 
 SDSS: blue; 
 %CROSSID: ...  
 NOTES: DB/A: spectroscopic classification \citep{eisenstein06}. % \citep{2006ApJS..167...40E}.

\item[101644.47+161343.5 --]
 %WISE: ...  
 SDSS: blue-white; red-orange source at $\approx$2\farcs6;
 CROSSID: star at 2\farcs5 ($g=22.07$, $i=19.13$).
 %NOTES: ... 

\item[103047.25+443859.3 --]
 %WISE: ...  
 SDSS: blue-white.
 %CROSSID: ...  
 %NOTES: ... 

\item[104052.58+284856.7 --]
 WISE: background structure around target source but FWHM is consistent with other point sources; 
 SDSS: blue; %no additional sources within 7\farcs8  
 %CROSSID: ...  
 NOTES: WD1038+290 (LP316-487).

\item[104933.58+022451.7 --]
 %WISE: ...  
 SDSS: blue; barely resolved red source at $\approx$1\farcs3.
 %CROSSID: ...  
 %NOTES: ... 

\item[111021.03+304737.4 --]
 %WISE: ...  
 SDSS: blue;  
 %CROSSID: ...  
 NOTES: LP318-7, WD+M9 from \citet{rebassa10}.

\item[111424.65+334123.7 --]
 %WISE: ...  
 SDSS: blue-white; %no additional sources within 7\farcs8  
 %CROSSID: ...  
 NOTES: LP264-22.

\item[112010.94+320619.6 --]
 WISE: photometry possibly contaminated by faint blended source within 7\farcs8;
 SDSS: blue-white; very faint, possibly extended source at $\approx$2\farcs6.
 %CROSSID: ...  
 %NOTES: ... 

\item[112541.71+422334.7 --]
 %WISE: ...  
 SDSS: blue with red blended source at $\approx$3\arcsec;
 %CROSSID: ...  
 NOTES: GD308.

\item[113022.52+313933.4 --]
 WISE: photometry possibly contaminated by faint source within 7\farcs8;
 SDSS: blue;  
 CROSSID: two galaxies at 5\farcs6 and 7\arcsec ($g=23.81$, $22.60$ and $i=22.57$, $21.28$, respectively);
 NOTES: late type M star companion? Classified previously as WD+M \citep{heller09,rebassa10}.

\item[113039.09-004023.0 --]
 %WISE: ...  
 SDSS: blue; %no additional sources within 7\farcs8  
 %CROSSID: ...  
 NOTES: currently identified as a QSO in SIMBAD.

\item[114827.96+153356.9 --]
 %WISE: ...  
 SDSS: blue-white.  
 %CROSSID: ...  
 %NOTES: ... 

\item[115612.99+323302.5 --]
 %WISE: ...  
 SDSS: blue-white; very faint source at $\approx$4\farcs6.
 %CROSSID: ...  
 %NOTES: ... 

\item[115814.51+000458.7 --]
 %WISE: no indication of photometric contamination; target source FWHM consistent with other point sources  
 SDSS: yellow; %no additional sources within 7\farcs8  
 CROSSID: galaxy at 7\farcs1 ($g=26.27$, $i=24.36$); 
 NOTES: $T_{\rm eff}=4350$~K DC WD \citep{kilic09}.

\item[120144.90+505315.0 --]
 WISE: photometry likely contaminated by blended source within 7\farcs8;
 SDSS: blue; additional blended source is not obvious, possibly faint extended source (background galaxy?). %no other additional sources within 7\farcs8  
 %CROSSID: ...  
 %NOTES: ... 

\item[125847.31+233844.2 --]
 WISE: photometry possibly contaminated by faint blended source within 7\farcs8; 
 SDSS: blue-white, possibly slightly extended; very faint source at $\approx$6\farcs2; 
 CROSSID: galaxy at 6\farcs2 ($g=23.35$, $i=21.44$).
 %NOTES: ... 

\item[142559.72+365800.7 --]
 %WISE: ...  
 SDSS: blue; %no additional sources within 7\farcs8  
 %CROSSID: ...  
 NOTES: WD+M spectroscopic binary \citep{heller09}.

\item[142833.77+440346.1 --]
 %WISE: ...  
 SDSS: blue-white; %no additional sources within 7\farcs8  
 %CROSSID: ...  
 NOTES: WD1426+442 (G200-42), a DZ WD \citep{eisenstein06}.

\item[143144.83+375011.8 --]
 %WISE: ...  
 SDSS: white-blue; %no additional sources within 7\farcs8  
 %CROSSID: ...  
 NOTES: a DQ WD \citep{koester06b}.

\item[144307.83+340523.5 --]
 %WISE: ...  
 SDSS: blue; %no additional sources within 7\farcs8  
 %CROSSID: ...  
 NOTES: WD+M spectroscopic binary with H-alpha and possibly other Balmer emission \citep{heller09}.

\item[150152.59+443316.6 --]
 %WISE: ...  
 SDSS: red blended(?) source;
 %CROSSID: ...  
 NOTES: classified as DA+M/dM \citep{eisenstein06,silvestri06,heller09,rebassa10}.

\item[154221.86+553957.2 --]
 %WISE: ... 
 SDSS: blue/red blend;
 %CROSSID: ... 
 NOTES: WD 1541+558; known optically resolved \citep{heller09} DA+M binary \citep{eisenstein06}.  % \citep{2009A&A...496..191H} % \citep{2006ApJS..167...40E}

\item[154833.29+353733.0 --]
 %WISE: ...  
 SDSS: blue-white; barely resolved, yellow source at $\approx$1\farcs9; two additional faint sources between $\approx$3--7\farcs8;
 CROSSID: star at 4\farcs7 ($g=23.10$, $i=21.71$); galaxy at 7\farcs0 ($g=23.40$, $i=22.06$); galaxy at 8\farcs4 ($g=24.09$, $i=23.19$).
 %NOTES: ... 

\item[160153.23+273547.1 --]
 WISE: photometry likely contaminated by multiple nearby sources; no indication that deblending was used (nb=1, na=0); 
 SDSS: blue-white; multiple faint, red sources between $\approx$3--7\farcs8;
 CROSSID: star at 3\farcs1 ($g=25.06$, $i=21.91$); galaxy at 5\farcs2 ($g=23.07$, $i=21.47$); galaxy at 5\farcs5 ($g=23.31$, $i=21.23$); galaxy at 7\farcs1 ($g=24.54$, $i=22.11$); galaxy at 8\farcs0 ($g=23.12$, $i=19.90$).
 %NOTES: ... 

\item[164216.62+225627.8 --]
 %WISE: ... 
 SDSS: pale blue; 
 %CROSSID: ... 
 NOTES: DA spectroscopic classification \citep{eisenstein06}. % \citep{2006ApJS..167...40E}.

\item[165629.94+400330.2 --]
 %WISE: ...  
 SDSS: blue-white; %no additional sources within 7\farcs8  
 %CROSSID: ...  
 NOTES: WD1654+401.

\item[172633.51+530300.7 --]
 WISE: two sources at 15\arcsec; passive deblending used but no indication of photometric contamination (nb=3, na=0, target FWHM is consistent with other point sources); 
 SDSS: blue; barely resolved blue source at $\approx$2\arcsec; 
 CROSSID: star at 1\farcs7 ($g=19.59$, $i=18.98$);
 NOTES: currently identified as a QSO in SIMBAD.

\item[221652.14+005312.8 --]
 WISE: photometry possibly contaminated by faint blended source within 7\farcs8;
 SDSS: blue-white; %no additional sources within 7\farcs8  
 %CROSSID: ...  
 NOTES: a DC WD \citep{eisenstein06}.

\item[223401.66-010016.3 --]
 %WISE: ...  
 SDSS: target appears extended (component separation of $\approx$1\arcsec), both components white.
 %CROSSID: ...  
 %NOTES: ... 

\end{description}

\subsection{Indeterminate Excess Candidates}

\begin{description}

\item[000410.42-034008.5 --]
  WISE: photometry likely contaminated by blended source within 7\farcs8;
  SDSS: blue; faint, white source at $\approx$6\farcs0;
  CROSSID: galaxy at 6\farcs1 ($g=22.57$, $i=21.22$); 
  NOTES: LP644-30.
 
\item[000641.08+273716.6 --]
  %WISE: ...  
  SDSS: images not available; 
  CROSSID: galaxy at 1\farcs6 ($g=20.68$, $i=18.58$); star at 7\arcsec ($g=23.36$, $i=20.73$);
  NOTES: WISE photometry significantly brighter than best fitting disk model, suggests probable contamination from spatially coincident source(s). 
 
\item[001306.21+005506.3 --]
  WISE: target corresponds to faint source in crowded region with multiple close, bright sources; photometry likely contaminated; detection only in the i3o catalog;
  SDSS: blue; several very faint sources within 7\farcs8; brighter red star or galaxy and red galaxy at $\approx$11\farcs3 and $\approx$18\arcsec, respectively;
  CROSSID: galaxy at 7\farcs2 ($g=22.59$, $i=21.51$).
  %NOTES: ... 
 
\item[005438.84-095219.7 --]
  WISE: photometry likely contaminated by blended source within 7\farcs8;
  SDSS: blue; blended, red, possibly extended (galaxy?) source barely visible;
  CROSSID: galaxy at 7\farcs8 ($g=23.75$, $i=21.86$);
  NOTES: WD0052-101 (PHL3101, LP706-59).
 
\item[011616.94-094347.9 --]
  WISE: photometry possibly contaminated by faint blended source within 7\farcs8;
  SDSS: blue-white. % no additional sources within 7\farcs8  
  %CROSSID: ...  
  %NOTES: ... 
 
\item[020227.39+141124.5 --]
  WISE: W2 photometry possibly contaminated by nearby source that is not detected in W1;
  SDSS: blue; % no additional sources within 7\farcs8  
  %CROSSID: ...  
  NOTES: WD0159+139.
 
\item[025801.20-005400.0 --]
  %WISE: ...  
  SDSS: blue; % no additional sources within 7\farcs8  
  %CROSSID: ...  
  NOTES: WD0255-010.
 
\item[073018.35+411320.4 --]
  WISE: photometry likely contaminated by blended source within 7\farcs8; possible contamination from nearby diffraction spike (ccflag = dd00); 
  SDSS: blue; faint sources at $\approx$2\farcs2 (white), $\approx$5\farcs8 (red), and $\approx$6\farcs4 (very faint, yellow, possibly extended);
  CROSSID: star at 1\farcs9 ($g=20.27$, $i=19.29$); galaxy at 5\farcs5 ($g=24.19$, $i=20.75$); galaxy at 6\farcs2 ($g=23.39$, $i=21.29$).  
  %NOTES: ... 
 
\item[073707.99+411227.4 --]
  WISE: photometry likely contaminated by blended source within 7\farcs8;
  SDSS: blue; sources at $\approx$5\farcs3 (white) and $\approx$7\farcs6 (faint, red);
  CROSSID: star at 5\farcs2 ($g=18.48$, $i=17.96$); star at 7\farcs7 ($g=23.07$, $i=20.39$).
  %NOTES: ... 
 
\item[074631.42+173448.1 --]
  WISE: structured background, but target source FWHM is consistent with other point sources;
  SDSS: blue-white; % no additional sources within 7\farcs8  
  CROSSID: star at 5\farcs7 ($g=23.78$, $i=21.78$).
  %NOTES: ... 
 
\item[075144.05+223004.8 --]
  WISE: photometry possibly contaminated by faint blended source within 7\farcs8;
  SDSS: blue; faint, red source at $\approx$3\farcs3;
  CROSSID: galaxy at 3\farcs3 ($g=22.36$, $i=20.55$).
  %NOTES: ... 
 
\item[083632.99+374259.3 --]
  %WISE: ...  
  SDSS: blue-white. % no additional sources within 7\farcs8  
  %CROSSID: ...  
  %NOTES: ... 
 
\item[085650.57+275118.0 --]
  WISE: passive deblending used (nb=3, na=0); target source FWHM consistent with other point sources;
  SDSS: blue; faint, extended, white source at $\approx$2\farcs5.
  %CROSSID: ...  
  %NOTES: ... 
 
\item[090911.36+501559.4 --]
  %WISE: ...  
  SDSS: blue; faint, yellow source at $\approx$3\farcs4;
  CROSSID: star at 3\farcs2 ($g=22.94$, $i=21.12$); galaxy at 7\farcs5 ($g=23.67$, $i=22.79$).
  %NOTES: ... 
 
\item[091312.73+403628.8 --]
  WISE: photometry possibly contaminated by bright source at $\approx$8\arcsec;
  SDSS: blue; additional source at $\approx$8\arcsec; barely resolved white source at $\approx$2\farcs1;
  CROSSID: galaxy at 8\farcs6 ($g=22.35$, $i=22.30$); galaxy at 8\farcs8 ($g=24.61$, $i=21.94$);  
  NOTES: WDJ0913+4036, a ZZ Ceti WD with no evidence for planetary companions via pulsation timing \citep{mullally08}.
 
\item[091356.83+404734.6 --]
  %WISE: ...  
  SDSS: blue-white; resolved, yellow source within $\approx$7\farcs8;
  CROSSID: star at 6\farcs1 ($g=21.31$, $i=19.98$);
  NOTES: WD0910+410 (LP210-58). 
 
\item[094422.33+552756.2 --]
  %WISE: ...  
  SDSS: blue-white;  
  %CROSSID: ...  
  NOTES: Poor fit to SED model, probable photometric contamination.
 
\item[101007.88+615515.7 --]
  WISE: photometry possibly contaminated by faint blended source within 7\farcs8;
  SDSS: blue; blended, red source at $\approx$2\farcs3.
  %CROSSID: ...  
  %NOTES: ... 
 
\item[101951.55+290100.6 --]
  WISE: photometry likely contaminated by blended source;
  SDSS: blue-white.  
  %CROSSID: ...  
  %NOTES: ... 
 
\item[102100.91+564644.7 --]
  WISE: photometry possibly contaminated by blended source within 7\farcs8;
  SDSS: blue; barely resolved white source at $\approx$1\farcs7.
  %CROSSID: ...  
  %NOTES: ... 
 
\item[102915.97+300251.5 --]
  WISE: mosaic images not available;
  SDSS: blue-white; faint, red sources at $\approx$8\farcs4 and $\approx$9\farcs5;
  CROSSID: galaxy at 8\farcs4 ($g=22.69$, $i=21.71$).
  %NOTES: ... 
 
\item[103112.73+444729.9 --]
  WISE: photometry likely contaminated by faint blended source within 7\farcs8;
  SDSS: blue-white; very faint, possibly extended source at $\approx$3\farcs6.
  %CROSSID: ...  
  %NOTES: ... 
 
\item[104659.78+374556.7 --]
  WISE: photometry possibly contaminated by faint blended source within 7\farcs8;
  SDSS: blue. % no additional sources within 7\farcs8  
  %CROSSID: ...  
  %NOTES: ... 
 
\item[105824.34+512738.7 --]
  %WISE: ...  
  SDSS: blue. % no additional sources within 7\farcs8  
  %CROSSID: ...  
  %NOTES: ... 
 
\item[105827.97+293223.0 --]
  WISE: photometry likely contaminated by blended source within 7\farcs8;
  SDSS: blue; possible very faint source at $\approx$5\farcs6;
  CROSSID: galaxy at 5\farcs5 ($g=23.66$, $i=21.89$).
  %NOTES: ... 
 
\item[110745.39+651722.1 --]
  %WISE: ...  
  SDSS: white-blue;  
  %CROSSID: ...  
  NOTES: DA spectroscopic classification \citep{eisenstein06}.
 
\item[111603.77+494343.8 --]
  WISE: photometry possibly contaminated by faint source(s) within 7\farcs8;
  SDSS: blue; %no neighbors within 7\farcs8 
  %CROSSID: ...
  NOTES: DA spectroscopic classification (2006ApJS..167...40E).

\item[111609.81+284308.4 --]
  WISE: FWHM slightly extended relative to other point sources; possible contamination from diffraction spike of nearby bright star (i3o catalog detection only, so no cc\_flag information);
  SDSS: blue-white. % no additional sources within 7\farcs8  
  %CROSSID: ...  
  %NOTES: ... 
 
\item[111706.70+184312.4 --]
  %WISE: ...  
  SDSS: blue; very faint sources at $\approx$3\arcsec (possible) and $\approx$7\farcs8;
  CROSSID: star at 2\farcs9 ($g=25.10$, $i=23.21$); galaxy at 5\farcs1 ($g=24.69$, $i=23.39$); galaxy at 7\farcs5 ($g=23.29$, $i=21.95$).
  %NOTES: ... 
 
\item[111753.51+263856.2 --]
  %WISE: ...  
  SDSS: extended galaxy partially blended with target;
  CROSSID: galaxy at 2\farcs7 ($g=20.35$, $i=19.64$);
  NOTES: probable contamination from galaxy in SED.
 
\item[112105.79+375615.2 --]
  WISE: nearby faint source at $\approx$9\farcs5 but no apparent contamination (nb=1, na=0, target source FWHM consistent with other point sources);
  SDSS: blue-white. % no additional sources within 7\farcs8  
  %CROSSID: ...  
  %NOTES: ... 
 
\item[112310.05+584407.2 --]
  WISE: photometry likely contaminated by blended source within 7\farcs8;
  SDSS: blue-white; two faint, resolved sources within 7\farcs8;
  CROSSID: star at 4\farcs4 ($g=24.72$, $i=21.92$); galaxy at 4\farcs5 ($g=22.23$, $i=21.19$).
  %NOTES: ... 
 
\item[113630.78+315447.9 --]
  %WISE: ...  
  SDSS: blue. % no additional sources within 7\farcs8  
  %CROSSID: ...  
  %NOTES: ... 
 
\item[113728.31+204109.4 --]
  WISE: photometry likely contaminated by blended source within 7\farcs8;
  SDSS: blue-white; faint, white source at $\approx$5\farcs8;
  CROSSID: star at 5\farcs8 ($g=21.80$, $i=21.27$).
  %NOTES: ... 
 
\item[114701.01+574114.7 --]
  %WISE: ...  
  SDSS: blue-white;  
  %CROSSID: ...  
  NOTES: DA spectroscopic classification \citep{eisenstein06}.
 
\item[115745.89+063148.2 --]
  %WISE: ...  
  SDSS: blue. % no additional sources within 7\farcs8  
  %CROSSID: ...  
  %NOTES: ... 
 
\item[120504.19+160746.8 --]
  WISE: photometry likely contaminated by blended source within 7\farcs8;
  SDSS: blue. % no additional sources within 7\farcs8  
  %CROSSID: ...  
  %NOTES: ... 
 
\item[124256.48+431311.1 --]
  WISE: photometry likely contaminated by blended source within 7\farcs8;
  SDSS: blue. % no additional sources within 7\farcs8  
  %CROSSID: ...  
  %NOTES: ... 
 
\item[124359.69+161203.5 --]
  WISE: photometry possibly contaminated by blended source within 7\farcs8;
  SDSS: blue. % no additional sources within 7\farcs8  
  %CROSSID: ...  
  %NOTES: ... 
 
\item[125037.75+205334.0 --]
  %WISE: ...  
  SDSS: blue-white. % no additional sources within 7\farcs8  
  %CROSSID: ...  
  %NOTES: ... 
 
\item[125733.64+542850.5 --]
  WISE: photometry likely contaminated by blended source within 7\farcs8;
  SDSS: blue-white; faint, extended source (galaxy?) at $\approx$5\farcs5;
  CROSSID: galaxy at 6\farcs1 ($g=24.19$, $i=21.03$);
  NOTES: unresolved double WD binary, with one or both components possibly magnetic at a level of $\sim$0.1--1 MG \citep{kulkarni10}.
 
\item[130957.59+350947.2 --]
  WISE: target source FWHM slightly extended compared to other point sources; photometry possibly contaminated by blended sources within 7\farcs8;
  SDSS: blue; two very faint sources between $\approx$3--7\farcs8;
  %CROSSID: ...  
  NOTES: WD1307+354 (BG CVn, GD154), a ZZ Ceti WD; no low luminosity companion found by \citet{farihi05}; Spitzer-IRAC observations used to rule out unresolved companions with mass $>$10 M$_{\rm J}$ \citep{kilic09}.
 
\item[131951.00+643309.1 --]
  %WISE: ...  
  SDSS: blue; % no additional sources within 7\farcs8  
  %CROSSID: ...  
  NOTES: WD1318+648.
 
\item[133100.61+004033.5 --]
  %WISE: ...  
  SDSS: blue; % no additional sources within 7\farcs8  
  CROSSID: star at 8\farcs5 ($g=24.60$, $i=22.60$);
  NOTES: WD1328+009, indeterminate excess detected by \citet{girven11}
 
\item[134333.64+231403.3 --]
  %WISE: ...  
  SDSS: blue with possible green(!?) blend/extension; faint source at $\approx$7\farcs0;
  CROSSID: galaxy at 6\farcs8 ($g=22.45$, $i=21.18$).
  %NOTES: ... 
 
\item[140644.77+530353.1 --]
  %WISE: ...  
  SDSS: blue;  
  %CROSSID: ...  
  NOTES: $z$-band excess in SED is suggestive of faint companion rather than disk; DA spectroscopic classification \citep{eisenstein06}.
 
\item[140723.04+203918.5 --]
  WISE: photometry possibly contaminated by fainter blended source;
  SDSS: blue-white; two faint, white, possibly extended sources within 7\farcs8;
  CROSSID: star at 2\farcs6 ($g=22.82$, $i=21.84$);
  NOTES: classified as DQ in Preliminary DR7 WD Catalogue.
 
\item[140945.23+421600.6 --]
  WISE: target source FWHM slightly extended compared to other point sources; possible contamination from blended source within 7\farcs8;
  SDSS: blue-white; faint red source at $\approx$5\farcs5;
  CROSSID: galaxy at 5\farcs6 ($g=21.92$, $i=19.52$);
  NOTES: WD1407+425, DAZ WD observed with Spitzer-IRAC, showing no ``reliable'' IR disk emission \citep{farihi08}; no low luminosity companion found by \citep{farihi05}.
 
\item[141017.32+463450.1 --]
  %WISE: ...  
  SDSS: blue-white. % no additional sources within 7\farcs8  
  %CROSSID: ...  
  %NOTES: ... 
 
\item[141448.24+021257.7 --]
  WISE: photometry likely contaminated by multiple (3?) blended sources within 7\farcs8;
  SDSS: blue; two resolved, faint sources within 7\farcs8;
  CROSSID: galaxy at 5\farcs9 ($g=24.04$, $i=21.53$); galaxy at 7\farcs4 ($g=25.68$, $i=21.84$);
  NOTES: WD1412+024. Listed as candidate debris disk in \citet{steele11}, listed as candidate excess by \citet{girven11}.
 
\item[141632.82+111003.9 --]
  WISE: passive deblending of multiple sources (nb=3, na=0), one of which is very bright; however, FWHM of target is consistent with point source;
  SDSS: blue-white; % no additional sources within 7\farcs8  
  CROSSID: galaxy at 5\farcs8 ($g=23.26$, $i=20.95$); star at 8\farcs7 ($g=24.87$, $i=22.46$).
  %NOTES: ... 
 
\item[142539.74+010926.8 --]
  WISE: photometry possibly contaminated by faint blended source; % i3o detection only; 
  SDSS: blue-white; very faint, white source at $\approx$5\farcs5;  
  CROSSID: galaxy at 5\farcs6 ($g=26.02$, $i=21.40$);  
  NOTES: WD1423+013; DA spectroscopic classification \citep{eisenstein06}.
 
\item[143406.75+150817.8 --]
  WISE: photometry possibly contaminated by blended source within 7\farcs8;
  SDSS: blue; faint, white source at $\approx$3\farcs4;
  %CROSSID: ...  
  NOTES: WD1431+153; {\em not} identified as a DA+dM binary in \citet{koester09}.
 
\item[144754.40+420004.9 --]
  %WISE: ...  
  SDSS: blue-white; sources at $\approx$3\arcsec (faint, red) and $\approx$2\farcs4 (faint, possibly extended, red -- possible extent of this source is larger than the WD; possibly a planetary nebula?);
  CROSSID: star at 3\farcs5 ($g=24.45$, $i=21.52$).
  %NOTES: ... 
 
\item[144847.79+145645.7 --]
  %WISE: ...  
  SDSS: blue-white, with possible faint red blend; possible very faint, extended source at $\approx$3\farcs6 (extending to target);
  CROSSID: galaxy at 8\farcs8 ($g=23.00$, $i=21.74$).
  %NOTES: ... 
 
\item[154224.94+044959.7 --]
  %WISE: ...  
  SDSS: blue-white;  
  %CROSSID: ...  
  NOTES: WISE photometry significantly brighter than best fitting SED models, indicating probable contamination. 
 
\item[154729.96+065909.5 --]
  %WISE: ...  
  SDSS: blue; % no additional sources within 7\farcs8  
  CROSSID: galaxy at 7\farcs1 ($g=22.90$, $i=21.51$).
  %NOTES: ... 
 
\item[155811.46+312706.4 --]
  WISE: photometry likely contaminated by faint blended source within 7\farcs8;
  SDSS: blue. % no additional sources within 7\farcs8  
  %CROSSID: ...  
  %NOTES: ... 
 
\item[160241.44+332301.4 --]
  WISE: likely blended source within 7\farcs8;
  SDSS: blue-white; possible very faint, possibly extended source at $\approx$3\arcsec.
  %CROSSID: ...  
  %NOTES: ... 
 
\item[160401.49+463249.5 --]
  WISE: target is located in a large image artifact similar to a diffraction spike; only listed in the i3o catalog, so no cc\_flag information; this target most likely has contaminated photometry.
  SDSS: blue. % no additional sources within 7\farcs8  
  %CROSSID: ...  
  %NOTES: ... 
 
\item[160715.80+134312.3 --]
  WISE: photometry likely contaminated by blended source within 7\farcs8;
  SDSS: blue-white; faint sources at $\approx$3\arcsec and $\approx$7\farcs4;
  CROSSID: galaxy at 7\farcs3 ($g=22.24$, $i=20.74$).
  %NOTES: ... 
 
\item[160839.52+172336.9 --]
  %WISE: ...  
  SDSS: blue-white. % no additional sources within 7\farcs8  
  %CROSSID: ...  
  %NOTES: ... 
 
\item[162139.79+481241.6 --]
  WISE: photometry likely contaminated by blended source within 7\farcs8;
  SDSS: blue-white; faint, possibly extended, source at $\approx$5\farcs4;
  CROSSID: galaxy at 4\farcs8 ($g=21.79$, $i=20.04$).
  %NOTES: ... 
 
\item[162555.28+375920.6 --]
  %WISE: ...  
  SDSS: blue-white; barely resolved, blue-white source at $\approx$1\farcs8;
  CROSSID: star at 1\farcs7 ($g=18.33$, $i=18.26$).
  %NOTES: ... 
 
\item[170144.73+624304.4 --]
  WISE: fainter source at $\approx$8\farcs5 and target source FWHM is extended relative to other point sources, no deblending performed (nb=1, na=0); photometry is likely affected;
  SDSS: blue; source at $\approx$8\farcs5; % no additional sources within 7\farcs8  
  CROSSID: galaxy at 8\farcs2 ($g=20.95$, $i=20.05$);
  NOTES: WD1701+627.
 
\item[173434.54+333521.3 --]
  %WISE: ...  
  SDSS: blue;  
  CROSSID: two stars at 4\farcs.1 ($g=25.18$, $25.69$ and $i=24.71$, $24.68$, respectively); galaxy at 8\farcs8 ($g=22.38$, $i=21.20$).
  %NOTES: ... 
 
\item[192433.15+373416.9 --]
  WISE: possible contamination from nearby diffraction spike (cc\_flag=dd00);
  SDSS: images not available;
  CROSSID: stars at 2\farcs6 and 6\farcs1 ($g=22.61$, $24.02$ and $i=18.67$, $20.74$, respectively);
  NOTES: possible resolved DA+M, in Kepler FOV. % \citep{?????}.
 
\item[192542.00+631741.6 --]
  %WISE: ...  
  SDSS: images not available;
  CROSSID: star at 4\farcs9 ($g=20.42$, $i=19.73$).  
  %NOTES: ... 
 
\item[231725.28-084032.9 --]
  %WISE: ...  
  SDSS: blue; barely resolved red source at $\approx$2\farcs3.
  %CROSSID: ...  
  %NOTES: ... 
 
\end{description}

%%% FIGURES

\clearpage

\clearpage

\clearpage

\clearpage

\clearpage

\clearpage

\clearpage

\clearpage

\clearpage

\clearpage

\clearpage

\clearpage

\clearpage

\clearpage

\clearpage

\clearpage

\clearpage

\clearpage

\clearpage

\clearpage

\clearpage

\clearpage

SD

\clearpage

%%% TABLES
%\include{table_phot1}
\begin{deluxetable}{rcccccccccccc}
\tabletypesize{\scriptsize}
\rotate
\tablecolumns{13}
\tablewidth{0pc}
\tablecaption{SDSS Photometry and System Parameters of DR7 WDs Detected by WISE (excerpt) \label{tab:phot1}}
\tablehead{
\multicolumn{3}{l}{WIRED:} & \multicolumn{5}{l}{SDSS:} & & & & & \\ 
\colhead{\#} & \colhead{Name} & \colhead{Flag} & \colhead{$u$} & \colhead{$g$} & \colhead{$r$} & \colhead{$i$} & \colhead{$z$} & \colhead{Dist$_{\rm phot}$} & \colhead{T$_{\rm eff}$} & \colhead{Age}   & \colhead{$\log{g}$} & \colhead{Mass} \\
 & & & \colhead{(mag)} & \colhead{(mag)} & \colhead{(mag)} & \colhead{(mag)} & \colhead{(mag)} &  \colhead{(pc)} & \colhead{(K)} & \colhead{(Gyr)} & & \colhead{(M$_{\odot}$)} 
}
\startdata
       1 & 000356.93-050332.7 & 1 &  18.522(23) &  18.203(28) &  18.152(13) &  17.503(14) &  16.884(18) &     509 &   17678 & 0.03 &    7.23 & 0.32 \\
       2 & 000410.42-034008.5 & 3 &  17.458(40) &  16.928(15) &  16.759(13) &  16.719(15) &  16.736(17) &      51 &    6887 & 1.19 &    7.71 & 0.45 \\
       3 & 000504.91+243409.6 & 1 &  19.513(35) &  18.895(15) &  18.486(14) &  17.503(14) &  16.766(18) &     264 &    7597 & 0.41 &    5.00 & 0.20 \\
       4 & 000531.09-054343.2 & 1 &  17.283(16) &  16.728(14) &  16.600(14) &  15.807(14) &  15.099(12) &     223 &   16417 & 0.05 &    7.33 & 0.34 \\
       5 & 000641.08+273716.6 & 3 &  19.794(42) &  19.847(33) &  20.672(205) &  20.437(255) &  21.039(471) &    1081 &   22857 & 0.03 &    7.58 & 0.44 \\
       6 & 000651.91+284647.1 & 1 &  19.272(34) &  18.665(20) &  18.247(15) &  17.147(17) &  16.470(15) &     229 &    7533 & 0.42 &    5.00 & 0.20 \\
       7 & 001247.18+001048.7 & 1 &  20.694(87) &  20.191(25) &  19.637(22) &  18.613(25) &  17.946(33) &     483 &    6300 & 0.69 &   10.00 & 0.19 \\
       8 & 001306.21+005506.3 & 3 &  19.849(49) &  19.385(24) &  19.409(18) &  19.580(32) &  19.624(94) &     318 &   11036 & 0.44 &    7.94 & 0.57 \\
       9 & 001324.33-085021.4 & 1 &  19.726(45) &  19.750(25) &  19.676(25) &  19.089(28) &  18.562(41) &     227 &    6318 & 0.88 &    9.99 & 0.17 \\
      10 & 001339.20+001924.3 & 4 &  15.769(21) &  15.373(25) &  15.425(29) &  15.492(17) &  15.669(18) &      35 &    9419 & 1.06 &    8.17 & 0.71 \\
\enddata
\tablecomments{The full table is available online. Flags denote target classifications as follows: 0 = WD+BD, 1 = WD+M star, 2 = WD+dust disk, 3 = indeterminate IR excess, 4 = naked WD.  Photometry is preliminary DR7 psfMag values.}
\tablenotetext{a}{WD listed as IR excess in \citet{steele11}}
\tablenotetext{b}{WD listed as IR excess in \citet{girven11}}
\end{deluxetable}

\begin{deluxetable}{rccccccccccc}
\tabletypesize{\scriptsize}
\rotate
\tablecolumns{12}
\tablewidth{0pc}
\tablecaption{Near-Infrared Photometry of DR7 WDs Detected by WISE (excerpt) \label{tab:phot2}}
\tablehead{
\multicolumn{3}{l}{WIRED:} & \multicolumn{6}{l}{UKIDSS:} & \multicolumn{3}{l}{2MASS:} \\ 
\colhead{\#} & \colhead{Name} & \colhead{Flag} & \colhead{catalog} & \colhead{$Z$} & \colhead{$Y$} & \colhead{$J$} & \colhead{$H$} & \colhead{$K_{\rm s}$} & \colhead{$J$} & \colhead{$H$} & \colhead{$K_{\rm s}$} \\
 & & & & \colhead{(mag)} & \colhead{(mag)} & \colhead{(mag)} & \colhead{(mag)} & \colhead{(mag)} & \colhead{(mag)} & \colhead{(mag)} & \colhead{(mag)} 
}
\startdata
       1 & 000356.93-050332.7 & 1 & \nodata & \nodata & \nodata & \nodata & \nodata & \nodata &  15.582(53) &  14.984(82) &  14.509(86) \\
       2 & 000410.42-034008.5 & 3 & \nodata & \nodata & \nodata & \nodata & \nodata & \nodata & \nodata & \nodata & \nodata \\
       3 & 000504.91+243409.6 & 1 & \nodata & \nodata & \nodata & \nodata & \nodata & \nodata &  15.366(49) &  14.726(75) &  14.408(76) \\
       4 & 000531.09-054343.2 & 1 & \nodata & \nodata & \nodata & \nodata & \nodata & \nodata &  13.768(21) &  13.112(30) &  12.890(33) \\
       5 & 000641.08+273716.6 & 3 & \nodata & \nodata & \nodata & \nodata & \nodata & \nodata & \nodata & \nodata & \nodata \\
       6 & 000651.91+284647.1 & 1 & \nodata & \nodata & \nodata & \nodata & \nodata & \nodata &  15.098(44) &  14.491(59) &  14.237(65) \\
       7 & 001247.18+001048.7 & 1 & LAS     & \nodata &  17.093(14) &  16.601(16) &  16.133(22) & \nodata &  16.432(105) &  15.876(166) &  15.425(243) \\
       8 & 001306.21+005506.3 & 3 & LAS     & \nodata &  19.514(95) &  19.571(135) & \nodata & \nodata & \nodata & \nodata & \nodata \\
       9 & 001324.33-085021.4 & 1 & \nodata & \nodata & \nodata & \nodata & \nodata & \nodata & \nodata & \nodata & \nodata \\
      10 & 001339.20+001924.3 & 4 & LAS     & \nodata &  15.206(4) &  15.148(6) & \nodata &  15.163(17) & \nodata & \nodata & \nodata \\
\enddata
\tablecomments{The full table is available online. Flags denote target classifications as follows: 0 = WD+BD, 1 = WD+M star, 2 = WD+dust disk, 3 = indeterminate IR excess, 4 = naked WD.  Photometry values without uncertainties are upper limits.  The UKIDSS catalogs are the Large Area Survey (LAS) and the Galactic Clusters Survey (GCS) from Data Release 5plus; there are no detected targets in the UKIDSS Galactic Plane Survey. All 2MASS photometry is from the All Sky Point Source Catalog.}
\end{deluxetable}

\begin{deluxetable}{rccccccc}
\tabletypesize{\scriptsize}
\rotate
\tablecolumns{8}
\tablewidth{0pc}
\tablecaption{WISE Infrared Photometry of DR7 WDs (excerpt) \label{tab:phot3}}
\tablehead{
\multicolumn{3}{l}{WIRED:} & \multicolumn{5}{l}{WISE:} \\ 
\colhead{\#} & \colhead{Name} & \colhead{Flag} & \colhead{catalog} & \colhead{$W1$} & \colhead{$W2$} & \colhead{$W3$} & \colhead{$W4$} \\
 & & & & \colhead{(mag)} & \colhead{(mag)} & \colhead{(mag)} & \colhead{(mag)} 
}
\startdata
       1 & 000356.93-050332.7 & 1 & i3o     &  14.494(34) &  14.386(63) &  12.617 &   9.171 \\
       2 & 000410.42-034008.5 & 3 & i3o     &  15.363(52) &  15.384(158) &  11.952 &   8.866 \\
       3 & 000504.91+243409.6 & 1 & i3o     &  14.236(31) &  14.201(60) &  12.826 &   9.261 \\
       4 & 000531.09-054343.2 & 1 & i3o     &  12.747(28) &  12.470(27) &  11.909 &   8.991 \\
       5 & 000641.08+273716.6 & 3 & i3o     &  15.539(54) &  15.235(111) &  12.472 &   9.516 \\
       6 & 000651.91+284647.1 & 1 & i3o     &  14.055(31) &  13.788(44) &  12.645 &   9.443 \\
       7 & 001247.18+001048.7 & 1 & i3o     &  15.737(60) &  15.477(149) &  12.679 &   9.492 \\
       8 & 001306.21+005506.3 & 3 & i3o     &  17.166(205) &  15.858 &  12.565 &   8.745 \\
       9 & 001324.33-085021.4 & 1 & i3o     &  16.042(86) &  15.556(177) &  12.550 &   9.140 \\
      10 & 001339.20+001924.3 & 4 & i3o     &  15.194(46) &  15.169(115) &  12.513 &   9.385 \\
\enddata
\tablecomments{The full table is available online. Flags denote target classifications as follows: 0 = WD+BD, 1 = WD+M star, 2 = WD+dust disk, 3 = indeterminate IR excess, 4 = naked WD.  Photometry values without uncertainties are upper limits.  See text for a description of the WISE catalogs.}
\end{deluxetable}

\begin{deluxetable}{lrrrrrrl}
\tabletypesize{\footnotesize}
\rotate
\tablecolumns{8}
\tablewidth{0pc}
\tablecaption{Targets-of-Interest with {\em Spitzer} IRAC Photometry \label{tab:spitzcomp}}
\tablehead{
\colhead{Name} & \colhead{$W1$} & \colhead{IRAC-1} & \colhead{Diff} & \colhead{$W2$} & \colhead{IRAC-2} & \colhead{Diff} & \colhead{Notes} \\
 & \colhead{($\mu$Jy)} & \colhead{($\mu$Jy)} & \colhead{\%(W-S)} & \colhead{($\mu$Jy)} & \colhead{($\mu$Jy)} & \colhead{\%(W-S)} & 
}
\startdata
  030253.09$-$010833.7 & 267.1(10.8) & 231(12)                  &    14.5 & 202.4(15.7) & 199(10)                  &    1.7  & GD 40 (1) \\
  084539.17+225728.0   & 267.1(10.6) & \nodata                  & \nodata & 237.3(19.5) & 248(20)                  & $-$4.4  & WD 0842+231 (2) \\
                       &             & 271(23)\tablenotemark{a} &  $-$1.4 &             & 218(18)\tablenotemark{a} &    8.5  & (3) \\
  104341.53+085558.2   & 54.3(8.8)   & 34(6.5)                  &    46.0 & \nodata     &  24(6)                   & \nodata & WD 1041+091 (2) \\
% *104635.22+594628.7  & 80.7(8.3)   &  85.81(1.26)             &  $-$6.1 & \nodata     &  57.45(1.18)             & \nodata & SWIRE3\_J104635.30+594627.9 \\
  122859.93+104032.9   & 181.9(10.9) & 235(10.6)                & $-$25.5 & 217.0(20.2) & 235(9.6)                 & $-$8.0  & (4) \\
  124359.69+161203.5   &  61.9(7)    & \nodata                  & \nodata & $<67.4$     &  54.2(3.3)               & \nodata & LBQS 1241+1628 (5), this work \\
  130957.59+350947.2   & 266.6(10.6) & 191.8(6.3)               &    32.6 & 147.4(13.7) & 118.4(6.7)               &   21.8  & WD 1307+354 (6) \\
  140945.23+421600.6   & 386.1(12.1) & 292(15)                  &    27.7 & 214.6(12.7) & 159(8)                   &   29.8  & WD 1407+425 (7) \\
  145806.53+293727.0   & 516.0(16.2) & 357(18)                  &    36.4 & 353.9(16.9) & 222(11)                  &   45.8  & WD 1455+298 (7) \\
  161717.04+162022.3   & 104.2(7.7)  & 108(5.8)                 &  $-$3.6 &  96.9(13.9) &  95(6.5)                 &    2.0  & (2) \\
% *225123.66+293945.4  & 873.1(23.3) & 777.3(24.2)              &    11.6 & 525.9(19.9) & 509.0(16.5)              &    3.3  & WD 2238+293 (8 - Kilic et al. (2009b) ) \\ 
\enddata
\tablenotetext{a}{Data from Akari, see (3).}
\tablecomments{(1) \citet{jura07}; (2) \citet{brinkworth11}; (3) \citet{farihi10}; (4) \citet{brinkworth09}; (5) \citet{berg92}; (6) \citet{kilic09}; (7) \citet{farihi08b}.}
\end{deluxetable}

\begin{deluxetable}{rcccccccc}	
\tabletypesize{\scriptsize}	
\tablecolumns{9}	
\tablewidth{0pc}	
\tablecaption{WD+M Star Binary Candidates \label{tab:m}}	
\tablehead{	
\colhead{\#} & \colhead{WIRED Name} & \colhead{Type} & \colhead{Mass}        & \colhead{Age}   & \colhead{D$_{\rm phot,WD}$} & \colhead{D$_{\rm phot,M}$} & \colhead{Spectral Type} & \colhead{$\chi^2$} \\	
             &                      &                & \colhead{(M$_\odot$)} & \colhead{(Gyr)} & \colhead{(pc)}              & \colhead{(pc)}             &                         & 	
}	
\startdata	
       1 & 000356.93-050332.7                  &   DAM &   0.32 &   0.03 &   509 &   509 &   M3$^{+  1.0}_{-  1.0}$ &   19.8\\	%  &   18.8\\
       3 & 000504.91+243409.6                  &   DAM &   0.20 &   0.41 &   239 &   264 &   M4$^{+  1.0}_{-  1.0}$ &   32.2\\	%  &   32.2\\
       4 & 000531.09-054343.2                  &   DAM &   0.34 &   0.05 &   223 &   223 &   M3$^{+  1.0}_{-  1.0}$ &   65.7\\	%  &   65.7\\
       6 & 000651.91+284647.1                  &   DAM &   0.20 &   0.42 &   211 &   229 &   M4$^{+  1.0}_{-  1.0}$ &   32.0\\	%  &   32.0\\
       7 & 001247.18+001048.7                  &   DAM &   0.19 &   0.69 &   262 &   483 &   M4$^{+  1.0}_{-  1.0}$ &   23.0\\	%  &   23.0\\
       9 & 001324.33-085021.4                  &  DC:M &   0.17 &   0.88 &   227 &   227 &   M7$^{+  4.0}_{-  1.0}$ &   35.9\\	%  &   35.9\\
      11 & 001359.39-110838.6                  &  DA:M &   0.41 &   0.00 &  2255 &  2255 &   M0$^{+ 26.0}_{-  0.0}$ & 1081\\	%  & 1081.5\\
      12 & 001736.90+145101.9\tablenotemark{a} &   DCM &   0.17 &   0.89 &   111 &   111 &   M8$^{+  2.0}_{-  1.0}$ &  426\\	%  &  426.5\\
      15 & 002157.90-110331.6                  &   DAM &   0.19 &   0.67 &   136 &   136 &   M5$^{+  1.0}_{-  1.0}$ &   30.8\\	%  &   30.8\\
      18 & 002620.41+144409.5\tablenotemark{a} &   DAM &   0.42 &   0.57 &   109 &   196 &   M4$^{+  1.0}_{-  1.0}$ &   83.3\\	%  &   83.3\\
\enddata	
\tablenotetext{a}{WD has a predicted $W1$ photospheric flux density $>50$ $\mu$Jy and is part of the flux limited sample.}	
\end{deluxetable}

\begin{deluxetable}{rcccccccc}
\tabletypesize{\scriptsize}
\tablecolumns{9}
\tablewidth{0pc}
\tablecaption{WD+BD Binary Candidates \label{tab:bd}}
\tablehead{
\colhead{\#} & \colhead{WIRED Name} & \colhead{Type} & \colhead{Mass}        & \colhead{Age}   & \colhead{D$_{\rm phot,WD}$} & \colhead{D$_{\rm phot,BD}$} & \colhead{Spectral Type} & \colhead{$\chi^2$} \\
             &                      &                & \colhead{(M$_\odot$)} & \colhead{(Gyr)} & \colhead{(pc)}              & \colhead{(pc)}              &                         & 
}
\startdata
      53 & 012532.02+135403.6                  &  DCM: &  0.17 &  0.76 &   281 &   281 &   L5$^{+ 11.0}_{-  3.0}$ &   7.0\\
      58 & 013532.97+144555.9\tablenotemark{a} &    DA &  0.39 &  0.78 &    73 &   115 &   L5$^{+  3.0}_{-  3.0}$ &  16.1\\
      59 & 013553.72+132209.2                  &  DAZ: &  0.20 &  0.47 &   290 &   290 &   L0$^{+ 16.0}_{-  3.0}$ &  26.3\\
     157 & 033444.86-011253.8                  &    DC &  1.12 &  1.54 &   202 &   202 &   L4$^{+  1.0}_{-  1.0}$ &   1.9\\
     187 & 064607.86+280510.1                  &   DAH &  1.22 &  0.41 &   192 &   192 &   L2$^{+  6.0}_{-  2.0}$ &   6.8\\
     269 & 081113.73+144150.6                  &   DAM &  0.46 &  1.00 &   149 &   149 &   L4$^{+  1.0}_{-  1.0}$ &   1.1\\
     305 & 082412.27+175155.8                  &   DAH &  0.86 &  3.00 &   191 &   191 &   L4$^{+  1.0}_{-  4.0}$ &  10.3\\
     319 & 083038.79+470247.0\tablenotemark{a} &   DAM &  0.19 &  0.69 &    89 &   123 &   L4$^{+  4.0}_{-  5.0}$ &  13.1\\
     326 & 083254.38+313904.2\tablenotemark{a} &    DA &  0.65 &  2.28 &    61 &   219 &   L0$^{+ 11.0}_{-  5.0}$ &   2.8\\
     392 & 085930.41+103241.1                  &  DAM: &  1.14 &  0.41 &   306 &   306 &   L1$^{+  4.0}_{-  2.0}$ &   7.2\\
     451 & 092233.13+050640.0                  & D(AH) &  0.22 &  0.11 &   504 &   504 &   L4$^{+ 12.0}_{-  7.0}$ &  15.7\\
     475 & 093821.34+342035.6                  &    DA &  0.37 &  0.70 &   201 &   223 &   L3$^{+  1.0}_{-  4.0}$ &   0.6\\
     520 & 100128.30+415001.6                  &    DA &  0.54 &  1.12 &   151 &   151 &   L4$^{+  1.0}_{-  2.0}$ &   1.8\\
     533 & 100646.07+413306.5                  & DBAH: &  0.63 &  0.38 &   192 &   192 &   L0$^{+  5.0}_{-  1.0}$ &  11.3\\
     553 & 101644.47+161343.5                  &    DA &  0.64 &  1.49 &   117 &   117 &   L5$^{+  1.0}_{-  1.0}$ &   1.2\\
     577 & 103047.25+443859.3                  &    DA &  0.58 &  0.98 &   276 &   276 &   L1$^{+  4.0}_{-  1.0}$ &   2.6\\
     594 & 104052.58+284856.7                  & DBAM: &  0.68 &  0.19 &   127 &   170 &   L1$^{+  4.0}_{-  4.0}$ &   1.8\\
     618 & 104933.58+022451.7                  &    DA &  0.28 &  0.69 &   228 &   228 &   L2$^{+  3.0}_{-  1.0}$ &   6.3\\
     664 & 111021.03+304737.4                  &   DAM &  0.41 &  0.68 &   106 &   110 &   L0$^{+  1.0}_{-  1.0}$ &   1.1\\
     670 & 111424.65+334123.7                  &   DAM &  0.59 &  1.03 &    92 &   101 &   L0$^{+  1.0}_{-  1.0}$ &   1.2\\
     690 & 112010.94+320619.6                  &    DA &  1.01 &  1.73 &   268 &   268 &   L4$^{+  1.0}_{-  3.0}$ &   1.2\\
     700 & 112541.71+422334.7\tablenotemark{a} &    DA &  0.75 &  0.95 &    61 &   125 &   L3$^{+  7.0}_{- 11.0}$ &   4.6\\
     707 & 113022.52+313933.4                  &   DAM &  0.72 &  0.77 &   173 &   173 &   L1$^{+  4.0}_{-  1.0}$ &   4.2\\
     709 & 113039.09-004023.0                  &    DC &  1.01 &  1.69 &   147 &   147 &   L5$^{+  6.0}_{-  2.0}$ &  23.5\\
     748 & 114827.96+153356.9                  &   DAH &  1.22 &  1.89 &   202 &   202 &   L4$^{+  1.0}_{-  2.0}$ &   1.5\\
     756 & 115612.99+323302.5                  &    DC &  0.20 &  0.26 &   203 &   349 &   L0$^{+  1.0}_{-  1.0}$ &   0.0\\
     758 & 115814.51+000458.7\tablenotemark{a} &    DC &  0.17 &  0.89 &    88 &   248 &   L1$^{+  9.0}_{-  7.0}$ &   8.7\\
     765 & 120144.90+505315.0                  &    DA &  1.00 &  2.01 &   137 &   176 &   L2$^{+  1.0}_{-  1.0}$ &   0.0\\
     875 & 125847.31+233844.2                  & DAHM: &  0.86 &  2.18 &   128 &   144 &   L4$^{+  1.0}_{-  2.0}$ &   1.5\\
    1056 & 142559.72+365800.7                  &   DAM &  1.00 &  2.64 &    75 &    75 &   T0$^{+  3.0}_{-  3.0}$ &  14.2\\
    1061 & 142833.77+440346.1\tablenotemark{a} &    DZ &  0.17 &  0.89 &    71 &    71 &   T6$^{+  1.0}_{- 10.0}$ &  19.2\\
    1066 & 143144.83+375011.8                  &    DQ &  0.19 &  0.69 &   190 &   190 &   L8$^{+  8.0}_{-  5.0}$ &   6.1\\
    1091 & 144307.83+340523.5                  &   DAM &  1.08 &  2.84 &    70 &   133 &   L8$^{+  8.0}_{- 18.0}$ &  61.0\\
    1135 & 150152.59+443316.6                  &   DAM &  0.99 &  1.83 &   181 &   181 &   L1$^{+  4.0}_{-  1.0}$ &  12.4\\
    1224 & 154221.86+553957.2                  &   DAM &  0.95 &  1.48 &   149 &   175 &   L0$^{+ 16.0}_{- 10.0}$ & 108.3\\
    1236 & 154833.29+353733.0                  &    DA &  0.68 &  1.27 &   150 &   163 &   L3$^{+  2.0}_{-  1.0}$ &   0.5\\
    1271 & 160153.23+273547.1\tablenotemark{a} &    DA &  0.58 &  1.51 &    66 &   157 &   L4$^{+  6.0}_{-  1.0}$ &   5.2\\
    1345 & 164216.62+225627.8                  &    DA &  0.74 &  0.47 &   114 &   114 &   L0$^{+  2.0}_{-  1.0}$ &  11.1\\
    1360 & 165629.94+400330.2                  &   DAM &  0.56 &  1.47 &   145 &   176 &   L3$^{+  2.0}_{-  2.0}$ &   2.2\\
    1400 & 172633.51+530300.7                  &    DC &  1.19 &  4.57 &    35 &    35 &   T1$^{+  1.0}_{-  2.0}$ &  46.5\\
    1463 & 221652.14+005312.8                  &    DC &  0.17 &  0.75 &   267 &   267 &   L5$^{+  3.0}_{-  3.0}$ &   3.6\\
    1479 & 223401.66-010016.3                  &  DAM: &  0.19 &  0.67 &   314 &   314 &   L4$^{+ 12.0}_{-  3.0}$ &   4.3\\
\enddata
\tablenotetext{a}{WD has a predicted $W1$ photospheric flux density $>50$ $\mu$Jy and is part of the flux limited sample.}
\end{deluxetable}

\begin{deluxetable}{rccccccccc}
\tabletypesize{\scriptsize}
\tablecolumns{10}
\tablewidth{0pc}
\tablecaption{WD+Dust Disk Candidates \label{tab:disk}}
\tablehead{
\colhead{\#} & \colhead{WIRED Name} & \colhead{Type} & \colhead{T$_{\rm eff}$} & \colhead{Mass}        & \colhead{Age}   & \colhead{D$_{\rm phot}$} & \colhead{R$_{\rm in}$}   & \colhead{i}     & \colhead{$\chi^2$} \\
             &                      &                & \colhead{(K)}           & \colhead{(M$_\odot$)} & \colhead{(Gyr)} & \colhead{(pc)}           & \colhead{(R$_{\rm WD}$)} & \colhead{(deg)} & 
}
\startdata
      43 & 011055.06+143922.2\tablenotemark{a} &    DA  &  9200 &   1.04 &   2.28 &    44 &   21 &    0 &   4.3\\
      54 & 012929.99+003411.2                  &    DA  & 16525 &   0.47 &   0.10 &   628 &    9 &    0 &  18.0\\
     104 & 024602.66+002539.2                  &    DA  & 14815 &   0.57 &   0.19 &   168 &   26 &    0 &   5.0\\
     107 & 025049.44+343651.0                  &    DA  & 21790 &   0.65 &   0.06 &   426 &   13 &    0 &  41.0\\
     124 & 030253.09-010833.7\tablenotemark{a} &    DB  & 15551 &   0.92 &   0.44 &    51 &   10 &   83 &  14.8\\
     134 & 031343.07-001623.3                  &   DAH  &  7579 &   1.27 &   3.07 &    66 &    3 &    0 &  39.7\\
     274 & 081308.51+480642.3                  &    DA  & 32727 &   0.59 &   0.01 &   279 &   55 &   45 &  19.0\\
     298 & 082125.22+153000.0                  &    DA  & 14074 &   0.55 &   0.21 &   519 &    7 &    0 &   4.8\\
     313 & 082624.40+062827.6                  &    DA  & 16149 &   0.61 &   0.15 &   207 &   12 &    0 &   9.4\\
     349 & 084303.98+275149.6                  &  DAE:  & 10430 &   1.34 &   1.85 &    65 &    5 &    0 &  16.0\\
     358 & 084539.17+225728.0\tablenotemark{a} & DB\_DB & 18621 &   0.56 &   0.09 &   111 &   10 &   80 &   0.7\\
     387 & 085742.05+363526.6                  &    DA  & 28932 &   0.55 &   0.01 &   552 &   19 &   59 &   1.3\\
     397 & 090344.14+574958.9                  &    DA  & 21668 &   0.59 &   0.05 &   862 &   13 &    0 &   1.4\\
     404 & 090522.93+071519.1                  &    DA  &  8693 &   0.70 &   1.31 &   134 &   10 &    0 &   3.3\\
     406 & 090611.00+414114.3                  &    DA  & 47637 &   0.65 &   0.00 &   469 &   38 &   59 &  10.0\\
     423 & 091411.11+434332.9                  &    DA  & 22621 &   0.56 &   0.04 &   820 &   14 &    0 &   3.8\\
     459 & 092528.22+044952.4                  &    DA  & 10261 &   0.80 &   1.03 &   125 &    9 &    0 &   5.1\\
     485 & 094127.02+062113.7                  &  DBH:  & 22878 &   0.43 &   0.03 &   706 &   14 &   50 &   5.7\\
     506 & 095337.97+493439.7                  &    DB  & 23109 &   0.56 &   0.04 &   605 &   20 &    9 &   1.0\\
     522 & 100145.03+364257.3                  &    DA  & 36977 &   0.33 &   0.00 &  2440 &   27 &    0 &  12.7\\
     541 & 101117.29+354004.8                  &    DA  & 13383 &   0.70 &   0.37 &   293 &    7 &    0 &   5.2\\
     589 & 103757.04+035023.6\tablenotemark{a} &    DC  &  5600 &   1.19 &   4.57 &    28 &    2 &    0 &  67.4\\
     605 & 104341.53+085558.2                  &    DA  & 17622 &   0.60 &   0.11 &   179 &   38 &    0 &   6.1\\
     725 & 113748.30-002714.6                  &    DA  & 17715 &   0.50 &   0.08 &   583 &   10 &    0 &   1.7\\
     747 & 114758.61+283156.2                  &    DA  & 12290 &   0.70 &   0.46 &   134 &   13 &   21 &   3.5\\
     793 & 122220.88+395553.9                  &    DA  & 11602 &   0.37 &   0.24 &   350 &    6 &    0 &   1.5\\
     801 & 122859.93+104032.9                  & DAZE:  & 22642 &   0.76 &   0.09 &   128 &   17 &   80 &   8.3\\
     815 & 123432.63+560643.0                  & DB\_DB & 13567 &   1.04 &   0.96 &   133 &    7 &   56 &   3.6\\
     843 & 124455.15+040220.6                  &    DA  & 65969 &   0.71 &   0.00 &  1096 &   59 &   56 &  10.0\\
     903 & 131641.73+122543.8                  &    DA  &  7444 &   0.48 &   1.06 &   165 &    6 &    0 &   2.8\\
     909 & 131849.24+501320.6                  &    DA  & 13305 &   0.59 &   0.27 &   481 &    6 &    0 &   5.2\\
     938 & 133212.85+100435.2                  &    DA  &  6979 &   0.66 &   1.99 &   127 &    6 &    0 &   6.1\\
     973 & 134800.05+282355.1                  &    DA  & 21594 &   0.56 &   0.05 &   923 &   13 &    0 &   9.0\\
    1025 & 141351.95+353429.6                  &    DA  & 12317 &   0.36 &   0.19 &   551 &    6 &    0 &  15.0\\
    1104 & 144823.67+444344.3                  &    DA  & 18188 &   0.37 &   0.05 &   267 &   41 &   39 &   0.1\\
    1127 & 145806.53+293727.0\tablenotemark{a} &    DA  &  7266 &   0.54 &   1.26 &    29 &   13 &   50 &   3.5\\
    1139 & 150347.29+615847.4                  &    DB  & 18006 &   0.53 &   0.10 &   693 &   10 &    0 &  39.1\\
    1150 & 150701.98+324545.1                  &    DA  &  7177 &   0.63 &   1.69 &   114 &    7 &    9 &   0.7\\
    1159 & 151200.04+494009.7                  &    DA  & 19527 &   0.49 &   0.05 &   532 &   22 &    0 &   2.1\\
    1168 & 151747.51+342209.7                  &    DA  & 22067 &   0.46 &   0.03 &   845 &   13 &    0 &   1.6\\
    1194 & 153017.00+470852.4                  &    DA  & 15479 &   0.39 &   0.09 &   703 &    8 &    0 &   4.2\\
    1198 & 153149.04+025705.0                  &   DAH  &  6557 &   0.19 &   0.62 &   247 &    3 &   86 &   4.0\\
    1213 & 153725.71+515126.9                  &   DBA  & 24926 &   0.73 &   0.06 &   199 &   16 &   65 &   7.0\\
    1220 & 154038.67+450710.0                  &    DA  &  8824 &   0.62 &   0.91 &   154 &    5 &    0 &   3.4\\
    1248 & 155206.11+391817.2                  &    DA  & 20040 &   0.53 &   0.05 &   945 &   12 &    0 &  13.1\\
    1251 & 155359.87+082131.3                  &    DA  & 16519 &   0.78 &   0.25 &   115 &   40 &    0 &   1.3\\
    1266 & 155955.27+263519.2\tablenotemark{a} &    DA  & 11890 &   0.78 &   0.63 &    67 &   17 &   74 &   2.9\\
    1308 & 161717.04+162022.3                  &    DA  & 12907 &   0.70 &   0.41 &   118 &    6 &   74 &   2.8\\
    1352 & 165012.47+112457.1                  &    DA  & 47909 &   0.47 &   0.00 &  1746 &   49 &    0 &   0.2\\
    1361 & 165747.02+624417.4                  &    DA  & 14241 &   0.55 &   0.20 &   351 &    8 &    0 &   1.0\\
    1467 & 222030.69-004107.3\tablenotemark{a} &    DA  &  7610 &   0.49 &   1.01 &    76 &    4 &   65 &   1.7\\
    1489 & 224626.38-005909.2                  &    DA  &  8717 &   0.59 &   0.86 &   115 &    9 &    0 &   1.9\\
% ultracool WD, moved to "naked WD" classification
%    1501 & 230645.72+212859.3                  & DAZCO  & 93333 &   0.52 &   0.00 &  1638 &   94 &   83 & 218.0\\
\enddata
\tablenotetext{a}{WD has a predicted $W1$ photospheric flux density $>50$ $\mu$Jy and is part of the flux limited sample.}
\end{deluxetable}

\begin{deluxetable}{rcccccccccc}
\tabletypesize{\scriptsize}
%\rotate
\tablecolumns{11}
\tablewidth{0pc}
\tablecaption{WDs with Indeterminate IR Excesses \label{tab:ind}}
\tablehead{
\colhead{\#} & \colhead{WIRED Name} & \colhead{Type} & \colhead{Mass}        & \colhead{Age}   & \colhead{D$_{\rm phot,WD}$} & \colhead{D$_{\rm phot,comp}$} & \colhead{Spectral Type} & \colhead{$\chi^2_{\rm comp}$} & \colhead{R$_{\rm in}$} & \colhead{$\chi^2_{\rm disk}$} \\
             &                      &                & \colhead{(M$_\odot$)} & \colhead{(Gyr)} & \colhead{(pc)}              & \colhead{(pc)}                &                         &                               & \colhead{(R$_{\rm WD}$)} & 
}
\startdata
       2 & 000410.42-034008.5\tablenotemark{a} &    DA &   0.45 &   1.19 &    51 &    51 &   T2 &   5.5 &   14 &   7.5 \\
       5 & 000641.08+273716.6                  &    DA &   0.44 &   0.03 &  1081 &  1081 &   M4 &  65.6 &   14 &  65.5 \\
       8 & 001306.21+005506.3                  &    DA &   0.57 &   0.44 &   318 &   318 &   L8 &  10.2 &    8 &   8.1 \\
      32 & 005438.84-095219.7\tablenotemark{a} &    DA &   0.62 &   0.94 &    63 &    63 &   T1 &   4.2 &   17 &   5.0 \\
      49 & 011616.94-094347.9                  &    DA &   0.63 &   2.32 &   104 &   104 &   L9 &   0.3 &    5 &   0.4 \\
      68 & 020227.39+141124.5                  &    DA &   1.11 &   2.01 &   175 &   175 &   L8 &  10.2 &    5 &   8.7 \\
     119 & 025801.20-005400.0                  &    DA &   0.57 &   0.71 &   132 &   132 &   T6 &   4.0 &   12 &   2.1 \\
     196 & 073018.35+411320.4                  &    DA &   0.53 &   0.16 &   133 &   133 &   L6 &   6.4 &   35 &   6.4 \\
     203 & 073707.99+411227.4\tablenotemark{a} &    DA &   0.75 &   0.73 &    51 &    51 &   T3 &   3.6 &   37 &   5.1 \\
     214 & 074631.42+173448.1                  &    DA &   0.97 &   2.00 &    66 &    66 &   T0 &   3.6 &   10 &   3.6 \\
     230 & 075144.05+223004.8                  &    DA &   0.57 &   0.06 &   175 &   175 &   L8 &   6.7 &   44 &   4.7 \\
     334 & 083632.99+374259.3                  &    DA &   0.66 &   1.56 &   116 &   116 &   L8 &   4.0 &    6 &   3.9 \\
     385 & 085650.57+275118.0                  &    DA &   0.56 &   0.07 &   428 &   428 &   M7 &  10.9 &   12 &  12.6 \\
     412 & 090911.36+501559.4                  &    DA &   0.83 &   1.40 &   123 &   123 &   L6 &   4.5 &    6 &   3.2 \\
     421 & 091312.73+403628.8                  &    DA &   0.62 &   0.42 &   153 &   153 &   L8 &   5.0 &   18 &   5.0 \\
     422 & 091356.83+404734.6\tablenotemark{a} &    DA &   0.38 &   1.09 &    64 &    64 &   T2 &   1.4 &   12 &   1.6 \\
     489 & 094422.33+552756.2                  &    DA &   0.51 &   0.30 &   389 &   389 &   M7 &  30.3 &    6 &  31.6 \\
     539 & 101007.88+615515.7                  &    DA &   0.79 &   2.72 &    94 &    94 &   L6 &   1.4 &    3 &   1.5 \\
     557 & 101951.55+290100.6                  &    DB &   0.63 &   0.06 &   413 &   413 &   M7 &   7.3 &   14 &   8.3 \\
     560 & 102100.91+564644.7                  & DBAQH &   0.67 &   0.09 &   312 &   312 &   L0 &   3.9 &   35 &   4.5 \\
     574 & 102915.97+300251.5                  &    DA &   0.52 &   1.03 &   153 &   153 &   L5 &   5.0 &    4 &   2.8 \\
     578 & 103112.73+444729.9                  &    DA &   0.52 &   1.08 &   109 &   109 &   T1 &   1.3 &   10 &   1.0 \\
     613 & 104659.78+374556.7                  &    DA &   0.57 &   0.06 &   183 &   183 &   L5 &   1.7 &   44 &   2.2 \\
     633 & 105824.34+512738.7                  &    DA &   0.69 &   0.42 &   213 &   213 &   L4 &   2.3 &    7 &   1.8 \\
     634 & 105827.97+293223.0                  &    DA &   0.60 &   0.17 &   229 &   229 &   L5 &   1.8 &   27 &   1.3 \\
     649 & 110745.39+651722.1                  &    DA &   0.79 &   1.21 &   321 &   321 &   L4 &   3.9 &    5 &   5.1 \\
     674 & 111603.77+494343.8                  &    DA &   0.51 &   0.00 &   668 &   668 &   M6 &   2.8 &   35 &   4.9 \\
     676 & 111609.81+284308.4                  &    DA &   0.41 &   0.89 &   173 &   173 &   L6 &   6.9 &    5 &   5.9 \\
     679 & 111706.70+184312.4                  &    DA &   0.56 &   0.10 &   209 &   209 &   L3 &   2.1 &   10 &   3.5 \\
     683 & 111753.51+263856.2                  &    DA &   0.60 &   1.21 &   125 &   125 &   L7 &   5.2 &    8 &   4.9 \\
     693 & 112105.79+375615.2\tablenotemark{a} &    DA &   0.81 &   1.33 &    33 &    33 &   T4 &   3.5 &   27 &   3.4 \\
     695 & 112310.05+584407.2                  &    DA &   0.56 &   1.80 &    97 &    97 &   L7 &   1.7 &    5 &   1.0 \\
     721 & 113630.78+315447.9                  &    DA &   0.60 &   0.12 &   305 &   305 &   M9 &   3.0 &   10 &   4.4 \\
     724 & 113728.31+204109.4\tablenotemark{a} &    DA &   0.50 &   1.04 &    76 &    76 &   T1 &   0.7 &   11 &   0.7 \\
     744 & 114701.01+574114.7                  &    DA &   0.63 &   0.44 &   337 &   337 &   M9 &   4.9 &    6 &   4.3 \\
     757 & 115745.89+063148.2\tablenotemark{a} &    DA &   0.47 &   0.86 &    83 &    83 &   T1 &   0.3 &   14 &   1.1 \\
     769 & 120504.19+160746.8\tablenotemark{a} &    DA &   0.86 &   1.36 &    69 &    69 &   T6 &   3.2 &   24 &   3.0 \\
     838 & 124256.48+431311.1                  &    DA &   0.91 &   0.87 &   209 &   209 &   L5 &   3.0 &    6 &   3.0 \\
     841 & 124359.69+161203.5                  &    DA &   0.52 &   1.06 &   108 &   108 &   L9 &   2.2 &    9 &   2.1 \\
     855 & 125037.75+205334.0                  &    DA &   0.70 &   0.63 &   252 &   252 &   L1 &   2.8 &    5 &   3.1 \\
     870 & 125733.64+542850.5\tablenotemark{a} &    DA &   1.03 &   2.98 &    32 &    32 &   T3 &   3.9 &   11 &   3.7 \\
     891 & 130957.59+350947.2\tablenotemark{a} &    DA &   0.78 &   0.79 &    40 &    40 &   T4 &   4.0 &   37 &   5.1 \\
     910 & 131951.00+643309.1                  &    DA &   0.59 &   0.02 &   177 &   177 &   L8 &   3.9 &   37 &   2.8 \\
     933 & 133100.61+004033.5                  &    DA &   0.56 &   0.14 &   418 &   418 &   M9 &   4.1 &   12 &   2.8 \\
     962 & 134333.64+231403.3                  &    DA &   0.63 &   0.62 &   147 &   147 &   L5 &   1.7 &    5 &   2.0 \\
    1011 & 140644.77+530353.1                  &    DA &   0.58 &   0.05 &   773 &   773 &   M6 &   4.9 &   13 &   5.2 \\
    1012 & 140723.04+203918.5                  &    DQ &   1.35 &   1.88 &    53 &    53 &   T1 &   6.2 &    5 &   6.8 \\
    1016 & 140945.23+421600.6\tablenotemark{a} &    DA &   0.77 &   1.11 &    30 &    30 &   T4 &   4.4 &   37 &   3.0 \\
    1018 & 141017.32+463450.1                  &    DA &   0.47 &   0.75 &   164 &   164 &   L7 &   0.8 &   10 &   0.8 \\
    1027 & 141448.24+021257.7                  &    DA &   0.55 &   1.04 &   111 &   111 &   L7 &   0.9 &    9 &   0.7 \\
    1031 & 141632.82+111003.9\tablenotemark{a} &    DA &   0.75 &   1.16 &    66 &    66 &   T6 &   4.3 &    9 &   2.5 \\
    1054 & 142539.74+010926.8                  &    DA &   0.48 &   0.04 &   672 &   672 &   M9 &   8.5 &   13 &   6.5 \\
    1071 & 143406.75+150817.8\tablenotemark{a} &    DA &   0.62 &   0.25 &    81 &    81 &   L7 &   1.7 &    8 &   2.3 \\
    1100 & 144754.40+420004.9                  &    DA &   0.52 &   0.79 &   136 &   136 &   L4 &   3.4 &    4 &   3.3 \\
    1106 & 144847.79+145645.7                  &    DA &   0.49 &   1.37 &    98 &    98 &   T1 &   2.2 &    7 &   2.1 \\
    1225 & 154224.94+044959.7                  &    DA &   0.44 &   0.21 &   641 &   641 &   M6 &  10.1 &    7 &  12.3 \\
    1233 & 154729.96+065909.5                  &    DZ &   0.20 &   0.25 &   199 &   199 &   T1 &   6.9 &    5 &   7.3 \\
    1264 & 155811.46+312706.4                  &    DA &   0.59 &   0.25 &   229 &   229 &   L5 &   3.9 &   19 &   3.6 \\
    1274 & 160241.44+332301.4                  &    DA &   0.39 &   0.87 &   122 &   122 &   T0 &   1.9 &   10 &   1.9 \\
    1278 & 160401.49+463249.5                  &    DA &   0.66 &   0.37 &   201 &   201 &   L5 &   5.7 &   14 &   5.6 \\
    1284 & 160715.80+134312.3                  &    DA &   0.99 &   2.43 &    63 &    63 &   T2 &   1.2 &   11 &   0.9 \\
    1288 & 160839.52+172336.9                  &    DA &   0.81 &   2.33 &    78 &    78 &   T1 &   0.4 &    9 &   0.3 \\
    1313 & 162139.79+481241.6                  &    DA &   0.97 &   2.95 &    90 &    90 &   L7 &   2.3 &    4 &   1.8 \\
    1325 & 162555.28+375920.6                  &    DA &   0.63 &   1.86 &    82 &    82 &   T1 &   0.6 &    4 &   0.7 \\
    1364 & 170144.73+624304.4\tablenotemark{a} &    DA &   0.71 &   1.35 &    52 &    52 &   T2 &   3.0 &    5 &   3.3 \\
    1409 & 173434.54+333521.3                  &    DA &   0.75 &   0.36 &   244 &   244 &   L0 &   8.2 &    8 &   8.6 \\
    1424 & 192433.15+373416.9                  &    DA &   0.61 &   0.63 &   154 &   154 &   L0 &   9.6 &    5 &  11.5 \\
    1426 & 192542.00+631741.6                  &    DA &   0.56 &   0.01 &   311 &   311 &   M9 &   7.8 &   51 &   6.3 \\
    1504 & 231725.28-084032.9                  &    DA &   0.30 &   0.79 &   124 &   124 &   L6 &   0.9 &    6 &   0.8 \\
\enddata
\tablenotetext{a}{WD has a predicted $W1$ photospheric flux density $>50$ $\mu$Jy and is part of the flux limited sample.}
\end{deluxetable}

\end{document}